\documentclass[pdflatex,sn-mathphys-num]{sn-jnl}

\usepackage{graphicx}%
\usepackage{multirow}%
\usepackage{amsmath,amssymb,amsfonts}%
\usepackage{amsthm}%
\usepackage{mathrsfs}%
\usepackage[title]{appendix}%
\usepackage{xcolor}%
\usepackage{textcomp}%
\usepackage{manyfoot}%
\usepackage{booktabs}%
\usepackage{algorithm}%
\usepackage{algorithmicx}%
\usepackage{algpseudocode}%
\usepackage{listings}%

\usepackage{enumitem}
\usepackage{graphicx}
\usepackage{subcaption}
\usepackage{url}
\usepackage{setspace}
\usepackage{adjustbox}

\theoremstyle{thmstyleone}%
\theoremstyle{thmstyletwo}%

\theoremstyle{thmstylethree}%

\begin{document}

\setstretch{1.2}

\newcommand{\hi}{H\,{\sc i}}
\newcommand{\hii}{H\,{\sc ii}}
\newcommand{\nii}{N\,{\sc ii}}
\newcommand{\lya}{Ly$\alpha$}
\newcommand{\jwst}{{\em JWST}}

\raggedbottom

\title[\textcolor{white}{X}]{\Large A massive, neutral gas reservoir permeating a galaxy proto-cluster after the reionization era}

\author[1,2,3]{Kasper~E.~Heintz}
\author[4,5,6]{Jake~S.~Bennett}
\author[3,1,2]{Pascal~A.~Oesch}
\author[1,2]{Albert~Sneppen}
\author[7]{Douglas~Rennehan}
\author[6,8]{Joris~Witstok}
\author[9]{Renske~Smit}
\author[1,2]{Simone~Vejlgaard}
\author[1,2]{Chamilla~Terp}
\author[10]{Umran~S.~Koca}
\author[1,2]{Gabriel~B.~Brammer}
\author[11,1,2]{Kristian~Finlator}
\author[12]{Matthew~J.~Hayes}
\author[5,6]{Debora~Sijacki}
\author[13]{Rohan~P.~Naidu}
\author[14]{Jorryt~Matthee}
\author[15,1]{Francesco~Valentino}
\author[16]{Nial~R.~Tanvir}
\author[17]{Páll~Jakobsson}
\author[1,2]{Peter~Laursen}
\author[1,2]{Darach~J.~Watson}
\author[18]{Romeel~Davé}
\author[18]{Laura~C.~Keating}
\author[3]{Alba~Covelo-Paz}


\affil[1]{\small Cosmic Dawn Center (DAWN), Denmark}
\affil[2]{\small Niels Bohr Institute, University of Copenhagen, Jagtvej 128, DK-2200 Copenhagen N, Denmark}
\affil[3]{\small Department of Astronomy, University of Geneva, Chemin Pegasi 51, 1290 Versoix, Switzerland}
\affil[4]{\small Center for Astrophysics, Harvard University, 60 Garden Street, Cambridge, MA 02138, USA}
\affil[5]{\small Institute of Astronomy, University of Cambridge, Madingley Road, Cambridge CB3 0HA, UK}
\affil[6]{\small Kavli Institute for Cosmology Cambridge, University of Cambridge, Madingley Road, Cambridge CB3 0HA, UK}
\affil[7]{\small Center for Computational Astrophysics, Flatiron Institute, 162 Fifth Ave, New York, NY 10010, USA} 
\affil[8]{\small Cavendish Laboratory, University of Cambridge, 19 JJ Thomson Avenue, Cambridge CB3 0HE, UK}
\affil[9]{\small Astrophysics Research Institute, Liverpool John Moores University, 146 Brownlow Hill, Liverpool L3 5RF, UK}
\affil[10]{\small California Institute of Technology, Pasadena, CA 91125, USA}
\affil[11]{\small New Mexico State University, Las Cruces, 88003 NM, USA}
\affil[12]{\small Stockholm University, Department of Astronomy and Oskar Klein Centre for Cosmoparticle Physics, AlbaNova University Centre, SE-10691, Stockholm, Sweden}
\affil[13]{\small MIT Kavli Institute for Astrophysics and Space Research, Cambridge, MA 02139, USA}
\affil[14]{\small Institute of Science and Technology Austria (ISTA), Am Campus 1, 3400 Klosterneuburg, Austria}
\affil[15]{\small European Southern Observatory, Karl-Schwarzschild-Str. 2, 85748 Garching, Germany}
\affil[16]{\small Department of Physics \& Astronomy and Leicester Institute of Space \& Earth Observation, University of Leicester, University Road, Leicester LE1 7RH, UK}
\affil[17]{\small Centre for Astrophysics and Cosmology, Science Institute, University of Iceland, Dunhagi 5, 107 Reykjavik, Iceland}
\affil[18]{\small Institute for Astronomy, Royal Observatory, University of Edinburgh, Edinburgh EH9 3HJ, UK}


\abstract{\bf 
Galaxy clusters are the most massive, gravitationally-bound structures in the Universe, emerging through hierarchical structure formation of large-scale dark matter and baryon overdensities. Early galaxy ``proto-clusters'' are believed to be important physical drivers of the overall cosmic star-formation rate density and serve as ``hotspots'' for the reionization of the intergalactic medium \cite{Chiang17}.  
Our understanding of the formation of these structures at the earliest cosmic epochs is, however, limited to sparse observations of their galaxy members \cite{Overzier16,Harikane19,Hu21,Helton23,Morishita23}, or based on phenomenological models and cosmological simulations \cite{Lovell21,Springel21}.  
Here we report the detection of a massive neutral, atomic hydrogen (\hi) gas reservoir permeating a galaxy proto-cluster at redshift $z=5.4$, observed one billion years after the Big Bang. The presence of this cold gas is revealed by strong damped Lyman-$\alpha$ absorption features observed in several background galaxy spectra taken with \jwst/NIRSpec in close on-sky projection. 
While overall the sightlines probe a large range in \hi\ column densities, $N_{\rm HI} = 10^{21.7}-10^{23.5}$\,cm$^{-2}$, they are similar across nearby sightlines, demonstrating that they probe the same dense, neutral gas. 
This observation of a massive, large-scale overdensity of cold neutral gas challenges current large-scale cosmological simulations and has strong implications for the reionization topology of the Universe. 
}


\maketitle

\section*{Main Text}\label{sec1}

The neutral, atomic hydrogen (\hi) gas content in the early Universe has historically been probed via absorption-spectroscopy of the ground-state Lyman-$\alpha$ (\lya) transition in bright quasar or gamma-ray burst sightlines at redshifts $z=2-6$ \cite{Wolfe05,Prochaska09,Noterdaeme12,Fynbo09,Tanvir19,Heintz23_GRB}. More recently with the advent and exquisite near-infrared sensitivity and capabilities of the {\em James Webb Space Telescope} (\jwst), extremely strong damped \lya\ absorption (DLA) features have been detected in spectra of entire galaxies at redshifts $z>8$ \cite{Heintz24_DLAs}, corresponding to $600$~Myr after the Big Bang. From a search for and characterization of these DLA galaxies based on spectroscopic observations with the \emph{JWST} Near-Infrared Spectrograph (NIRSpec) Prism-mode \cite{Heintz24} we identified ten galaxies at $z=5.4-6.0$, all showing extreme DLA features. One particular source was identified to have the majority of the absorbing gas in its distant foreground \cite{Terp24}, with an absorption-redshift consistent with an overdense region of foreground galaxies at $z\approx 5.4$ reported in the same field \cite{Helton23} and in close projected distance to the background galaxy. We thus sought to investigate the DLA features observed in the ten background galaxies in more detail and characterize the baryonic components of this rare, massive galaxy proto-cluster. Charting the abundance and distribution of the cold, dense gas around these early galaxy overdensities provides key insight into their formation, star-formation history and their impact on the ionizing photon output believed to drive the large-scale reionization of the Universe.    

To characterize the galaxy members of the proto-cluster first detected in the field by Helton et al. \cite{Helton23} we use \jwst\ NIRCam slitless grism observations obtained through the ``First Reionization Epoch Spectroscopically Complete Observations'' (FRESCO) survey  (GO-1895, PI: Oesch; \cite{Oesch23}). This survey is designed to cover two extragalactic legacy fields with unprecedented near-infrared spectroscopic depth at $4-5\mu$m in the F444W filter. We identify the same overdensity, including $\gtrsim 25$ galaxies at $z=5.35-5.40$ within a projected physical distance of $1\,$pMpc (Methods), centred approximately at the celestial coordinates R.A. = $3^{\rm h}32^{\rm m} 20^{\rm s}$, Decl. = $-27^\circ 49^\prime 50^{\prime \prime}$. The redshift of the proto-cluster members are accurately measured by first imposing a photometric prior and then requiring at least $5\sigma$ detections of H$\alpha$, with occasional simultaneous detections of [\nii], which is covered throughout redshifts $z=4.9-6.6$ by design. The identified proto-cluster galaxy members span a large range in brightness and mass, with absolute UV magnitudes from $M_{\rm UV} = -18.0$ up to $-21.5$\,mag (Methods). The total estimated halo mass of the galaxy proto-cluster system is $\log (M_{\rm halo}/M_\odot) = 12.6-12.8$, obtained through stellar-to-halo mass abundance matching \cite{Helton23}. 

\begin{figure}
\centering
\includegraphics[width=\textwidth]{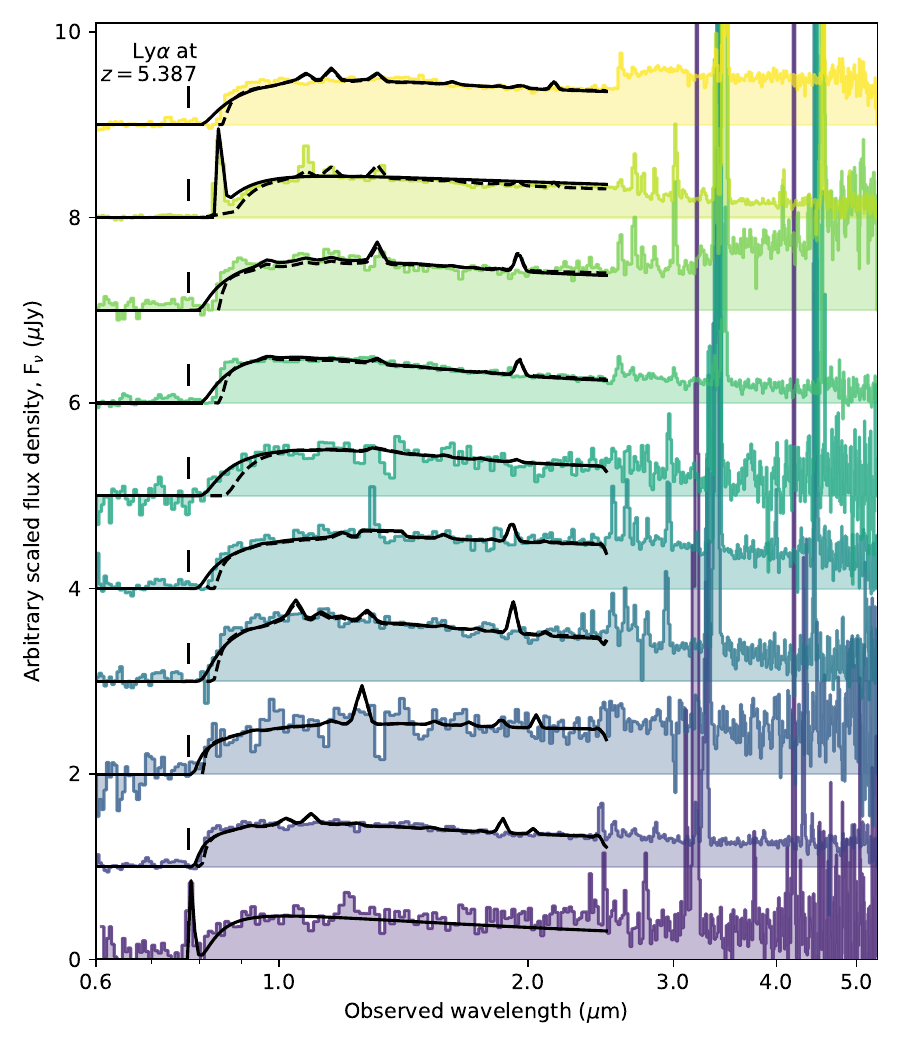}
\caption{\jwst/NIRSpec Prism Spectra of the ten galaxies with apparent strong DLAs, in close on-sky proximity to the massive galaxy proto-cluster at redshift $z=5.387$. The best-fit models for each spectrum are shown by the solid black lines, using the high-redshift galaxy template spectra as the intrinsic continuum models, and the models with fixed $z_{\rm abs}=z_{\rm gal}$ are shown by the dashed black lines. }
\label{fig:fig1}
\end{figure}

The ten galaxies in the background to this galaxy proto-cluster with strong DLA features have all been observed with the \jwst/NIRSpec Prism configuration, mainly as part of the \jwst\ Advanced Deep Extragalactic Survey (JADES, see Table~1 and Methods). The reduced and processed 1D spectra used in this study were obtained from the DAWN \jwst\ Archive (DJA)\footnote{\url{https://dawn-cph.github.io/dja/}} \cite{Heintz24} and are all photometrically-calibrated to the available \jwst/NIRCam photometry and corrected from slitloss according to the exact source position in the shutter (Methods). 
The final reduced 1D spectra for each of the ten sources with strong DLAs in the field are shown in Fig.~\ref{fig:fig1}. 
For each galaxy, we determine their redshifts based on the multitude of nebular emission lines detected in the spectra. Since we can pin-point the exact background-galaxy redshifts, we are able to accurately model the \lya\ transmission of each individual source. 
We model the intrinsic galaxy continuum at rest-frame UV wavelengths ($1200-2600\,\AA$) covered by the \jwst/NIRSpec Prism spectra using three approaches: assuming either a standard spectral power-law, $F_\lambda \propto \lambda^{-\beta}$, a representative galaxy template model, or via direct spectro-photometric modelling of the spectral energy distribution, all convolved with the spectral resolution of the NIRSpec Prism configuration (Methods). We find that the results are insensitive within 0.1\,dex to the exact choice of the three model versions of the intrinsic spectra. This is likely because the broad \lya\ absorption trough and strong damping feature dominates over small potential variations in the intrinsic spectra. Throughout the work, we report the results using the high-redshift galaxy spectral templates \cite{Larson23} as the intrinsic continuum model. 

We model the optical depth of \lya\ and quantify the \hi\ column density, $N_{\rm HI}$, with a Voigt profile, 
following the approximation of \cite{TepperGarcia06}. For large \hi\ column densities, the Lorentzian wings dominate the shape of the observed \lya\ absorption line profile such that the optical depth, $\tau$, is only sensitive to the column density of neutral atomic hydrogen. We further assume that there is a negligible damping effect from the neutral hydrogen fraction of the intergalactic medium (IGM) at these redshifts, since the large-scale reionization of the Universe is expected to be fully complete by then \cite{Bosman22}. Even in the unlikely scenario where these galaxies probe a fully neutral IGM, the observed \lya\ damping features are not consistent with any IGM transmission models since the derived column densities produce stronger damping wings than expected for a fully neutral IGM (Methods). We first model the DLA features assuming that the majority of the absorption originates in the interstellar or circumgalactic medium of the background galaxies, $z_{\rm abs}=z_{\rm gal}$. However, as demonstrated in Fig.~\ref{fig:fig1}, none of the target background galaxies have DLA profiles consistent with the majority of the neutral, atomic hydrogen gas being located at the source location. Generally, the spectra show DLA features with very shallow roll-overs, inconsistent with a DLA line profile with any given \hi\ column density at the source redshift.   

\begin{figure}
\centering
\includegraphics[width=14cm]{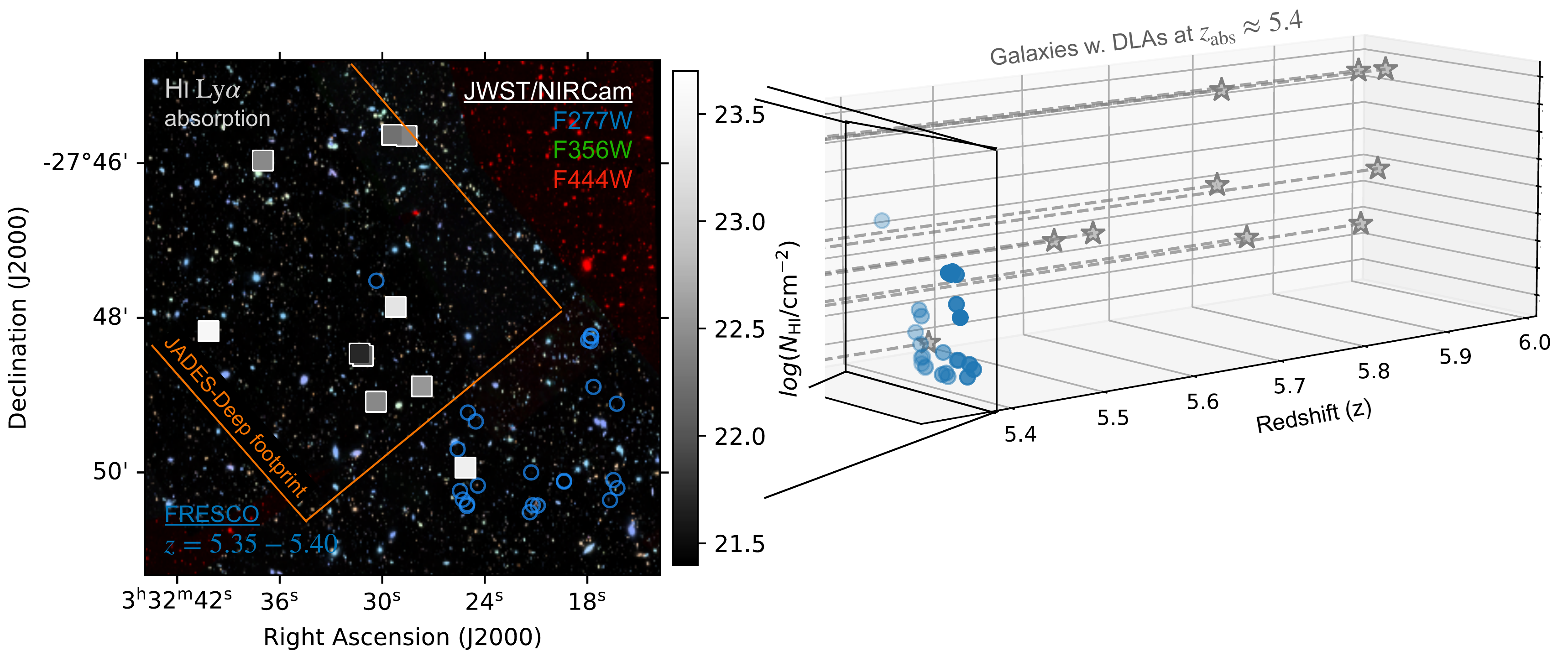}
\caption{Spatial and redshift distribution of the galaxy proto-cluster members and the background galaxies with strong DLAs. The cluster members identified from the FRESCO survey at $z=5.35-5.40$ are shown by open blue circles, and the background galaxies with grey star symbols. Their impact parameters are shown in the left \jwst/NIRCam image, color-coded according to the \hi\ column densities derived at the cluster redshift. }
\label{fig:fig2}
\end{figure}

The model posteriors on $z_{\rm abs}$ seem to independently converge on a foreground, lower-redshift, solution and for the majority of cases exclude the galaxy redshifts at $>3\sigma$ confidence (Methods). 
To quantify whether the cluster redshift provides a better solution to the models, we integrate over the likelihood levels from the Nested Sampling analysis to determine the Bayes factor based on the fit evidence for both models, assuming either $z_{\rm abs}=z_{\rm gal}$ or $z_{\rm abs} = 5.387$, the median of the galaxy proto-cluster redshift. For seven out of the total ten galaxies with strong DLA features, we determine Bayes factors of $>100$ as summarized in Table~\ref{tab:props}, thus with very strong evidence in favor of the galaxy proto-cluster redshift over the galaxy redshifts. For two out of ten, we determine Bayes factors $77$ and $99$, still showing very strong evidence for the cluster redshift. For one case, the galaxy at $z_{\rm gal}=5.6166$ with ID 13877, we do not find evidence for preferring one model over the other, though the close association and on-sky proximity to another sightline probing similar foreground neutral gas abundances favors the cluster-redshift solution. At the proto-cluster redshift, we measure substantial integrated \hi\ column densities in the range $N_{\rm HI} = 10^{21.7}-10^{23.5}\,$cm$^{-2}$, revealing the presence of an extended neutral, dense region permeating and surrounding the galaxy proto-cluster extending out to $\sim 2$\,pMpc as illustrated in Fig.~\ref{fig:fig2}. 

\begin{figure}
\centering
\includegraphics[width=\textwidth]{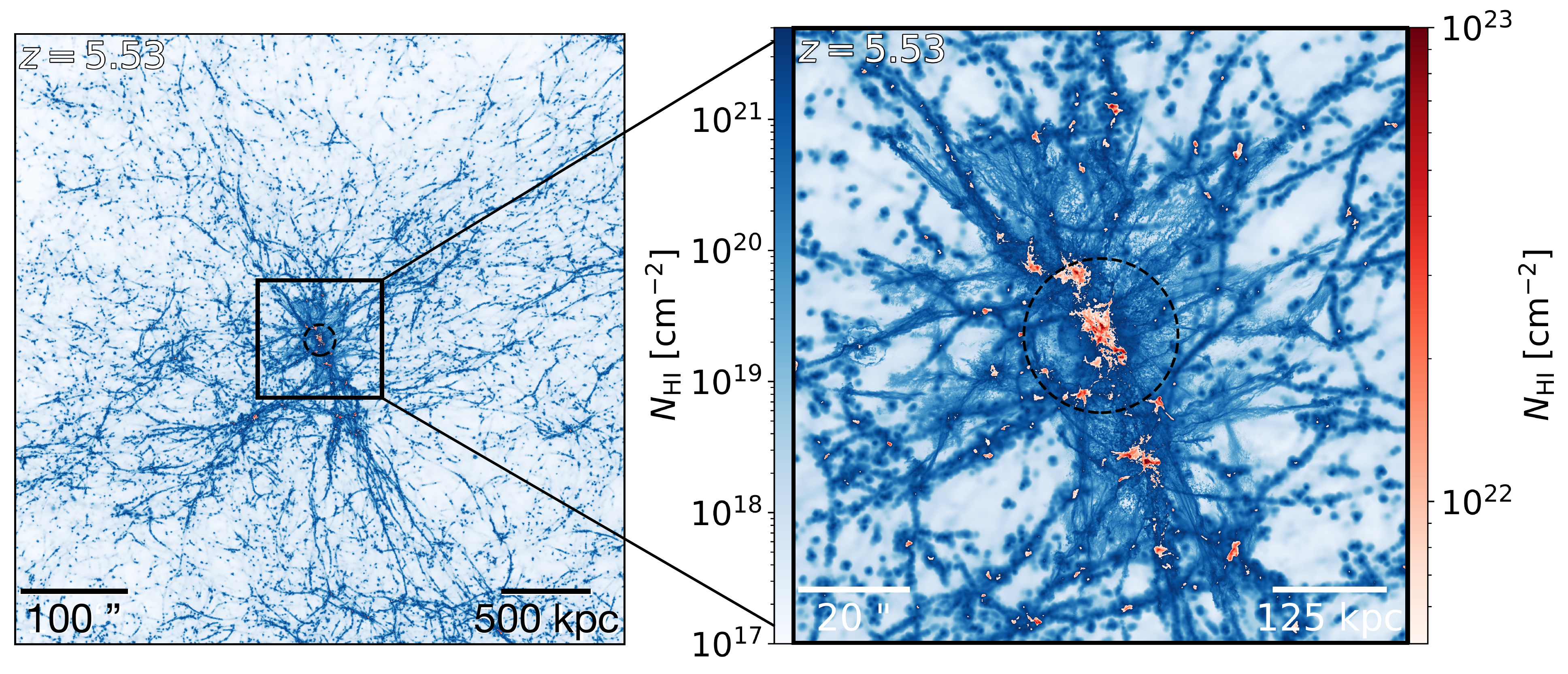}
\caption{Simulated density maps of \hi\ column density, $N_{\rm HI}$, for a proto-cluster with halo mass, $M_{\rm h,200} = 5.5\times 10^{12}\,M_\odot$ at $z=5.53$ \citep{Bennett20}. The width of the image on the left is $20\times R_{200}$ ($\sim 1.6$\,pMpc), representing approximately the scales of the probed galaxy absorption sightlines. The dashed circle represents the radius at which the spherically averaged mass density reached 200 times the critical mass density ($R_{200}$).}
\label{fig:fig3}
\end{figure}

An independent observational test of the hypothesis of a foreground neutral gas reservoir relies on the spatial distribution of sightlines and their corresponding \hi~column densities. The patchy and filamentary structure found in simulations would suggests little-to-no correlation for random sightlines drawn on Mpc or even hundreds of kpc scales, while scales below 100\,kpc should trace similar parts of the foreground absorber (see Fig.~\ref{fig:fig3} and Methods). Indeed, this is the pattern displayed by the spatial distribution of the observed column-densities with: i) no overall radial structure, but instead patchy regions with variations of $N_{\rm HI}$ up to 1.8 orders in magnitude, and ii) large degree of coherence between galaxies with small projected distance (see Methods and Extended Figs.~4 and 5). Specifically, the two galaxy-pairs with smallest angular separation (ID; 13577 and 13877, 13618 and 13620) while separated by several Mpc in radial distance are within respectively $20$ and $70$~kpc in their projected angular diameter distance at $z=5.387$. These galaxies display similar column densities (i.e. consistent within a factor of 2.6 and 1.3), which is more homogeneous than the two order-of-magnitude variations seen across sightlines -- and up to ten-fold smaller fractional variation than the median variation between sightlines with projected distance above $100$~kpc. This strongly disfavors the interpretation of multiple DLAs along similar line-of-sights as these systems are distinct (in absolute distance) but must be hidden behind a similar gas-reservoir. The similarity in inferred column density is, however, naturally explained by the small projected distance at the proto-cluster redshift. An alternative scenario, given that the median absorption redshift is $z_{\rm abs} = 5.53^{+0.19}_{-0.23}$, where the uncertainties denote the 16th to 84th percentiles of the distribution (Methods), could be that the galaxy sightlines are probing a separate, physically disassociated proto-cluster at $z\approx 5.5$. This would, however, require that all the galaxy members are below the line-detection limit of FRESCO, in contrast to the cluster at $z=5.4$, since no overdensity of galaxies are observed in that region and redshift interval, or that we are observing it in a young, nascent stage. 

To determine an approximate upper bound on the mass of the cold, neutral gas connected to this proto-cluster we assume that it can be physically modelled as a simple circular disk in the source plane, with $M_{\rm HI,tot} = \langle N_{\rm HI}\rangle \pi R^2_{\rm HI,max}f_{\rm cov,HI}$. For the derived mean $\langle N_{\rm HI}\rangle = 10^{22.5}$\,cm$^{-2}$ and maximum observed radial size of the \hi\ gas, $R^2_{\rm HI,max} = 2$\,pMpc, and assuming a 10\% covering fraction $f_{\rm cov,HI} = 0.1$, this yields $M_{\rm HI,tot} = 3\times 10^{14}\,M_\odot$. This corresponds to an overdensity of $\delta = 10^{4}$ from the cosmic baryonic average at $z\approx 5$ \cite{Walter20}, three orders of magnitude in excess of the predicted galaxy overdensity \cite{Helton23}. Due to the limited set of galaxy sightlines, likely only covering one part of the predicted filament structure or cold gas stream, and the expected large density variations of the gas it is impossible to observationally constrain the full covering fraction, $f_{\rm cov,HI}$, both in the line-of-sight and projected on the sky. 
 
To understand the covering fraction, infall and the physical state of the cold gas surrounding these early galaxy proto-clusters, we employ a range of simulations (Methods) that allow us to probe the neutral gas permeating such massive galaxy overdensities \cite{Bennett20,Rennehan24}. Such high-column density systems are not captured in reionization simulations designed to model the IGM, as they use simplified galaxy formation models which turns all dense gas into stars and thereby only predict the large-scale ionization fields around these massive halos \cite{Keating24}. Other cosmological radiation hydrodynamic simulations \cite{Finlator18,Witten24} do not sample large enough simulation boxes and therefore cannot probe the correct sample of halo assembly histories that would produce those baryonic mass overdensities \cite{Rennehan24}. 

Fig.~\ref{fig:fig3} shows \hi\ column densities of the neutral gas permeating a simulated galaxy proto-cluster at $z=5.5$ from a specific set of simulation designed to probe these overdensities, with a halo mass $M_{\rm h,200} = 5.5\times 10^{12}\,M_\odot$ \cite{Bennett20}. This was selected to match the estimated halo mass of the observed proto-cluster \cite{Helton23}. Projected over the entire depth of the zoom region of the simulation ($\sim 2$\,pMpc), similar to the observed scales probed by the galaxy absorption sightlines, we extract a total \hi\ gas mass of $M_{\rm HI} = 1.4\times 10^{13}\,M_\odot$ or $M_{\rm HI,R_{200}} = 4.6\times 10^{11}\,M_\odot$ within the radius at which the spherically averaged mass density reached 200 times the critical mass density ($R_{200}$). For this particular zoom-in simulation, only a small fraction $\approx 2\%$ of sightlines are expected to probe column densities $N_{\rm HI} > 10^{21}\,$cm$^{-2}$, with less than a permille at $N_{\rm HI} > 10^{22}\,$cm$^{-2}$. However, if we modify the simulation to the extreme and assume that all the gas cells are fully neutral, we find that nearly every sightline probes gas with column densities $N_{\rm HI} > 10^{21}$\,cm$^{-2}$ though still rarely with $N_{\rm HI} > 10^{22}$\,cm$^{-2}$ (0.1\% of the pixels). This shows that while there is sufficient gas around such a halo to agree with the observational inference, matching the number of high \hi\ columns found would require a significantly higher neutral fraction than is seen in the fiducial simulation. 

Considering other simulations that specifically targets rare, massive galaxy proto-clusters ($M_{\rm vir}\sim 10^{14}\,M_\odot$ at $z=2$) \cite{Rennehan24}, we find that the covering fraction of the cold, neutral gas can be up to $15\%$ at $z=5.4$ with several sightlines that probe similarly dense gas ($N_{\rm HI}>10^{21.7}\,$cm$^{-2}$) in multiple filaments as observed here (Methods). Compared to the default simulation, these $100$ proto-clusters have almost an order-of-magnitude higher halo mass at the target redshift $z = 5.4$. 
Importantly, we show that the viewing angle and exact column density cut impacts the covering fraction significantly. If we consider the sightlines with column densities above $N_{\rm HI}>10^{21}\,\mathrm{cm}^{-2}$, the mean covering fraction increases dramatically (with wide scatter) across all of the simulated proto-clusters -- but still depends heavily on viewing angle. Therefore, correctly sampling the assembly history distribution and capturing ``extremely rare" objects is paramount in interpreting the observations. By comparing our results with these rare overdensities recovered in simulations suggest that we are observing the formation and assembly of the primary baryonic component into one of the most massive structures in the present-day Universe.

This study provides the first direct observational evidence for dense, neutral hydrogen gas environment connected to a massive galaxy proto-cluster at $z=5.4$ shortly after the end of the reionization era, just one billion years after the Big Bang. There has been growing evidence for an increasing cold gas component in the overall baryon budget of galaxy proto-clusters with redshifts \cite{Angelinelli23}, and past simulations showed that early galaxy-cluster structures should be dominated by cold-gas accretion \cite{Daddi22}. Observations of galaxies and quasars in overdensities at lower redshifts, $z\sim 2-3$, have also hinted at dense, extended neutral gas environments but at substantially lower column densities than observed here. Extending this type of study to the reionization epoch have further previously been limited by the small volume population of bright quasars and the lack of sensitive near-infrared spectroscopy to cover the redshifted rest-frame continuum around the \lya\ transition at redshifts $z\gtrsim 5$. With the advent of \jwst, we are now able to identify large numbers of galaxy proto-clusters at rest-frame optical wavelengths and map their \lya\ properties. As demonstrated here, this provides the required sensitivity to chart the large-scale cold, neutral gas directly in absorption using the more numerous population of background galaxies as cosmic backlights. This is particularly important given that even extremely deep observations targeting the \lya\ emission from the neutral gas on large scales around early galaxy or quasar overdensities are not sensitive enough to detect this gas in emission \cite{Daddi21}. 
Our results highlight the often overlooked fact that regions of high ionization fraction is a necessary, but not sufficient condition for the escape of ionizing photons from galaxies or galaxy clusters \cite{Leonova22}; for such massive cold gas reservoirs as seen here, the total amount of neutral gas may still be high enough to impede escape. If such massive neutral gas reservoirs are common in and around early galaxy overdensities, this has strong implications for the cosmological evolution and topology of reionization. This would greatly change how the tomography of reionization is reconstructed in large cosmological simulations, largely driven by galaxy overdensities, and our understanding of this last major phase transition of the Universe.


\backmatter

\bmhead{Data availability}

The JWST imaging and spectroscopic data are publicly available on the JWST MAST archive at \url{https://mast.stsci.edu}. The data has been processed using public software codes; {\tt grizli} v. 1.9.11 \cite{Brammer23_grizli} and {\tt MsaExp} v. 0.6.17 \cite{Brammer23_msaexp}. The reduced spectroscopic data are all available on the DAWN \jwst\ Archive (DJA)\footnote{\url{https://dawn-cph.github.io/dja/}}.

\bmhead{Acknowledgments}
This work has received funding from the Swiss State Secretariat for Education, Research and Innovation (SERI) under contract number MB22.00072.
The Cosmic Dawn Center (DAWN) is funded by the Danish National Research Foundation under grant DNRF140.
The data products presented herein were retrieved from the DAWN JWST Archive (DJA). DJA is an initiative of the Cosmic Dawn Center, which is funded by the Danish National Research Foundation under grant DNRF140.
This work is based on observations made with the NASA/ESA/CSA James Webb Space Telescope. The data were obtained from the Mikulski Archive for Space Telescopes at the Space Telescope Science Institute, which is operated by the Association of Universities for Research in Astronomy, Inc., under NASA contract NAS 5-03127 for JWST. J.S.B. acknowledges support from the Simons Collaboration on ‘Learning the Universe’. J.S.B.'s simulations used resources from the Cambridge Service for Data Driven Discovery (CSD3) operated by the University of Cambridge Research Computing Service (www.csd3.cam.ac.uk), provided by Dell EMC and Intel using Tier-2 funding from the Engineering and Physical Sciences Research Council (capital grant EP/P020259/1).
K.F.'s simulation utilized resources from the New Mexico State University High Performance Computing Group, which is directly supported by the National Science Foundation (OAC-2019000), the Student Technology Advisory Committee, and New Mexico State University and benefits from inclusion in various grants (DoD ARO-W911NF1810454; NSF EPSCoR OIA-1757207; Partnership for the Advancement of Cancer Research, supported in part by NCI grants U54 CA132383 (NMSU)). K.F. also gratefully acknowledges support from the National Science Foundation (NSF) under Award Number 2006550. M.J.H. is fellow of the Knut \& Alice Wallenberg Foundation. D.S. acknowledges support from the Science and Technology Facilities Council (STFC). U.S.K. was partially funded by the Summer Undergraduate Research Fellowships program at Caltech.

\bmhead{Author contributions}

K.E.H. wrote the manuscript and led the analysis. G.B. reduced and extracted the photometric and spectroscopic data. R.S. and J.W. identified the first evidence for strong damped Lyman-$\alpha$ absorption in the target galaxies. P.A.O. led the FRESCO observations and analysis of the galaxy proto-cluster members. S.V., C.T., and U.K. performed the Lyman-$\alpha$ modelling of the spectra. J.B. and D.R. performed the simulations and extracted the relevant data. A.S. did the statistical clustering analysis. All authors contributed to the text and interpretations of the results. 

\bmhead{Competing interests}

The authors declare no competing interests. 

\clearpage
\newpage

\clearpage
\newpage

\clearpage
\newpage

\begin{table}[h]
\begin{minipage}{40pt}
\caption{Positions, source IDs, and line-of-sight absorption of the ten background galaxies.}%
\begin{tabular}{@{}lccccccc@{}}
\toprule
R.A. (J2000) & Decl. (J2000) & Prog. ID & Source ID & $z_{\rm gal}$ & $z_{\rm abs}$ & $\log (N_{\rm HI}/{\rm cm^{-2}})$ & Bayes factor \\
&&&&&&& ($z_{\rm abs}=5.387$ / \\
&&&&&&& $z_{\rm abs}=z_{\rm gal}$) \\
\midrule
53.167303  & $-27.802870$ & 1210 & 5113 & 5.8215 & $5.24\pm 0.16$ & $23.55\pm 0.20$ & 76  \\
53.154087 & $-27.766062$ & 1210 & 9842 & 5.8051 & $5.60\pm 0.10$ & $22.45\pm 0.05$ & 168 \\
53.121759 & $-27.797633$ & 1210 & 13176 & 5.9440 & $5.4\pm 0.10$ & $23.26\pm 0.02$ & 629 \\
53.1297219 & $-27.808177$ & 1210 & 13577 & 5.5711 & $5.46\pm 0.10$ & $22.11\pm 0.03$ & 347 \\
53.119112 & $-27.760802$ & 1210 & 13618 & 5.9482 & $5.77\pm 0.10$ & $22.38\pm 0.03$ & 368 \\
53.130574 & $-27.807725 $ & 1210 & 13877 & 5.6166 & $5.55\pm 0.15$ & $21.70\pm 0.25$ & 0.1 \\ 
53.11537 & $-27.814767$ & 1210 & 15099 & 5.7725 & $5.61\pm 0.10$ & $22.55\pm 0.05$ & 99 \\
53.126535 & $-27.818092$ & 1210 & 13704 & 5.9315 & $5.65\pm 0.10$ & $22.44\pm 0.05$ & 220 \\
53.12259 & $-27.760569$ & 1210 & 13620 & 5.9204 & $5.74\pm 0.10$ & $22.25\pm 0.03$ & 613 \\
53.104756 & $-27.832291$ & 2198 & 7807 & 5.3960 & $5.30\pm 0.10$ & $23.40\pm 0.10$ & -- \\
\botrule
\end{tabular}
\footnotetext{{\bf Notes.} The spectroscopic redshift, $z_{\rm gal}$, is derived from the nebular emission lines. $z_{\rm abs}$ lists the best-fit Ly$\alpha$ absorption redshift, where the column density, $N_{\rm HI}$, assumes the mean of the proto-cluster redshift $z_{\rm abs}=5.387$. The Bayes factor for this model over the intrinsic galaxy model is provided in the last column. ${\rm BF}>100$ is highly significant, $70-100$ is very significant. }
\label{tab:props}
\end{minipage}
\end{table}

\clearpage
\newpage

\clearpage
\newpage

\section*{Methods}

\bmhead{Cosmology}

Throughout this work, we assume a standard flat $\Lambda$CDM cosmological model based on the most recent {\it Planck} measurements \cite{Planck18}, with a Hubble constant $H_0 = 67.4\,{\rm km\,s^{-1}\,Mpc^{-1}}$, matter density, $\Omega_{\rm m} = 0.315$, and dark energy density, $\Omega_{\rm \Lambda} = 0.685$. Cosmological measurements such as the co-moving and proper distances, luminosity distances, and age of the Universe are based on the measured redshifts $z$ and computed using {\tt Astropy} \cite{astropy}. 

\bmhead{Observations and data reduction}

Nine out of the ten identified background galaxies were observed as part of the deep JWST Deep Extragalactic Survey (JADES, prog. ID: 1210) survey \cite{Bunker23}, and the additional one source observed as part of Prog. ID 2198 \cite{Barrufet24}, as summarized in Table~\ref{tab:props}. All galaxies were observed with the \jwst/NIRSpec low-resolution ($\mathcal{R}\approx 100$) Prism configuration \cite{Jakobsen22}, covering wavelengths from $0.6-5.3\,\mu$m. 

The ten galaxies were reduced and post-processed using {\tt MsaExp} v. 0.6.17 \cite{Brammer23_msaexp} and obtained from the DAWN \jwst\ Archive (DJA)\footnote{\url{https://dawn-cph.github.io/dja/}}. The details on the spectroscopic pipeline for the reductions and post-processing are outlined in Heintz et al. \cite{Heintz24}. Briefly, DJA utilizes the Stage 2 output from the Mikulski Archive for Space Telescopes (MAST) JWST archive and performs standard wave-length, flat-field and photometric calibrations on the individual NIRSpec exposure files. {\tt MsaExp} corrects the inverse-noise and the bias levels in individual exposures and perform a slit-loss correction based on the morphology and position with the shutter of the source. This correction greatly improves the overall flux calibration of the spectra, both in terms of colors and absolute flux densities, to within $\approx 15\%$ \cite{Heintz24_DLAs,Schaerer24}. The final 1D spectra were generated by optimally extracting the trace from the 2D spectra using an inverse-weighted sum of the 2D spectra in the dispersion direction \cite{Horne86}. The final 1D spectra are shown in Fig.~\ref{fig:fig1}.

The ten background sources were identified based on their broad apparent damped Lyman-$\alpha$ absorption (DLA) profile and their close on-sky proximity as part of the \jwst/NIRSpec PRImordial gas Mass AssembLy (PRIMAL) archival survey \cite{Heintz24}. The galaxies in this sample are all selected to have robust emission-line redshift measurements and sufficient signal-to-noise (${\rm S/N}>3$) of the continuum around the redshifted rest-frame ultraviolet and \lya\ wavelength region to ensure robust measurements of the damping wings. This is essential to disentangle the contributions from the intergalactic medium (IGM), ionized UV bubbles, or local dense neutral atomic hydrogen in the interstellar or circumgalactic medium of the galaxies. 

The observations of the cluster-member galaxies were obtained from the First Reionization Epoch Spectroscopically Complete Observations (FRESCO) survey, prog. ID: 1895 \cite{Oesch23}. FRESCO covered the two GOODS/CANDELS fields with deep ($\sim 2\,$hr) pointings using the NIRCam/grism F444W filter slitless spectroscopic configuration. This yields a spectral coverage over $3.8-5.0\,\mu$m with resolving power $\mathcal{R}=1600$. 
The redshifts of the cluster members were derived based on their rest-frame optical emission lines, primarily H$\alpha$ and occasionally [\nii] when detected. Complementary photometry in the target fields was used to derive the spectral energy distribution of each source, the photometric redshifts were used as priors for the line identifications and emission-line redshift measurements \cite{Oesch23}. 

\bmhead{Characteristics of the galaxy proto-cluster members}

The foreground galaxy proto-cluster were first identified by Helton et al. \cite{Helton23} based on slitless \jwst/NIRCam grism observations of the field from the FRESCO survey. Using the extracted H$\alpha$ catalog from this survey, we identify the same overdensity here, including $\gtrsim 25$ galaxies at $z=5.35-5.40$. The redshift distribution of the full set of sources from the FRESCO survey with significant, ${\rm S/N}>5$, detections of H$\alpha$ at $z=4.9-6.6$ is shown in Extended Fig.~1 with the galaxy overdensity highlighted. The average spectroscopic redshift of the large-scale structure is $z=5.387$. The galaxy cluster is centred at celestial coordinates R.A. = $3^{\rm h}32^{\rm m} 20^{\rm s}$, Decl. = $-27^\circ 49^\prime 50^{\prime \prime}$. The spatial and brightness distribution of the cluster members are shown in Extended Fig.~2. The majority of the sources are located within 1\,pMpc from the estimated center of the cluster. We observe no strong dependence of the spatial location of the sources on the absolute UV brightness, but note that the brightest members ($M_{\rm UV } < -20$\,mag) are all located within $\sim 1$\,pMpc whereas the faintest proto-cluster galaxies ($M_{\rm UV} > -19$\,mag) are found in the outskirts. This would be consistent with the predicted ``inside-out'' growth early galaxy proto-clusters \cite{Chiang17}.


\bmhead{Modelling the damped Lyman-$\alpha$ absorption feature}

To model the broad DLA features observed in each of the background galaxy spectra, we first considered the effects of a partly neutral intergalactic medium (IGM). The optical depth from the Gunn-Peterson effect due to absorption from \hi\ in the IGM can be described via a single redshift-dependent parameter, the average neutral hydrogen fraction $x_{\rm HI} = n_{\rm HI}/n_{\rm H,tot}$ \cite{MiraldaEscude98,McQuinn08}. Since the large-scale reionization of the IGM on average is determined to be complete, $x_{\rm HI} \rightarrow 0$, by redshift $z\approx 5.5-6$ \cite{McGreer15,Bosman22,Fan23} this likely does not contribute to the observed Ly$\alpha$ damping wings in the ten background galaxy spectra. To be conservative, we include a prescription for the IGM component following \cite{MiraldaEscude98} but using the approximation by \cite{Totani06} setting the maximum redshift to $z_{\rm gal}$ and integrate down to $z=5.4$. For all cases, however, the observed DLA broadening are much greater than expected for even a fully neutral IGM with $x_{\rm HI}=1$, corresponding to an equivalent \hi\ column density of $N_{\rm HI} = 10^{21.2}\,$cm$^{-2}$ \cite{Heintz24_DLAs}. To reach the observed column densities of $N_{\rm HI} \gtrsim 10^{22}-10^{23}\,$cm$^{-2}$ requires overdensities $\sim 3-10\times$ the global average, equivalent to $x_{\rm HI} = 3-10$. This latter estimate of the neutral gas mass overdensity is consistent with the predicted galaxy overdensity \cite{Helton23}. 

We modelled the DLA profiles with a Voigt profile, assuming a single-redshift origin, approximated as:
\begin{equation}
    \tau_{\rm DLA} = C a H(a,x) N_{\rm HI}
\end{equation}
Here $C$ is the photon absorption constant, $a$ is the damping parameter, and $H(a,x)$ is the Voigt-Hjerting function \cite{TepperGarcia06}, with $x$ representing the difference in rest-frame wavelength with respect to the resonant wavelength of \lya\ expressed in Doppler units. For strong DLA features where the Lorentzian wings dominate the shape of the absorption-line profile the optical depth is only sensitive to the column density of neutral atomic hydrogen, $N_{\rm HI}$ and does not depend on the Doppler broadening. This approximation is typically assumed in the study of foreground absorbers associated with galaxies in bright quasar or gamma-ray burst sightlines \cite{Wolfe05,Prochaska09,Noterdaeme12,Fynbo09,Tanvir19,Heintz23_GRB}, and have more recently been applied to single high-redshift galaxies as well \cite{Heintz24_DLAs,Umeda23,DEugenio23,Hainline24,Carniani24}. It assumes that a dominating fraction of the absorbing gas is located at the approximate same redshift when integrating over the line-of-sight column density, in contrast to typical IGM approximations that integrate over longer redshift intervals. The former approximation best represents the scenario of a massive foreground neutral gas reservoir connected to a single galaxy overdensity, but the small variations in the spectra could be due to a radial distribution along the line of sight ($z_{\rm abs} \pm 0.1$) of the foreground gas. 

We model the intrinsic continuum spectral energy distribution (SED) underlying the DLA component using three different prescriptions: {\it i}) A rest-frame UV spectral shape following a simple power-law of the form $F_{\lambda} \propto \lambda^{-\beta}$, {\it ii}) A set of galaxy template spectra designed to match the blue rest-frame UV colors of high-redshift galaxies \cite{Larson23}, and {\it iii}) Direct modelling of the SED using {\tt Eazy} \cite{Brammer08_Eazy}, but excluding photometric filter bandpasses that include the redshifted Lyman-$\alpha$ edge. We find that the results on the derived $N_{\rm HI}$ are generally insensitive to the exact choice of the underlying intrinsic continuum SEDs and dust attenuation ($\Delta \log N({\rm HI/cm^{-2}}) \lesssim 0.1$) since the small variations in the intrinsic spectral shape is negligible compared to the observed broad damping wings which mainly drive the column density estimations. For the following analysis we use the high-redshift galaxy template spectra, specifically the recommended {\sc binc100z001age6.cloudy} model with weak intrinsic Lyman-$\alpha$ emission and young stellar population. The exact choice of the galaxy template does not have a larger impact on the results either. Following Heintz et al. \cite{Heintz24_DLAs} we sampled the posterior probability distributions of the DLA \hi\ column densities using the {\tt MultiNest} algorithm \cite{Feroz08} implemented in the {\tt PyMultiNest}-package \cite{Buchner14}. This approach determines the marginal likelihood (evidence of the fit) via a transformation of the likelihood space into a one-dimensional probability distribution. The output results are derived by sampling this posterior distribution, weighted by the assumed prior probability distribution. By integrating the likelihood levels we calculate the evidence of the fit used for model comparison. 

For each background galaxy, we modelled the spectrum using the galaxy templates as the intrinsic SEDs with the added visual extinction, $A_V$, the average IGM neutral fraction, $x_{\rm HI}$, and the \hi\ column density, $N_{\rm HI}$ as free parameters. We used flat, conservative priors on $N_{\rm HI}$ in the range $\log (N_{\rm HI}/{\rm cm}^{-2}) = 18-24$. Motivated by the single-case study of Terp et al. \cite{Terp24}, we first leave the absorption redshift, $z_{\rm abs}$, as a free parameter in the fit. The resulting posterior distribution functions (PDFs) are shown in Extended Fig.~3. In the majority of cases, excluding the galaxy with source ID 13877 at $z_{\rm gal}=5.6166$, the PDFs all prefer a lower-redshift solution, excluding the background-galaxy redshift solution at $>3\sigma$ confidence. We note also that the galaxy with source ID 7807 at $z_{\rm gal} = 5.3960$ is likely embedded within the galaxy proto-cluster and therefore by selection probe slightly foreground gas to the mean cluster-redshift, though the PDF is still consistent at $1\sigma$ confidence with the mean. From the joint PDFs of the individual redshift solutions we find combined, median absorption redshift of $z_{\rm abs} = 5.53^{+0.19}_{-0.23}$ where the uncertainties denote the 16th to 84th percentiles of the distribution. The galaxy proto-cluster redshift is thus consistent within $1\sigma$ confidence to the derived absorption redshift.
Further, when performing the modelling using a prior on $z_{\rm abs} = 5.387$ we find that the Bayes factor of the two models suggest no significant difference. We therefore set $z_{\rm abs} = 5.387$ in the following analysis, and report the Bayes factor over the galaxy redshift in Table~\ref{tab:props}. For all cases but the galaxy with source ID 13877 at $z_{\rm gal}=5.6166$, the foreground galaxy-cluster redshift solution is preferred at high significance. With this assumption, we derive \hi\ column densities in the range $N_{\rm HI} = 10^{21.7}-10^{23.5}$\,cm$^{-2}$. In Extended Fig.~4 we show the impact parameters of the background galaxy sightlines from the estimated proto-cluster center, size-coded according to the foreground \hi\ column density and color-coded according to galaxy redshift. We find no dependence on the inferred $N_{\rm HI}$ with background galaxy redshift and no apparent radial dependence on the derived $N_{\rm HI}$, though note a strong coherence between close on-sky galaxy-sightlines. The lack of radial dependence is in any case not expected given the large density variations of the filamentary structures permeating these massive galaxy overdensities predicted from simulations \cite{Bennett20,Rennehan24}, as also highlighted in the section below. 

To statistically investigate the internal coherence between sightlines, we show in Extended Fig.~5 the ratio of inferred \hi\, column densities and the projected distance between pairs of objects. The two pair of objects with the smallest projected distance (20 and 70 kpc apart at $z\approx 5.4$) display equivalent column densities suggesting they probe similarly dense foreground regions. In physical space, these pairs are clearly distinct with each galaxy being separated by tens of co-moving Mpc. Comparing to the observed distribution of $N_\mathrm{H\,I}$-ratios for all objects at projected distances, $250\, \mathrm{kpc} < d < 1\, \mathrm{Mpc}$, these nearest pairs are outliers given their observed degree of coherence. The KS-test statistic gives a 9\% probability of randomly getting two column-densities as similar as that contained by the two nearest pairs (i.e. all objects within 100 kpc). Further, given the 5 pairs within 250 kpc projected distance, the probability of randomly getting the observed similar column-densities is less than 2.5\%. 

There is now mounting evidence in the literature that a higher prevalence of DLAs in galaxy spectra are associated with overdensities. Observations from the ground of galaxy or quasar overdensities at redshifts $z\sim 2-3$ have revealed large, neutral gas reservoirs connected to the cluster members \cite{Frye08,Cai17,Lee18,Hayashino19,Newman22}. At higher redshifts towards the reionization era, the first evidence came from the serendipitous detection of a massive foreground absorber in close on-sky proximity to one of the most massive proto-clusters at $z>5$ discovered to date \cite{Terp24}. This scenario was required to explain the simultaneous detection of strong Lyman-$\alpha$ emission {\em and} DLA absorption (source ID 13176 with $z_{\rm gal} = 5.944$ also studied here), though it have also been reported as a potential galaxy dominated by nebular-continuum emission at rest-frame UV and optical wavelengths \cite{Cameron23}. The second such object with potential nebular-continuum emission at $z=7.8$ from the latter study was, however, also later associated with a galaxy overdensity \cite{Chen24}. There is also recent evidence for an excess fraction of DLAs identified from photometry-only in independent galaxy fields associated with galaxy overdensities \cite{Li24}, lending strong credibility to this scenario. The results presented here is the first conclusive, large-scale detection of the cold, neutral medium connected to a high-redshift galaxy proto-cluster. 

\bmhead{Predictions from simulations}

In order to illustrate how the large, spatially-coherent neutral regions in the post-reionization Universe such as those suggested by Extended Fig.~5 challenge models, we compare the coherent length scale with its associated column density in Extended Fig.~6.
For a uniform medium with overdensity $\delta$, co-moving size $L$, and neutral fraction $f_{\rm HI}$, the neutral gas column density is
\begin{equation}
    N_{\rm HI} = f_{\rm HI} \langle n_{\rm H} \rangle L (1+\delta) ~,
\end{equation}
where $\langle n_{\rm H}\rangle$ is the mean hydrogen number density. This relationship is shown for several trial values of $f_{\rm HI}(1+\delta)$. Adopting our measured range of \hi\ columns and projected separations yields the blue error bar and limit. Overdensities in the range $f_{\rm HI} (1+\delta) = 10^{3}-10^{4}$ are required to match the column density distribution, consistent with that derived from the simple geometric assumption about the spatial distribution of the gas.

Orange points with error bars indicate the expected trend at $z=5.5$ from the cosmological radiation hydrodynamic Technicolor Dawn simulations, which simultaneously track galaxy evolution and \hi\ reionization~\citep{Finlator18}. The measured combination of \hi\ column density and coherent length scale do not arise in the simulation. However, as the simulation spans only 24.35\,cMpc, it does not include the large, more rare density fluctuations required to reach the halo masses estimated for this galaxy overdensity. We note additionally that the inferred length scale of coherent \hi\ absorption may be inaccurate because our galaxies were \emph{selected} to show strong \hi\ absorption. A more complete characterization of the distribution of \hi\ columns within this region would be needed to test whether the inferred overdensities are indeed overestimated by 2--3 orders of magnitude.

To provide a more direct theoretical comparison to the observations, we employ a modified version of the \textsc{fable} suite of simulations, tailored to follow the evolution of galaxy groups and clusters \cite{Henden18}. We run a zoom simulation of a halo with mass $M_{200} = 5.5\times10^{12} \,\mathrm{M}_\odot$ at $z=5.5$. This object grows into a large cluster at $z=0$, with a mass $M_{200} = 1.30 \times 10^{15}\,\mathrm{M}_\odot$ consistent with the Coma cluster, and was taken from a parent dark matter only simulation, the (4.1\,Gpc)$^3$ Millennium XXL \cite{Angulo12}. The Lagrangian region of this zoom -- the region within which all dark matter particles will end up within $5\times R_{500}$ of the $z=0$ cluster -- is aspherical but has a diameter of at least $7.5\,$pMpc at $z=5.5$. All \hi\ maps are then made projecting along the $z$-axis through the entire depth of this Lagrangian region.

In this simulation we also use the shock refinement scheme introduced in \cite{Bennett20}. This significantly increases gas mass resolution by a factor of 512 within $\approx 500$\,pkpc around the centre of the halo, with cells having a mass of $3.2\times10^4 \, \mathrm{M}_\odot$. Outside of this region the gas remains at the base resolution of the simulation, with cell masses of $1.64\times10^7 \, \mathrm{M}_\odot$. For full details of this scheme, see \cite{Bennett20}.
A full treatment of the neutral fraction of gas within the simulation would require radiative transfer. In this simulation we instead use the prescription of \cite{Bird2013} to estimate the \hi\ content of each gas cell, which is then summed to provide the column density maps in Fig.~\ref{fig:fig3}.

To make predictions for the assembly and later evolution of this galaxy proto-cluster, we extract the halo mass and baryonic mass components at a range of redshifts, $z=10,8,5.5,2,0$ corresponding to $0.47$, $0.64$, $1.0$, $3.3$, and $13.8$~Gyrs after the Big Bang as shown in Extended Fig.~7. We plot the mass of the total gas and the \hi\ gas component within the entire zoom region (1.6\,pMpc) and within $R<R_{200}$ and compare it to the halo mass and the stellar mass contained within $R<R_{200}$. At all redshifts, the neutral gas mass appear to only comprise $\approx 1-5\%$ of the total gas within the box. However, focusing on the central region within $R<R_{200}$, the gas is largely neutral until $z\approx 5.5$ at which point the gas becomes more ionized and eventually (by $z\approx 2$) processed into stars. Intriguingly, these simulations further show that the halo and stellar mass surpass the total and local neutral gas budget, respectively, at redshifts $z\sim 2-3$. This is coincident with the peak of the average cosmic star-formation rate density \cite{Madau14}. This is further evidence that the availability of pristine, neutral gas, or the lack thereof, is likely a key physical driver of the overall cosmic star formation history causing the eventual turnover and decline observed at the present day.

Similarly, we also consider the Manhattan Suite, a set of $100$ zoom-in simulations focused on galaxy clusters with $M_\mathrm{200,c} > 10^{14}\,\mathrm{M}_\mathrm{\odot}$ at $z = 2$, and baryonic particle resolution $2\times10^7 \, \mathrm{M}_\odot$ \cite{Rennehan24}. The suite is intentionally biased towards the most rapidly assembling, massive halos in the early Universe in order to mimic the observational selection bias towards highly star forming and over-dense structures at high redshift. By analyzing the $100$ most massive progenitor chains in the simulations at $z = 5.4$, we determined the covering fractions of neutral hydrogen column densities above $10^{21}\,\mathrm{cm}^{-2}$, $10^{21.3}\,\mathrm{cm}^{-2}$, $10^{21.5}\,\mathrm{cm}^{-2}$, and $10^{21.7}\,\mathrm{cm}^{-2}$ within $R_\mathrm{200,c}$, along the $3$ principal axes, and produced probability density functions of the covering fractions (see the bottom plot in Extended Fig.~8). While the proto-clusters in the Manhattan Suite are rare, not all of the viewing angles or progenitor histories produce significant sightlines above $10^{21.7}\,\mathrm{cm}^{-2}$. However, there is an extended tail in the distributions where the proto-cluster has up to $15\%$ covering fraction of neutral hydrogen, especially if we allow slightly lower column densities. Compared to the focused zoom, we stress the importance of a statistical sample of proto-clusters that are focused on the most massively assembling halos in order to capture the extreme gas densities that we present in this work. The proto-clusters in the Manhattan Suite will go on to form galaxy clusters with masses $M_\mathrm{200,c} \sim 10^{15}\,\mathrm{M}_\mathrm{\odot}$ at $z = 0$, which are the most massive structures today. 

Our results presented are in stark contrast to previous cosmological reionization simulations that do not capture the large neutral gas overdensities observed here, which require more detailed galaxy formation simulations. These large overdensities are, however, crucial to model in order to obtain a complete census of galaxy assembly and star-formation rates in galaxy proto-clusters and the effect of the more dense, massive neutral gas surrounding these overdensities on the expected reionization topology.  

\clearpage
\newpage

\begin{figure}
    \centering
    \includegraphics[width=0.9\textwidth]{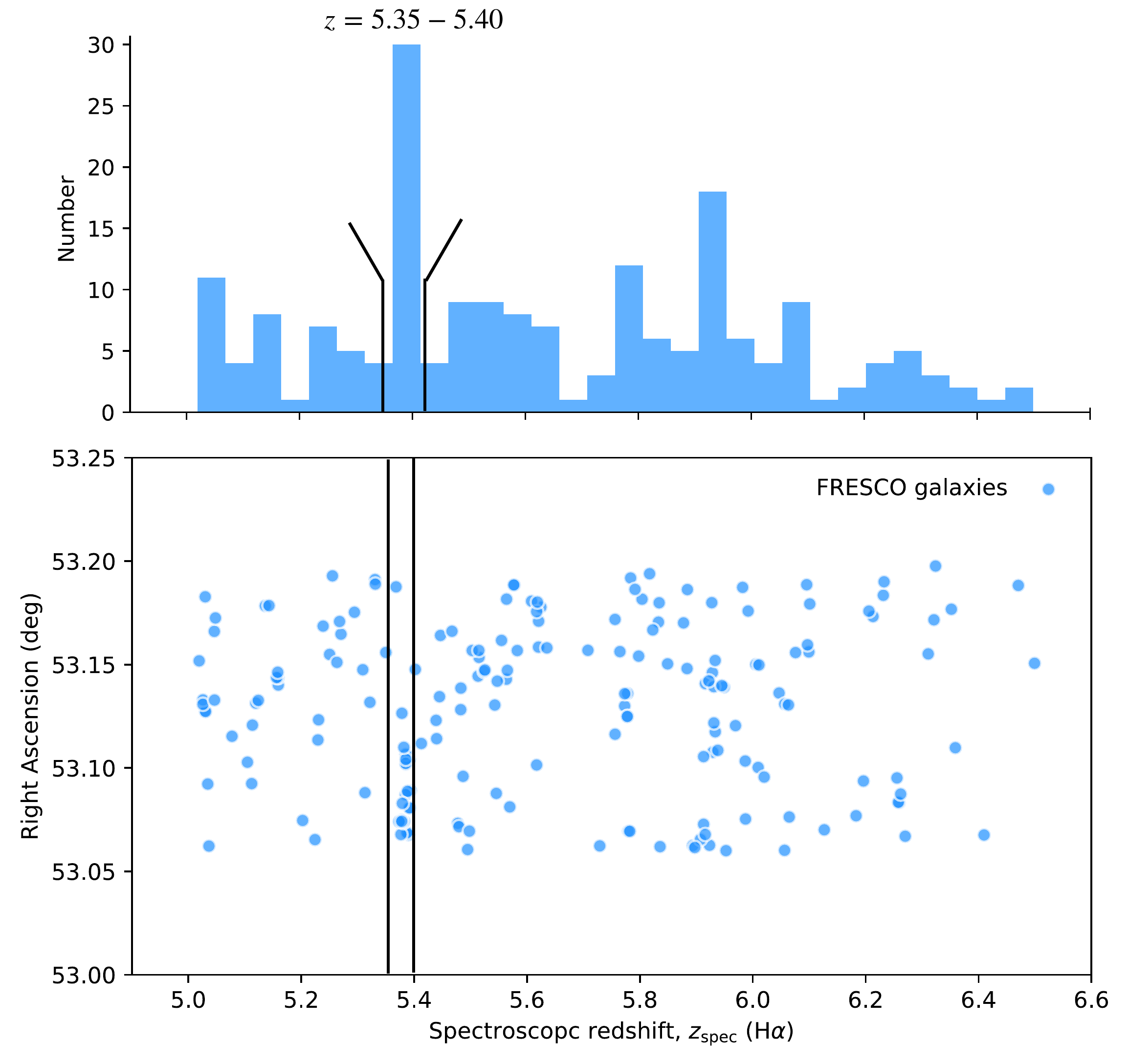}
    \label{fig:frescodist}
\end{figure}

\noindent {\bf Extended Fig.~1.} Redshift distribution of the sources in FRESCO with significant, ${\rm S/N}>5$, detections of H$\alpha$ at $z=4.9-6.6$. The galaxy overdensity identified in the survey at $z=5.35-5.40$ is highlighted.

\clearpage
\newpage

\begin{figure}
    \centering
    \includegraphics[width=0.8\textwidth]{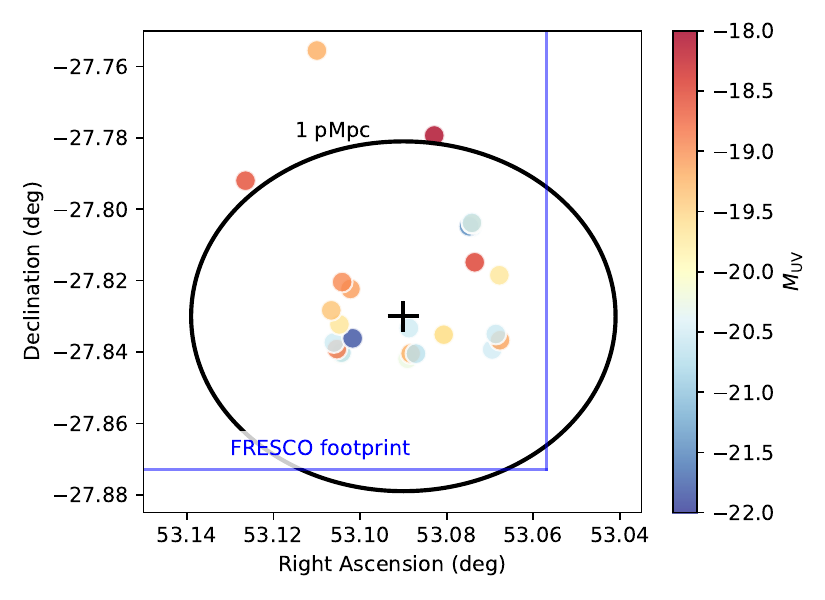}
    \label{fig:frescomuv}
\end{figure}

\noindent {\bf Extended Fig.~2.} Spatial distribution of the identified galaxy proto-cluster members at $z=5.35-5.40$. The sources are color-coded by their rest-frame UV absolute magnitude, $M_{\rm UV}$. The centroid of the cluster is marked by the cross, and the ellipse show the region within 1\,pMpc. The FRESCO survey footprint is marked by the blue line. 

\clearpage
\newpage

\begin{figure}
    \centering
    \includegraphics[width=0.8\textwidth]{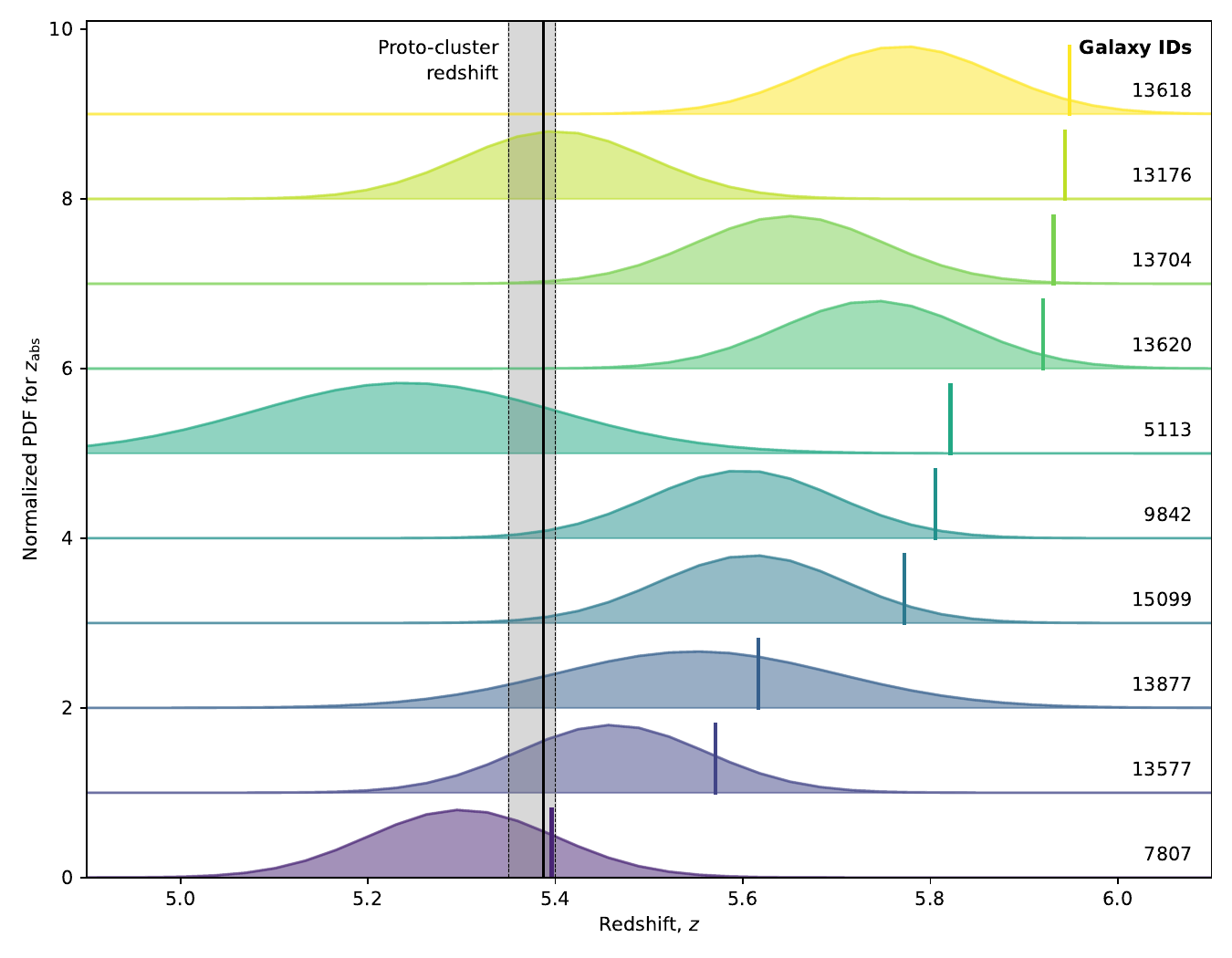}
    \label{fig:zabspost}
\end{figure}

\noindent {\bf Extended Fig.~3.} Posterior distributions of the DLA modelling leaving the absorption redshift, $z_{\rm abs}$, as a free parameter. The vertical bars indicate the background galaxy redshifts. Most galaxy sightlines prefer a foreground absorber and exclude a host-galaxy origin at $>3\sigma$ confidence.

\clearpage
\newpage

\begin{figure}
    \centering
    \includegraphics[width=0.8\textwidth]{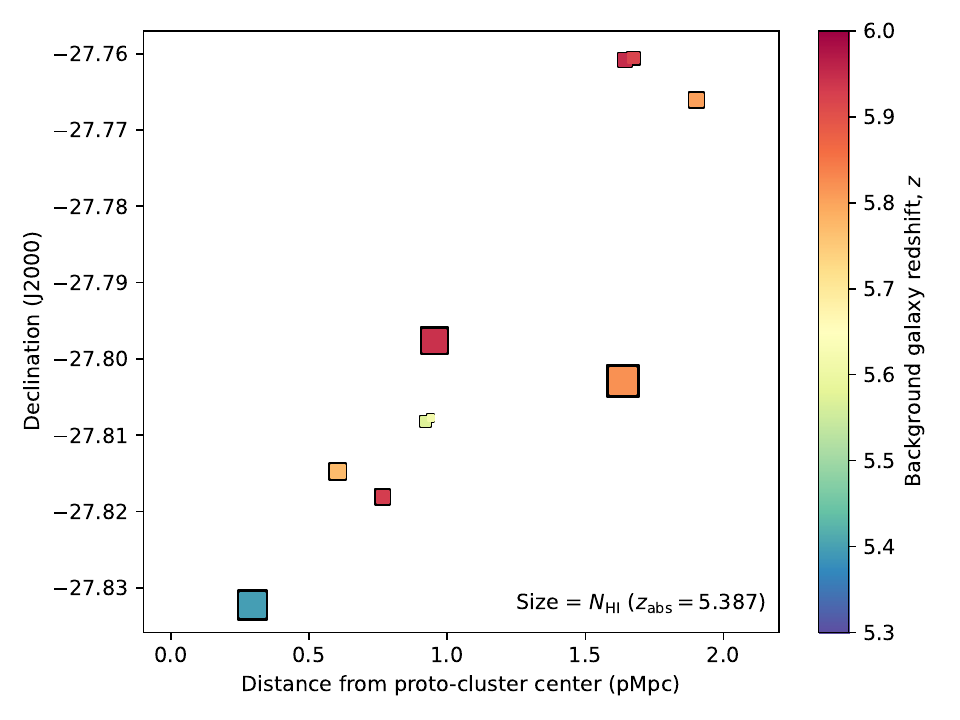}
    \label{fig:nhidist}
\end{figure}

\noindent {\bf Extended Fig.~4.} The spatial distribution of the galaxy sightlines as a function of impact parameter from the center of the proto-cluster, color-coded by the derived foreground \hi\ column densities. 

\clearpage
\newpage

\begin{figure}
    \centering
    \includegraphics[width=0.9\textwidth, viewport=22 20 680 520 ,clip=]{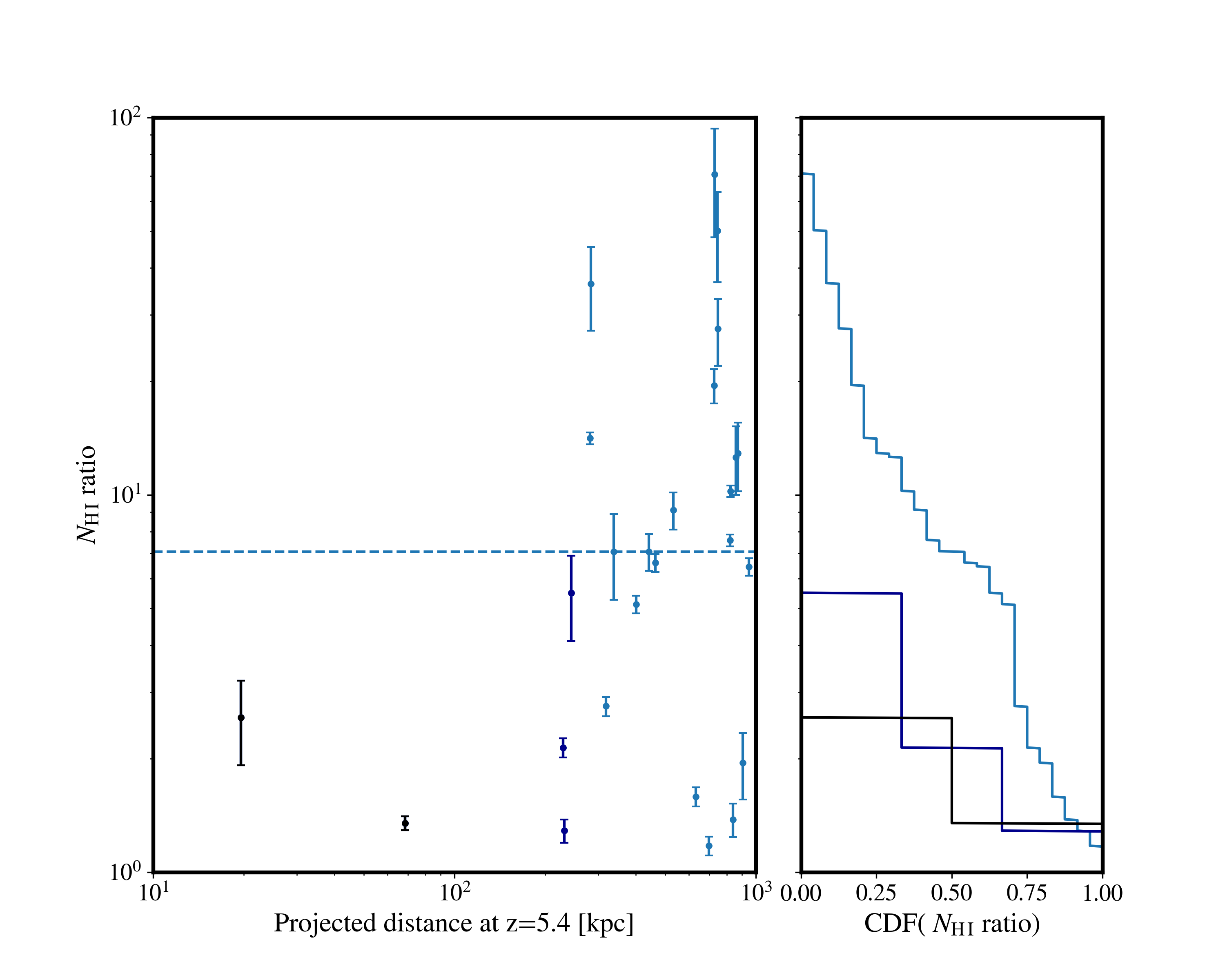}
    \label{fig:DLA_sightline}
\end{figure}

\noindent {\bf Extended Fig.~5.} Relative line-of-sight \hi\ column densities as a function of projected distance at $z=5.4$. The right hand panel indicates the CDF of $N_\mathrm{H\,I}$-ratios. The dashed line indicates the median of the observed distribution of $N_{\textit H\,I}$ ratio. The galaxies associated with the smallest on-sky projected distance display similar column-densities. Objects with projected distance $d<100$,$100<d<250$, $d>250$ kpc are indicated in black, dark-blue and blue respectively. 
\clearpage
\newpage

\begin{figure}
    \centering
    \includegraphics[width=0.9\textwidth]{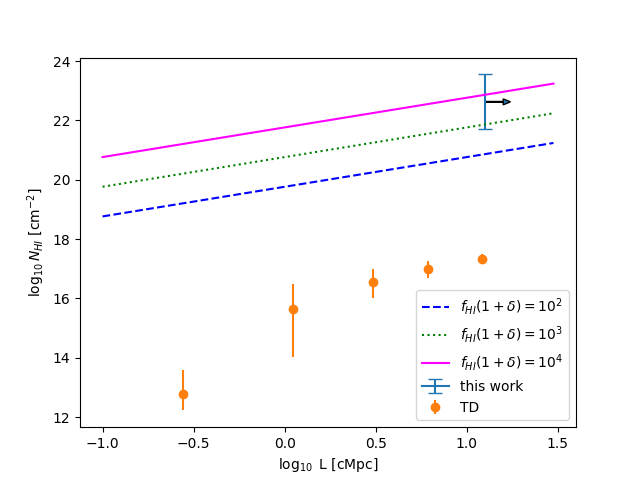}
    \label{fig:hioverdensity}
\end{figure}

\noindent {\bf Extended Fig.~6.} Relationship between neutral gas overdensity $f_{\rm HI} (1+\delta)$, column density, and co-moving coherent length scale. The blue error bar indicates our measurement; lines indicate the expected relationship for assuming a unity fill factor, and orange points indicate the trend obtained by subsampling the Technicolor Dawn (TD) simulation at $z=5.5$. The observations would imply overdensities $f_{\rm HI} (1+\delta)\approx 10^{3}-10^{4}$ if the probed sightlines represent the entire region. Such large, spatially-coherent, neutral regions do not arise in the simulation at $z=5.5$, confirming that they are quite rare.

\clearpage
\newpage

\begin{figure}
    \centering
    \includegraphics[width=0.9\textwidth]{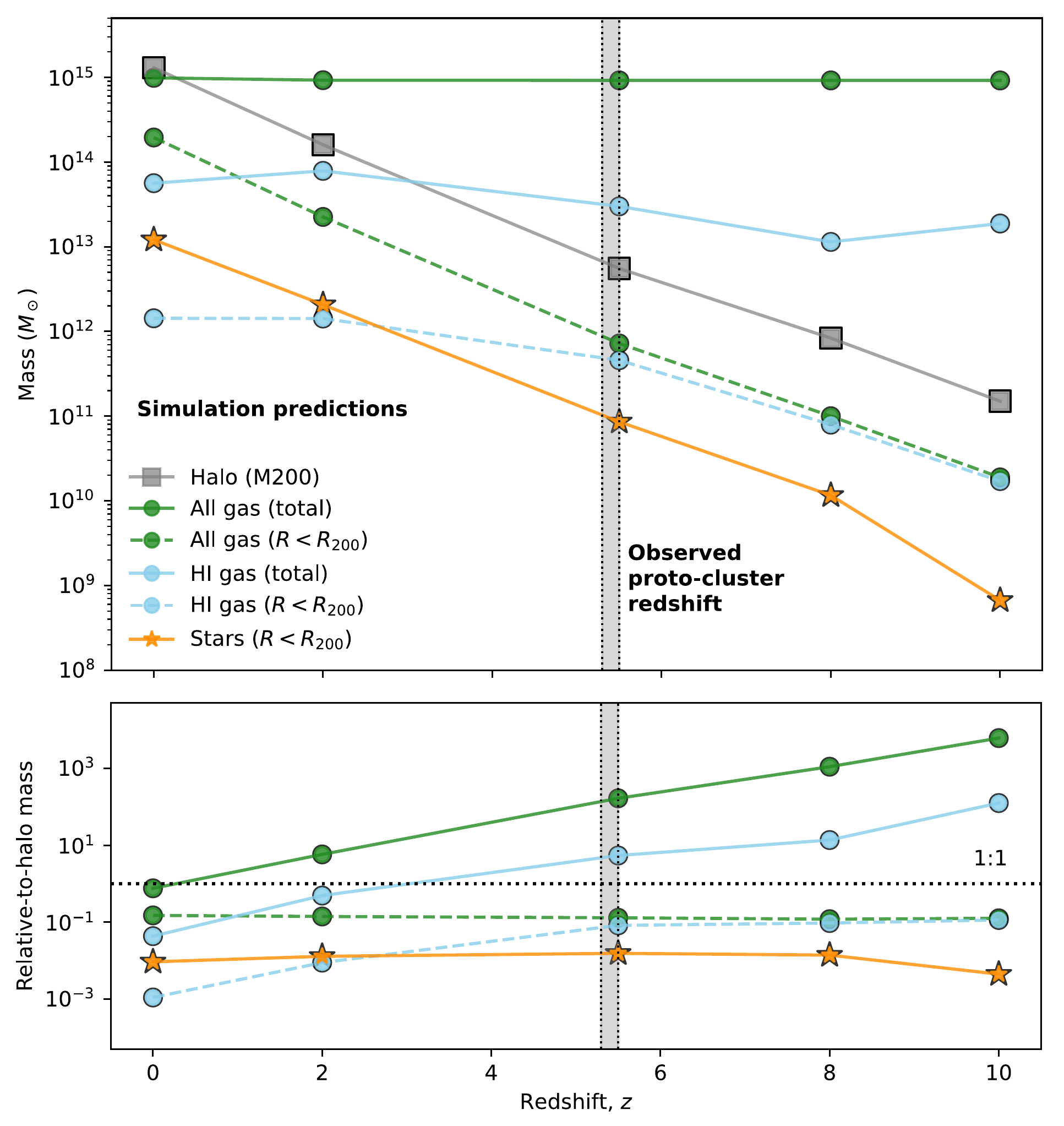}
    \label{fig:clusterz}
\end{figure}

\noindent {\bf Extended Fig.~7.} The simulated redshift evolution of the halo and baryonic mass in the proto-cluster at $z=5.4$. Solid lines show the total mass within the full simulation zoom region and the dashed lines the mass within $R<R_{200}$.

\clearpage
\newpage

\begin{figure}
    \centering
    \includegraphics[width=0.9\textwidth]{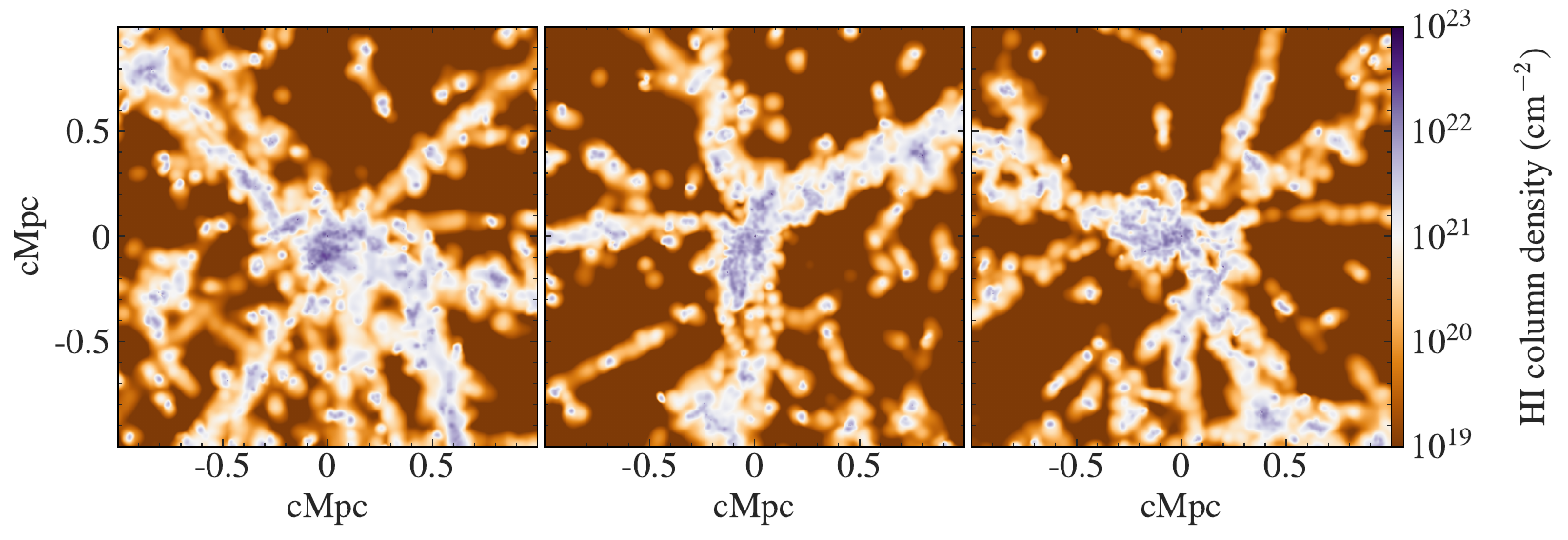}
    \includegraphics[width=0.7\textwidth]{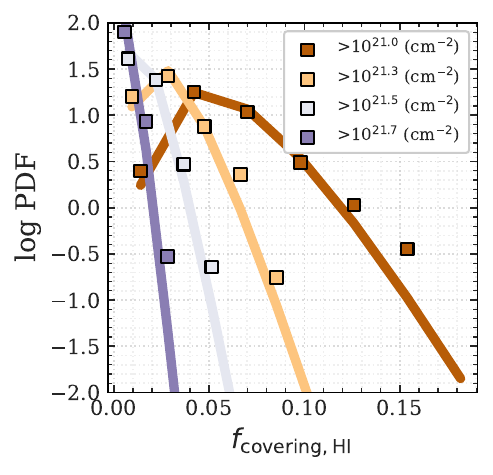}
    \label{fig:sims}
\end{figure}

\noindent {\bf Extended Fig.~8.} (top) An example projection of the neutral hydrogen density from the Manhattan Suite at $z = 5.4$, centered on the most massive structure in the simulation suite. The panels are three views along the three principal axes, and the coordinates are in co-moving Mpc.  There is spatially-extensive high column density ($>10^{21}\,\mathrm{cm}^{-2}$) gas through the halo. (bottom) The squares show the probability density function of covering fractions at a minimum column density threshold (given in the legend).  The curves are gamma function fits to the simulation data. The histograms of columns are produced from all $100$ unique proto-clusters in the Manhattan Suite at $z = 5.4$, with $3$ viewing angles each (along the principal axes). The transition region between $10^{21}\,\mathrm{cm}^{-2}$ and $10^{21.7}\,\mathrm{cm}^{-2}$ shows a wide variety of possible covering fractions depending on the proto-cluster and viewing angle.


\clearpage
\newpage

\setstretch{1.}


\newcommand{\actaa}{Acta Astron.}   
\newcommand{\araa}{Annu. Rev. Astron. Astrophys.}   
\newcommand{\areps}{Annu. Rev. Earth Planet. Sci.} 
\newcommand{\aar}{Astron. Astrophys. Rev.} 
\newcommand{\ab}{Astrobiol.}    
\newcommand{\aj}{Astron. J.}   
\newcommand{\ac}{Astron. Comput.} 
\newcommand{\apart}{Astropart. Phys.} 
\newcommand{\apj}{Astrophys. J.}   
\newcommand{\apjl}{Astrophys. J. Lett.}   
\newcommand{\apjs}{Astrophys. J. Suppl. Ser.}   
\newcommand{\ao}{Appl. Opt.}   
\newcommand{\apss}{Astrophys. Space Sci.}   
\newcommand{\aap}{Astron. Astrophys.}   
\newcommand{\aapr}{Astron. Astrophys. Rev.}   
\newcommand{\aaps}{Astron. Astrophys. Suppl.}   
\newcommand{\baas}{Bull. Am. Astron. Soc.}   
\newcommand{\caa}{Chin. Astron. Astrophys.}   
\newcommand{\cjaa}{Chin. J. Astron. Astrophys.}   
\newcommand{\cqg}{Class. Quantum Gravity}    
\newcommand{\epsl}{Earth Planet. Sci. Lett.}    
\newcommand{\frass}{Front. Astron. Space Sci.}    
\newcommand{\gal}{Galaxies}    
\newcommand{\gca}{Geochim. Cosmochim. Acta}   
\newcommand{\grl}{Geophys. Res. Lett.}   
\newcommand{\icarus}{Icarus}   
\newcommand{\jatis}{J. Astron. Telesc. Instrum. Syst.}  
\newcommand{\jcap}{J. Cosmol. Astropart. Phys.}   
\newcommand{\jgr}{J. Geophys. Res.}   
\newcommand{\jgrp}{J. Geophys. Res.: Planets}    
\newcommand{\jqsrt}{J. Quant. Spectrosc. Radiat. Transf.} 
\newcommand{\lrca}{Living Rev. Comput. Astrophys.}    
\newcommand{\lrr}{Living Rev. Relativ.}    
\newcommand{\lrsp}{Living Rev. Sol. Phys.}    
\newcommand{\memsai}{Mem. Soc. Astron. Italiana}   
\newcommand{\mnras}{Mon. Not. R. Astron. Soc.}   
\newcommand{\nat}{Nature} 
\newcommand{\nastro}{Nat. Astron.} 
\newcommand{\ncomms}{Nat. Commun.} 
\newcommand{\nphys}{Nat. Phys.} 
\newcommand{\na}{New Astron.}   
\newcommand{\nar}{New Astron. Rev.}   
\newcommand{\physrep}{Phys. Rep.}   
\newcommand{\pra}{Phys. Rev. A}   
\newcommand{\prb}{Phys. Rev. B}   
\newcommand{\prc}{Phys. Rev. C}   
\newcommand{\prd}{Phys. Rev. D}   
\newcommand{\pre}{Phys. Rev. E}   
\newcommand{\prl}{Phys. Rev. Lett.}   
\newcommand{\psj}{Planet. Sci. J.}   
\newcommand{\planss}{Planet. Space Sci.}   
\newcommand{\pnas}{Proc. Natl Acad. Sci. USA}   
\newcommand{\procspie}{Proc. SPIE}   
\newcommand{\pasa}{Publ. Astron. Soc. Aust.}   
\newcommand{\pasj}{Publ. Astron. Soc. Jpn}   
\newcommand{\pasp}{Publ. Astron. Soc. Pac.}   
\newcommand{\raa}{Res. Astron. Astrophys.} 
\newcommand{\rmxaa}{Rev. Mexicana Astron. Astrofis.}   
\newcommand{\sci}{Science} 
\newcommand{\sciadv}{Sci. Adv.} 
\newcommand{\solphys}{Sol. Phys.}   
\newcommand{\sovast}{Soviet Astron.}   
\newcommand{\ssr}{Space Sci. Rev.}   
\newcommand{\uni}{Universe} 


\bibliography{ref}


\begin{thebibliography}{64}
\ifx \bisbn   \undefined \def \bisbn  #1{ISBN #1}\fi
\ifx \binits  \undefined \def \binits#1{#1}\fi
\ifx \bauthor  \undefined \def \bauthor#1{#1}\fi
\ifx \batitle  \undefined \def \batitle#1{#1}\fi
\ifx \bjtitle  \undefined \def \bjtitle#1{#1}\fi
\ifx \bvolume  \undefined \def \bvolume#1{\textbf{#1}}\fi
\ifx \byear  \undefined \def \byear#1{#1}\fi
\ifx \bissue  \undefined \def \bissue#1{#1}\fi
\ifx \bfpage  \undefined \def \bfpage#1{#1}\fi
\ifx \blpage  \undefined \def \blpage #1{#1}\fi
\ifx \burl  \undefined \def \burl#1{\textsf{#1}}\fi
\ifx \doiurl  \undefined \def \doiurl#1{\url{https://doi.org/#1}}\fi
\ifx \betal  \undefined \def \betal{\textit{et al.}}\fi
\ifx \binstitute  \undefined \def \binstitute#1{#1}\fi
\ifx \binstitutionaled  \undefined \def \binstitutionaled#1{#1}\fi
\ifx \bctitle  \undefined \def \bctitle#1{#1}\fi
\ifx \beditor  \undefined \def \beditor#1{#1}\fi
\ifx \bpublisher  \undefined \def \bpublisher#1{#1}\fi
\ifx \bbtitle  \undefined \def \bbtitle#1{#1}\fi
\ifx \bedition  \undefined \def \bedition#1{#1}\fi
\ifx \bseriesno  \undefined \def \bseriesno#1{#1}\fi
\ifx \blocation  \undefined \def \blocation#1{#1}\fi
\ifx \bsertitle  \undefined \def \bsertitle#1{#1}\fi
\ifx \bsnm \undefined \def \bsnm#1{#1}\fi
\ifx \bsuffix \undefined \def \bsuffix#1{#1}\fi
\ifx \bparticle \undefined \def \bparticle#1{#1}\fi
\ifx \barticle \undefined \def \barticle#1{#1}\fi
\bibcommenthead
\ifx \bconfdate \undefined \def \bconfdate #1{#1}\fi
\ifx \botherref \undefined \def \botherref #1{#1}\fi
\ifx \url \undefined \def \url#1{\textsf{#1}}\fi
\ifx \bchapter \undefined \def \bchapter#1{#1}\fi
\ifx \bbook \undefined \def \bbook#1{#1}\fi
\ifx \bcomment \undefined \def \bcomment#1{#1}\fi
\ifx \oauthor \undefined \def \oauthor#1{#1}\fi
\ifx \citeauthoryear \undefined \def \citeauthoryear#1{#1}\fi
\ifx \endbibitem  \undefined \def \endbibitem {}\fi
\ifx \bconflocation  \undefined \def \bconflocation#1{#1}\fi
\ifx \arxivurl  \undefined \def \arxivurl#1{\textsf{#1}}\fi
\csname PreBibitemsHook\endcsname

\bibitem[\protect\citeauthoryear{{Chiang} et~al.}{2017}]{Chiang17}
\begin{barticle}
\bauthor{\bsnm{{Chiang}}, \binits{Y.-K.}},
\bauthor{\bsnm{{Overzier}}, \binits{R.A.}},
\bauthor{\bsnm{{Gebhardt}}, \binits{K.}},
\bauthor{\bsnm{{Henriques}}, \binits{B.}}:
\batitle{{Galaxy Protoclusters as Drivers of Cosmic Star Formation History in the First 2 Gyr}}.
\bjtitle{\apjl}
\bvolume{844}(\bissue{2}),
\bfpage{23}
(\byear{2017})
\doiurl{10.3847/2041-8213/aa7e7b}
{\href{https://arxiv.org/abs/1705.01634}{{arXiv:1705.01634}}}
{[astro-ph.GA]}
\end{barticle}
\endbibitem

\bibitem[\protect\citeauthoryear{{Overzier}}{2016}]{Overzier16}
\begin{barticle}
\bauthor{\bsnm{{Overzier}}, \binits{R.A.}}:
\batitle{{The realm of the galaxy protoclusters. A review}}.
\bjtitle{\aapr}
\bvolume{24}(\bissue{1}),
\bfpage{14}
(\byear{2016})
\doiurl{10.1007/s00159-016-0100-3}
{\href{https://arxiv.org/abs/1610.05201}{{arXiv:1610.05201}}}
{[astro-ph.GA]}
\end{barticle}
\endbibitem

\bibitem[\protect\citeauthoryear{{Harikane} et~al.}{2019}]{Harikane19}
\begin{barticle}
\bauthor{\bsnm{{Harikane}}, \binits{Y.}},
\bauthor{\bsnm{{Ouchi}}, \binits{M.}},
\bauthor{\bsnm{{Ono}}, \binits{Y.}},
\bauthor{\bsnm{{Fujimoto}}, \binits{S.}},
\bauthor{\bsnm{{Donevski}}, \binits{D.}},
\bauthor{\bsnm{{Shibuya}}, \binits{T.}},
\bauthor{\bsnm{{Faisst}}, \binits{A.L.}},
\bauthor{\bsnm{{Goto}}, \binits{T.}},
\bauthor{\bsnm{{Hatsukade}}, \binits{B.}},
\bauthor{\bsnm{{Kashikawa}}, \binits{N.}},
\bauthor{\bsnm{{Kohno}}, \binits{K.}},
\bauthor{\bsnm{{Hashimoto}}, \binits{T.}},
\bauthor{\bsnm{{Higuchi}}, \binits{R.}},
\bauthor{\bsnm{{Inoue}}, \binits{A.K.}},
\bauthor{\bsnm{{Lin}}, \binits{Y.-T.}},
\bauthor{\bsnm{{Martin}}, \binits{C.L.}},
\bauthor{\bsnm{{Overzier}}, \binits{R.}},
\bauthor{\bsnm{{Smail}}, \binits{I.}},
\bauthor{\bsnm{{Toshikawa}}, \binits{J.}},
\bauthor{\bsnm{{Umehata}}, \binits{H.}},
\bauthor{\bsnm{{Ao}}, \binits{Y.}},
\bauthor{\bsnm{{Chapman}}, \binits{S.}},
\bauthor{\bsnm{{Clements}}, \binits{D.L.}},
\bauthor{\bsnm{{Im}}, \binits{M.}},
\bauthor{\bsnm{{Jing}}, \binits{Y.}},
\bauthor{\bsnm{{Kawaguchi}}, \binits{T.}},
\bauthor{\bsnm{{Lee}}, \binits{C.-H.}},
\bauthor{\bsnm{{Lee}}, \binits{M.M.}},
\bauthor{\bsnm{{Lin}}, \binits{L.}},
\bauthor{\bsnm{{Matsuoka}}, \binits{Y.}},
\bauthor{\bsnm{{Marinello}}, \binits{M.}},
\bauthor{\bsnm{{Nagao}}, \binits{T.}},
\bauthor{\bsnm{{Onodera}}, \binits{M.}},
\bauthor{\bsnm{{Toft}}, \binits{S.}},
\bauthor{\bsnm{{Wang}}, \binits{W.-H.}}:
\batitle{{SILVERRUSH. VIII. Spectroscopic Identifications of Early Large-scale Structures with Protoclusters over 200 Mpc at z {\ensuremath{\sim}} 6-7: Strong Associations of Dusty Star-forming Galaxies}}.
\bjtitle{\apj}
\bvolume{883}(\bissue{2}),
\bfpage{142}
(\byear{2019})
\doiurl{10.3847/1538-4357/ab2cd5}
{\href{https://arxiv.org/abs/1902.09555}{{arXiv:1902.09555}}}
{[astro-ph.GA]}
\end{barticle}
\endbibitem

\bibitem[\protect\citeauthoryear{{Hu} et~al.}{2021}]{Hu21}
\begin{barticle}
\bauthor{\bsnm{{Hu}}, \binits{W.}},
\bauthor{\bsnm{{Wang}}, \binits{J.}},
\bauthor{\bsnm{{Infante}}, \binits{L.}},
\bauthor{\bsnm{{Rhoads}}, \binits{J.E.}},
\bauthor{\bsnm{{Zheng}}, \binits{Z.-Y.}},
\bauthor{\bsnm{{Yang}}, \binits{H.}},
\bauthor{\bsnm{{Malhotra}}, \binits{S.}},
\bauthor{\bsnm{{Barrientos}}, \binits{L.F.}},
\bauthor{\bsnm{{Jiang}}, \binits{C.}},
\bauthor{\bsnm{{Gonz{\'a}lez-L{\'o}pez}}, \binits{J.}},
\bauthor{\bsnm{{Prieto}}, \binits{G.}},
\bauthor{\bsnm{{Perez}}, \binits{L.A.}},
\bauthor{\bsnm{{Hibon}}, \binits{P.}},
\bauthor{\bsnm{{Galaz}}, \binits{G.}},
\bauthor{\bsnm{{Coughlin}}, \binits{A.}},
\bauthor{\bsnm{{Harish}}, \binits{S.}},
\bauthor{\bsnm{{Kong}}, \binits{X.}},
\bauthor{\bsnm{{Kang}}, \binits{W.}},
\bauthor{\bsnm{{Khostovan}}, \binits{A.A.}},
\bauthor{\bsnm{{Pharo}}, \binits{J.}},
\bauthor{\bsnm{{Valdes}}, \binits{F.}},
\bauthor{\bsnm{{Wold}}, \binits{I.}},
\bauthor{\bsnm{{Walker}}, \binits{A.R.}},
\bauthor{\bsnm{{Zheng}}, \binits{X.}}:
\batitle{{A Lyman-{\ensuremath{\alpha}} protocluster at redshift 6.9}}.
\bjtitle{Nature Astronomy}
\bvolume{5},
\bfpage{485}--\blpage{490}
(\byear{2021})
\doiurl{10.1038/s41550-020-01291-y}
{\href{https://arxiv.org/abs/2101.10204}{{arXiv:2101.10204}}}
{[astro-ph.GA]}
\end{barticle}
\endbibitem

\bibitem[\protect\citeauthoryear{{Helton} et~al.}{2023}]{Helton23}
\begin{botherref}
\oauthor{\bsnm{{Helton}}, \binits{J.M.}},
\oauthor{\bsnm{{Sun}}, \binits{F.}},
\oauthor{\bsnm{{Woodrum}}, \binits{C.}},
\oauthor{\bsnm{{Hainline}}, \binits{K.N.}},
\oauthor{\bsnm{{Willmer}}, \binits{C.N.A.}},
\oauthor{\bsnm{{Rieke}}, \binits{G.H.}},
\oauthor{\bsnm{{Rieke}}, \binits{M.J.}},
\oauthor{\bsnm{{Tacchella}}, \binits{S.}},
\oauthor{\bsnm{{Robertson}}, \binits{B.}},
\oauthor{\bsnm{{Johnson}}, \binits{B.D.}},
\oauthor{\bsnm{{Alberts}}, \binits{S.}},
\oauthor{\bsnm{{Eisenstein}}, \binits{D.J.}},
\oauthor{\bsnm{{Hausen}}, \binits{R.}},
\oauthor{\bsnm{{Bonaventura}}, \binits{N.R.}},
\oauthor{\bsnm{{Bunker}}, \binits{A.}},
\oauthor{\bsnm{{Charlot}}, \binits{S.}},
\oauthor{\bsnm{{Curti}}, \binits{M.}},
\oauthor{\bsnm{{Curtis-Lake}}, \binits{E.}},
\oauthor{\bsnm{{Looser}}, \binits{T.J.}},
\oauthor{\bsnm{{Maiolino}}, \binits{R.}},
\oauthor{\bsnm{{Willott}}, \binits{C.}},
\oauthor{\bsnm{{Witstok}}, \binits{J.}},
\oauthor{\bsnm{{Boyett}}, \binits{K.}},
\oauthor{\bsnm{{Chen}}, \binits{Z.}},
\oauthor{\bsnm{{Egami}}, \binits{E.}},
\oauthor{\bsnm{{Endsley}}, \binits{R.}},
\oauthor{\bsnm{{Hviding}}, \binits{R.E.}},
\oauthor{\bsnm{{Jaffe}}, \binits{D.T.}},
\oauthor{\bsnm{{Ji}}, \binits{Z.}},
\oauthor{\bsnm{{Lyu}}, \binits{J.}},
\oauthor{\bsnm{{Sandles}}, \binits{L.}}:
{The JWST Advanced Deep Extragalactic Survey: Discovery of an Extreme Galaxy Overdensity at $z = 5.4$ with JWST/NIRCam in GOODS-S}.
arXiv e-prints,
2302--10217
(2023)
\doiurl{10.48550/arXiv.2302.10217}
{\href{https://arxiv.org/abs/2302.10217}{{arXiv:2302.10217}}}
{[astro-ph.GA]}
\end{botherref}
\endbibitem

\bibitem[\protect\citeauthoryear{{Morishita} et~al.}{2023}]{Morishita23}
\begin{barticle}
\bauthor{\bsnm{{Morishita}}, \binits{T.}},
\bauthor{\bsnm{{Roberts-Borsani}}, \binits{G.}},
\bauthor{\bsnm{{Treu}}, \binits{T.}},
\bauthor{\bsnm{{Brammer}}, \binits{G.}},
\bauthor{\bsnm{{Mason}}, \binits{C.A.}},
\bauthor{\bsnm{{Trenti}}, \binits{M.}},
\bauthor{\bsnm{{Vulcani}}, \binits{B.}},
\bauthor{\bsnm{{Wang}}, \binits{X.}},
\bauthor{\bsnm{{Acebron}}, \binits{A.}},
\bauthor{\bsnm{{Bah{\'e}}}, \binits{Y.}},
\bauthor{\bsnm{{Bergamini}}, \binits{P.}},
\bauthor{\bsnm{{Boyett}}, \binits{K.}},
\bauthor{\bsnm{{Bradac}}, \binits{M.}},
\bauthor{\bsnm{{Calabr{\`o}}}, \binits{A.}},
\bauthor{\bsnm{{Castellano}}, \binits{M.}},
\bauthor{\bsnm{{Chen}}, \binits{W.}},
\bauthor{\bsnm{{De Lucia}}, \binits{G.}},
\bauthor{\bsnm{{Filippenko}}, \binits{A.V.}},
\bauthor{\bsnm{{Fontana}}, \binits{A.}},
\bauthor{\bsnm{{Glazebrook}}, \binits{K.}},
\bauthor{\bsnm{{Grillo}}, \binits{C.}},
\bauthor{\bsnm{{Henry}}, \binits{A.}},
\bauthor{\bsnm{{Jones}}, \binits{T.}},
\bauthor{\bsnm{{Kelly}}, \binits{P.L.}},
\bauthor{\bsnm{{Koekemoer}}, \binits{A.M.}},
\bauthor{\bsnm{{Leethochawalit}}, \binits{N.}},
\bauthor{\bsnm{{Lu}}, \binits{T.-Y.}},
\bauthor{\bsnm{{Marchesini}}, \binits{D.}},
\bauthor{\bsnm{{Mascia}}, \binits{S.}},
\bauthor{\bsnm{{Mercurio}}, \binits{A.}},
\bauthor{\bsnm{{Merlin}}, \binits{E.}},
\bauthor{\bsnm{{Metha}}, \binits{B.}},
\bauthor{\bsnm{{Nanayakkara}}, \binits{T.}},
\bauthor{\bsnm{{Nonino}}, \binits{M.}},
\bauthor{\bsnm{{Paris}}, \binits{D.}},
\bauthor{\bsnm{{Pentericci}}, \binits{L.}},
\bauthor{\bsnm{{Rosati}}, \binits{P.}},
\bauthor{\bsnm{{Santini}}, \binits{P.}},
\bauthor{\bsnm{{Strait}}, \binits{V.}},
\bauthor{\bsnm{{Vanzella}}, \binits{E.}},
\bauthor{\bsnm{{Windhorst}}, \binits{R.A.}},
\bauthor{\bsnm{{Xie}}, \binits{L.}}:
\batitle{{Early Results from GLASS-JWST. XIV. A Spectroscopically Confirmed Protocluster 650 Million Years after the Big Bang}}.
\bjtitle{\apjl}
\bvolume{947}(\bissue{2}),
\bfpage{24}
(\byear{2023})
\doiurl{10.3847/2041-8213/acb99e}
{\href{https://arxiv.org/abs/2211.09097}{{arXiv:2211.09097}}}
{[astro-ph.GA]}
\end{barticle}
\endbibitem

\bibitem[\protect\citeauthoryear{{Lovell} et~al.}{2021}]{Lovell21}
\begin{barticle}
\bauthor{\bsnm{{Lovell}}, \binits{C.C.}},
\bauthor{\bsnm{{Vijayan}}, \binits{A.P.}},
\bauthor{\bsnm{{Thomas}}, \binits{P.A.}},
\bauthor{\bsnm{{Wilkins}}, \binits{S.M.}},
\bauthor{\bsnm{{Barnes}}, \binits{D.J.}},
\bauthor{\bsnm{{Irodotou}}, \binits{D.}},
\bauthor{\bsnm{{Roper}}, \binits{W.}}:
\batitle{{First Light And Reionization Epoch Simulations (FLARES) - I. Environmental dependence of high-redshift galaxy evolution}}.
\bjtitle{\mnras}
\bvolume{500}(\bissue{2}),
\bfpage{2127}--\blpage{2145}
(\byear{2021})
\doiurl{10.1093/mnras/staa3360}
{\href{https://arxiv.org/abs/2004.07283}{{arXiv:2004.07283}}}
{[astro-ph.GA]}
\end{barticle}
\endbibitem

\bibitem[\protect\citeauthoryear{{Springel} et~al.}{2021}]{Springel21}
\begin{barticle}
\bauthor{\bsnm{{Springel}}, \binits{V.}},
\bauthor{\bsnm{{Pakmor}}, \binits{R.}},
\bauthor{\bsnm{{Zier}}, \binits{O.}},
\bauthor{\bsnm{{Reinecke}}, \binits{M.}}:
\batitle{{Simulating cosmic structure formation with the GADGET-4 code}}.
\bjtitle{\mnras}
\bvolume{506}(\bissue{2}),
\bfpage{2871}--\blpage{2949}
(\byear{2021})
\doiurl{10.1093/mnras/stab1855}
{\href{https://arxiv.org/abs/2010.03567}{{arXiv:2010.03567}}}
{[astro-ph.IM]}
\end{barticle}
\endbibitem

\bibitem[\protect\citeauthoryear{{Wolfe} et~al.}{2005}]{Wolfe05}
\begin{barticle}
\bauthor{\bsnm{{Wolfe}}, \binits{A.M.}},
\bauthor{\bsnm{{Gawiser}}, \binits{E.}},
\bauthor{\bsnm{{Prochaska}}, \binits{J.X.}}:
\batitle{{Damped Ly {\ensuremath{\alpha}} Systems}}.
\bjtitle{\araa}
\bvolume{43}(\bissue{1}),
\bfpage{861}--\blpage{918}
(\byear{2005})
\doiurl{10.1146/annurev.astro.42.053102.133950}
{\href{https://arxiv.org/abs/astro-ph/0509481}{{arXiv:astro-ph/0509481}}}
{[astro-ph]}
\end{barticle}
\endbibitem

\bibitem[\protect\citeauthoryear{{Prochaska} and {Wolfe}}{2009}]{Prochaska09}
\begin{barticle}
\bauthor{\bsnm{{Prochaska}}, \binits{J.X.}},
\bauthor{\bsnm{{Wolfe}}, \binits{A.M.}}:
\batitle{{On the (Non)Evolution of H I Gas in Galaxies Over Cosmic Time}}.
\bjtitle{\apj}
\bvolume{696}(\bissue{2}),
\bfpage{1543}--\blpage{1547}
(\byear{2009})
\doiurl{10.1088/0004-637X/696/2/1543}
{\href{https://arxiv.org/abs/0811.2003}{{arXiv:0811.2003}}}
{[astro-ph]}
\end{barticle}
\endbibitem

\bibitem[\protect\citeauthoryear{{Noterdaeme} et~al.}{2012}]{Noterdaeme12}
\begin{barticle}
\bauthor{\bsnm{{Noterdaeme}}, \binits{P.}},
\bauthor{\bsnm{{Petitjean}}, \binits{P.}},
\bauthor{\bsnm{{Carithers}}, \binits{W.C.}},
\bauthor{\bsnm{{P{\^a}ris}}, \binits{I.}},
\bauthor{\bsnm{{Font-Ribera}}, \binits{A.}},
\bauthor{\bsnm{{Bailey}}, \binits{S.}},
\bauthor{\bsnm{{Aubourg}}, \binits{E.}},
\bauthor{\bsnm{{Bizyaev}}, \binits{D.}},
\bauthor{\bsnm{{Ebelke}}, \binits{G.}},
\bauthor{\bsnm{{Finley}}, \binits{H.}},
\bauthor{\bsnm{{Ge}}, \binits{J.}},
\bauthor{\bsnm{{Malanushenko}}, \binits{E.}},
\bauthor{\bsnm{{Malanushenko}}, \binits{V.}},
\bauthor{\bsnm{{Miralda-Escud{\'e}}}, \binits{J.}},
\bauthor{\bsnm{{Myers}}, \binits{A.D.}},
\bauthor{\bsnm{{Oravetz}}, \binits{D.}},
\bauthor{\bsnm{{Pan}}, \binits{K.}},
\bauthor{\bsnm{{Pieri}}, \binits{M.M.}},
\bauthor{\bsnm{{Ross}}, \binits{N.P.}},
\bauthor{\bsnm{{Schneider}}, \binits{D.P.}},
\bauthor{\bsnm{{Simmons}}, \binits{A.}},
\bauthor{\bsnm{{York}}, \binits{D.G.}}:
\batitle{{Column density distribution and cosmological mass density of neutral gas: Sloan Digital Sky Survey-III Data Release 9}}.
\bjtitle{\aap}
\bvolume{547},
\bfpage{1}
(\byear{2012})
\doiurl{10.1051/0004-6361/201220259}
{\href{https://arxiv.org/abs/1210.1213}{{arXiv:1210.1213}}}
{[astro-ph.CO]}
\end{barticle}
\endbibitem

\bibitem[\protect\citeauthoryear{{Fynbo} et~al.}{2009}]{Fynbo09}
\begin{barticle}
\bauthor{\bsnm{{Fynbo}}, \binits{J.P.U.}},
\bauthor{\bsnm{{Jakobsson}}, \binits{P.}},
\bauthor{\bsnm{{Prochaska}}, \binits{J.X.}},
\bauthor{\bsnm{{Malesani}}, \binits{D.}},
\bauthor{\bsnm{{Ledoux}}, \binits{C.}},
\bauthor{\bsnm{{de Ugarte Postigo}}, \binits{A.}},
\bauthor{\bsnm{{Nardini}}, \binits{M.}},
\bauthor{\bsnm{{Vreeswijk}}, \binits{P.M.}},
\bauthor{\bsnm{{Wiersema}}, \binits{K.}},
\bauthor{\bsnm{{Hjorth}}, \binits{J.}},
\bauthor{\bsnm{{Sollerman}}, \binits{J.}},
\bauthor{\bsnm{{Chen}}, \binits{H.-W.}},
\bauthor{\bsnm{{Th{\"o}ne}}, \binits{C.C.}},
\bauthor{\bsnm{{Bj{\"o}rnsson}}, \binits{G.}},
\bauthor{\bsnm{{Bloom}}, \binits{J.S.}},
\bauthor{\bsnm{{Castro-Tirado}}, \binits{A.J.}},
\bauthor{\bsnm{{Christensen}}, \binits{L.}},
\bauthor{\bsnm{{De Cia}}, \binits{A.}},
\bauthor{\bsnm{{Fruchter}}, \binits{A.S.}},
\bauthor{\bsnm{{Gorosabel}}, \binits{J.}},
\bauthor{\bsnm{{Graham}}, \binits{J.F.}},
\bauthor{\bsnm{{Jaunsen}}, \binits{A.O.}},
\bauthor{\bsnm{{Jensen}}, \binits{B.L.}},
\bauthor{\bsnm{{Kann}}, \binits{D.A.}},
\bauthor{\bsnm{{Kouveliotou}}, \binits{C.}},
\bauthor{\bsnm{{Levan}}, \binits{A.J.}},
\bauthor{\bsnm{{Maund}}, \binits{J.}},
\bauthor{\bsnm{{Masetti}}, \binits{N.}},
\bauthor{\bsnm{{Milvang-Jensen}}, \binits{B.}},
\bauthor{\bsnm{{Palazzi}}, \binits{E.}},
\bauthor{\bsnm{{Perley}}, \binits{D.A.}},
\bauthor{\bsnm{{Pian}}, \binits{E.}},
\bauthor{\bsnm{{Rol}}, \binits{E.}},
\bauthor{\bsnm{{Schady}}, \binits{P.}},
\bauthor{\bsnm{{Starling}}, \binits{R.L.C.}},
\bauthor{\bsnm{{Tanvir}}, \binits{N.R.}},
\bauthor{\bsnm{{Watson}}, \binits{D.J.}},
\bauthor{\bsnm{{Xu}}, \binits{D.}},
\bauthor{\bsnm{{Augusteijn}}, \binits{T.}},
\bauthor{\bsnm{{Grundahl}}, \binits{F.}},
\bauthor{\bsnm{{Telting}}, \binits{J.}},
\bauthor{\bsnm{{Quirion}}, \binits{P.-O.}}:
\batitle{{Low-resolution Spectroscopy of Gamma-ray Burst Optical Afterglows: Biases in the Swift Sample and Characterization of the Absorbers}}.
\bjtitle{\apjs}
\bvolume{185}(\bissue{2}),
\bfpage{526}--\blpage{573}
(\byear{2009})
\doiurl{10.1088/0067-0049/185/2/526}
{\href{https://arxiv.org/abs/0907.3449}{{arXiv:0907.3449}}}
{[astro-ph.CO]}
\end{barticle}
\endbibitem

\bibitem[\protect\citeauthoryear{{Tanvir} et~al.}{2019}]{Tanvir19}
\begin{barticle}
\bauthor{\bsnm{{Tanvir}}, \binits{N.R.}},
\bauthor{\bsnm{{Fynbo}}, \binits{J.P.U.}},
\bauthor{\bsnm{{de Ugarte Postigo}}, \binits{A.}},
\bauthor{\bsnm{{Japelj}}, \binits{J.}},
\bauthor{\bsnm{{Wiersema}}, \binits{K.}},
\bauthor{\bsnm{{Malesani}}, \binits{D.}},
\bauthor{\bsnm{{Perley}}, \binits{D.A.}},
\bauthor{\bsnm{{Levan}}, \binits{A.J.}},
\bauthor{\bsnm{{Selsing}}, \binits{J.}},
\bauthor{\bsnm{{Cenko}}, \binits{S.B.}},
\bauthor{\bsnm{{Kann}}, \binits{D.A.}},
\bauthor{\bsnm{{Milvang-Jensen}}, \binits{B.}},
\bauthor{\bsnm{{Berger}}, \binits{E.}},
\bauthor{\bsnm{{Cano}}, \binits{Z.}},
\bauthor{\bsnm{{Chornock}}, \binits{R.}},
\bauthor{\bsnm{{Covino}}, \binits{S.}},
\bauthor{\bsnm{{Cucchiara}}, \binits{A.}},
\bauthor{\bsnm{{D'Elia}}, \binits{V.}},
\bauthor{\bsnm{{Gargiulo}}, \binits{A.}},
\bauthor{\bsnm{{Goldoni}}, \binits{P.}},
\bauthor{\bsnm{{Gomboc}}, \binits{A.}},
\bauthor{\bsnm{{Heintz}}, \binits{K.E.}},
\bauthor{\bsnm{{Hjorth}}, \binits{J.}},
\bauthor{\bsnm{{Izzo}}, \binits{L.}},
\bauthor{\bsnm{{Jakobsson}}, \binits{P.}},
\bauthor{\bsnm{{Kaper}}, \binits{L.}},
\bauthor{\bsnm{{Kr{\"u}hler}}, \binits{T.}},
\bauthor{\bsnm{{Laskar}}, \binits{T.}},
\bauthor{\bsnm{{Myers}}, \binits{M.}},
\bauthor{\bsnm{{Piranomonte}}, \binits{S.}},
\bauthor{\bsnm{{Pugliese}}, \binits{G.}},
\bauthor{\bsnm{{Rossi}}, \binits{A.}},
\bauthor{\bsnm{{S{\'a}nchez-Ram{\'\i}rez}}, \binits{R.}},
\bauthor{\bsnm{{Schulze}}, \binits{S.}},
\bauthor{\bsnm{{Sparre}}, \binits{M.}},
\bauthor{\bsnm{{Stanway}}, \binits{E.R.}},
\bauthor{\bsnm{{Tagliaferri}}, \binits{G.}},
\bauthor{\bsnm{{Th{\"o}ne}}, \binits{C.C.}},
\bauthor{\bsnm{{Vergani}}, \binits{S.}},
\bauthor{\bsnm{{Vreeswijk}}, \binits{P.M.}},
\bauthor{\bsnm{{Wijers}}, \binits{R.A.M.J.}},
\bauthor{\bsnm{{Watson}}, \binits{D.}},
\bauthor{\bsnm{{Xu}}, \binits{D.}}:
\batitle{{The fraction of ionizing radiation from massive stars that escapes to the intergalactic medium}}.
\bjtitle{\mnras}
\bvolume{483}(\bissue{4}),
\bfpage{5380}--\blpage{5408}
(\byear{2019})
\doiurl{10.1093/mnras/sty3460}
{\href{https://arxiv.org/abs/1805.07318}{{arXiv:1805.07318}}}
{[astro-ph.GA]}
\end{barticle}
\endbibitem

\bibitem[\protect\citeauthoryear{{Heintz} et~al.}{2023}]{Heintz23_GRB}
\begin{barticle}
\bauthor{\bsnm{{Heintz}}, \binits{K.E.}},
\bauthor{\bsnm{{De Cia}}, \binits{A.}},
\bauthor{\bsnm{{Th{\"o}ne}}, \binits{C.C.}},
\bauthor{\bsnm{{Krogager}}, \binits{J.-K.}},
\bauthor{\bsnm{{Yates}}, \binits{R.M.}},
\bauthor{\bsnm{{Vejlgaard}}, \binits{S.}},
\bauthor{\bsnm{{Konstantopoulou}}, \binits{C.}},
\bauthor{\bsnm{{Fynbo}}, \binits{J.P.U.}},
\bauthor{\bsnm{{Watson}}, \binits{D.}},
\bauthor{\bsnm{{Narayanan}}, \binits{D.}},
\bauthor{\bsnm{{Wilson}}, \binits{S.N.}},
\bauthor{\bsnm{{Arabsalmani}}, \binits{M.}},
\bauthor{\bsnm{{Campana}}, \binits{S.}},
\bauthor{\bsnm{{D'Elia}}, \binits{V.}},
\bauthor{\bsnm{{De Pasquale}}, \binits{M.}},
\bauthor{\bsnm{{Hartmann}}, \binits{D.H.}},
\bauthor{\bsnm{{Izzo}}, \binits{L.}},
\bauthor{\bsnm{{Jakobsson}}, \binits{P.}},
\bauthor{\bsnm{{Kouveliotou}}, \binits{C.}},
\bauthor{\bsnm{{Levan}}, \binits{A.}},
\bauthor{\bsnm{{Li}}, \binits{Q.}},
\bauthor{\bsnm{{Malesani}}, \binits{D.B.}},
\bauthor{\bsnm{{Melandri}}, \binits{A.}},
\bauthor{\bsnm{{Milvang-Jensen}}, \binits{B.}},
\bauthor{\bsnm{{M{\o}ller}}, \binits{P.}},
\bauthor{\bsnm{{Palazzi}}, \binits{E.}},
\bauthor{\bsnm{{Palmerio}}, \binits{J.}},
\bauthor{\bsnm{{Petitjean}}, \binits{P.}},
\bauthor{\bsnm{{Pugliese}}, \binits{G.}},
\bauthor{\bsnm{{Rossi}}, \binits{A.}},
\bauthor{\bsnm{{Saccardi}}, \binits{A.}},
\bauthor{\bsnm{{Salvaterra}}, \binits{R.}},
\bauthor{\bsnm{{Savaglio}}, \binits{S.}},
\bauthor{\bsnm{{Schady}}, \binits{P.}},
\bauthor{\bsnm{{Stratta}}, \binits{G.}},
\bauthor{\bsnm{{Tanvir}}, \binits{N.R.}},
\bauthor{\bsnm{{de Ugarte Postigo}}, \binits{A.}},
\bauthor{\bsnm{{Vergani}}, \binits{S.D.}},
\bauthor{\bsnm{{Wiersema}}, \binits{K.}},
\bauthor{\bsnm{{Wijers}}, \binits{R.A.M.J.}},
\bauthor{\bsnm{{Zafar}}, \binits{T.}}:
\batitle{{The cosmic buildup of dust and metals. Accurate abundances from GRB-selected star-forming galaxies at 1.7 < z < 6.3}}.
\bjtitle{\aap}
\bvolume{679},
\bfpage{91}
(\byear{2023})
\doiurl{10.1051/0004-6361/202347418}
{\href{https://arxiv.org/abs/2308.14812}{{arXiv:2308.14812}}}
{[astro-ph.GA]}
\end{barticle}
\endbibitem

\bibitem[\protect\citeauthoryear{{Heintz} et~al.}{2024a}]{Heintz24_DLAs}
\begin{barticle}
\bauthor{\bsnm{{Heintz}}, \binits{K.E.}},
\bauthor{\bsnm{{Watson}}, \binits{D.}},
\bauthor{\bsnm{{Brammer}}, \binits{G.}},
\bauthor{\bsnm{{Vejlgaard}}, \binits{S.}},
\bauthor{\bsnm{{Hutter}}, \binits{A.}},
\bauthor{\bsnm{{Strait}}, \binits{V.B.}},
\bauthor{\bsnm{{Matthee}}, \binits{J.}},
\bauthor{\bsnm{{Oesch}}, \binits{P.A.}},
\bauthor{\bsnm{{Jakobsson}}, \binits{P.}},
\bauthor{\bsnm{{Tanvir}}, \binits{N.R.}},
\bauthor{\bsnm{{Laursen}}, \binits{P.}},
\bauthor{\bsnm{{Naidu}}, \binits{R.P.}},
\bauthor{\bsnm{{Mason}}, \binits{C.A.}},
\bauthor{\bsnm{{Killi}}, \binits{M.}},
\bauthor{\bsnm{{Jung}}, \binits{I.}},
\bauthor{\bsnm{{Hsiao}}, \binits{T.Y.-Y.}},
\bauthor{\bsnm{{Abdurro'uf}}},
\bauthor{\bsnm{{Coe}}, \binits{D.}},
\bauthor{\bsnm{{Arrabal Haro}}, \binits{P.}},
\bauthor{\bsnm{{Finkelstein}}, \binits{S.L.}},
\bauthor{\bsnm{{Toft}}, \binits{S.}}:
\batitle{{Strong damped Lyman-{\ensuremath{\alpha}} absorption in young star-forming galaxies at redshifts 9 to 11}}.
\bjtitle{Science}
\bvolume{384}(\bissue{6698}),
\bfpage{890}--\blpage{894}
(\byear{2024})
\doiurl{10.1126/science.adj0343}
{\href{https://arxiv.org/abs/2306.00647}{{arXiv:2306.00647}}}
{[astro-ph.GA]}
\end{barticle}
\endbibitem

\bibitem[\protect\citeauthoryear{{Heintz} et~al.}{2024b}]{Heintz24}
\begin{botherref}
\oauthor{\bsnm{{Heintz}}, \binits{K.E.}},
\oauthor{\bsnm{{Brammer}}, \binits{G.B.}},
\oauthor{\bsnm{{Watson}}, \binits{D.}},
\oauthor{\bsnm{{Oesch}}, \binits{P.A.}},
\oauthor{\bsnm{{Keating}}, \binits{L.C.}},
\oauthor{\bsnm{{Hayes}}, \binits{M.J.}},
\oauthor{\bsnm{{Abdurro'uf}}},
\oauthor{\bsnm{{Arellano-C{\'o}rdova}}, \binits{K.Z.}},
\oauthor{\bsnm{{Carnall}}, \binits{A.C.}},
\oauthor{\bsnm{{Christiansen}}, \binits{C.R.}},
\oauthor{\bsnm{{Cullen}}, \binits{F.}},
\oauthor{\bsnm{{Dav{\'e}}}, \binits{R.}},
\oauthor{\bsnm{{Dayal}}, \binits{P.}},
\oauthor{\bsnm{{Ferrara}}, \binits{A.}},
\oauthor{\bsnm{{Finlator}}, \binits{K.}},
\oauthor{\bsnm{{Fynbo}}, \binits{J.P.U.}},
\oauthor{\bsnm{{Flury}}, \binits{S.R.}},
\oauthor{\bsnm{{Gelli}}, \binits{V.}},
\oauthor{\bsnm{{Gillman}}, \binits{S.}},
\oauthor{\bsnm{{Gottumukkala}}, \binits{R.}},
\oauthor{\bsnm{{Gould}}, \binits{K.}},
\oauthor{\bsnm{{Greve}}, \binits{T.R.}},
\oauthor{\bsnm{{Hardin}}, \binits{S.E.}},
\oauthor{\bsnm{{Y. -Y Hsiao}}, \binits{T.}},
\oauthor{\bsnm{{Hutter}}, \binits{A.}},
\oauthor{\bsnm{{Jakobsson}}, \binits{P.}},
\oauthor{\bsnm{{Killi}}, \binits{M.}},
\oauthor{\bsnm{{Khosravaninezhad}}, \binits{N.}},
\oauthor{\bsnm{{Laursen}}, \binits{P.}},
\oauthor{\bsnm{{Lee}}, \binits{M.M.}},
\oauthor{\bsnm{{Magdis}}, \binits{G.E.}},
\oauthor{\bsnm{{Matthee}}, \binits{J.}},
\oauthor{\bsnm{{Naidu}}, \binits{R.P.}},
\oauthor{\bsnm{{Narayanan}}, \binits{D.}},
\oauthor{\bsnm{{Pollock}}, \binits{C.}},
\oauthor{\bsnm{{Prescott}}, \binits{M.}},
\oauthor{\bsnm{{Rusakov}}, \binits{V.}},
\oauthor{\bsnm{{Shuntov}}, \binits{M.}},
\oauthor{\bsnm{{Sneppen}}, \binits{A.}},
\oauthor{\bsnm{{Smit}}, \binits{R.}},
\oauthor{\bsnm{{Tanvir}}, \binits{N.R.}},
\oauthor{\bsnm{{Terp}}, \binits{C.}},
\oauthor{\bsnm{{Toft}}, \binits{S.}},
\oauthor{\bsnm{{Valentino}}, \binits{F.}},
\oauthor{\bsnm{{Vijayan}}, \binits{A.P.}},
\oauthor{\bsnm{{Weaver}}, \binits{J.R.}},
\oauthor{\bsnm{{Wise}}, \binits{J.H.}},
\oauthor{\bsnm{{Witstok}}, \binits{J.}}:
{The JWST-PRIMAL Legacy Survey. A JWST/NIRSpec reference sample for the physical properties and Lyman-$\alpha$ absorption and emission of $\sim 500$ galaxies at $z=5.5-13.4$}.
arXiv e-prints,
2404--02211
(2024)
\doiurl{10.48550/arXiv.2404.02211}
{\href{https://arxiv.org/abs/2404.02211}{{arXiv:2404.02211}}}
{[astro-ph.GA]}
\end{botherref}
\endbibitem

\bibitem[\protect\citeauthoryear{{Terp} et~al.}{2024}]{Terp24}
\begin{botherref}
\oauthor{\bsnm{{Terp}}, \binits{C.}},
\oauthor{\bsnm{{Heintz}}, \binits{K.E.}},
\oauthor{\bsnm{{Watson}}, \binits{D.}},
\oauthor{\bsnm{{Brammer}}, \binits{G.}},
\oauthor{\bsnm{{Carnall}}, \binits{A.}},
\oauthor{\bsnm{{Witstok}}, \binits{J.}},
\oauthor{\bsnm{{Smit}}, \binits{R.}},
\oauthor{\bsnm{{Vejlgaard}}, \binits{S.}}:
{Uncovering the physical origin of the prominent Lyman-$\alpha$ emission and absorption in GS9422 at $z = 5.943$}.
arXiv e-prints,
2404--06543
(2024)
\doiurl{10.48550/arXiv.2404.06543}
{\href{https://arxiv.org/abs/2404.06543}{{arXiv:2404.06543}}}
{[astro-ph.GA]}
\end{botherref}
\endbibitem

\bibitem[\protect\citeauthoryear{{Oesch} et~al.}{2023}]{Oesch23}
\begin{barticle}
\bauthor{\bsnm{{Oesch}}, \binits{P.A.}},
\bauthor{\bsnm{{Brammer}}, \binits{G.}},
\bauthor{\bsnm{{Naidu}}, \binits{R.P.}},
\bauthor{\bsnm{{Bouwens}}, \binits{R.J.}},
\bauthor{\bsnm{{Chisholm}}, \binits{J.}},
\bauthor{\bsnm{{Illingworth}}, \binits{G.D.}},
\bauthor{\bsnm{{Matthee}}, \binits{J.}},
\bauthor{\bsnm{{Nelson}}, \binits{E.}},
\bauthor{\bsnm{{Qin}}, \binits{Y.}},
\bauthor{\bsnm{{Reddy}}, \binits{N.}},
\bauthor{\bsnm{{Shapley}}, \binits{A.}},
\bauthor{\bsnm{{Shivaei}}, \binits{I.}},
\bauthor{\bsnm{{van Dokkum}}, \binits{P.}},
\bauthor{\bsnm{{Weibel}}, \binits{A.}},
\bauthor{\bsnm{{Whitaker}}, \binits{K.}},
\bauthor{\bsnm{{Wuyts}}, \binits{S.}},
\bauthor{\bsnm{{Covelo-Paz}}, \binits{A.}},
\bauthor{\bsnm{{Endsley}}, \binits{R.}},
\bauthor{\bsnm{{Fudamoto}}, \binits{Y.}},
\bauthor{\bsnm{{Giovinazzo}}, \binits{E.}},
\bauthor{\bsnm{{Herard-Demanche}}, \binits{T.}},
\bauthor{\bsnm{{Kerutt}}, \binits{J.}},
\bauthor{\bsnm{{Kramarenko}}, \binits{I.}},
\bauthor{\bsnm{{Labbe}}, \binits{I.}},
\bauthor{\bsnm{{Leonova}}, \binits{E.}},
\bauthor{\bsnm{{Lin}}, \binits{J.}},
\bauthor{\bsnm{{Magee}}, \binits{D.}},
\bauthor{\bsnm{{Marchesini}}, \binits{D.}},
\bauthor{\bsnm{{Maseda}}, \binits{M.}},
\bauthor{\bsnm{{Mason}}, \binits{C.}},
\bauthor{\bsnm{{Matharu}}, \binits{J.}},
\bauthor{\bsnm{{Meyer}}, \binits{R.A.}},
\bauthor{\bsnm{{Neufeld}}, \binits{C.}},
\bauthor{\bsnm{{Prieto Lyon}}, \binits{G.}},
\bauthor{\bsnm{{Schaerer}}, \binits{D.}},
\bauthor{\bsnm{{Sharma}}, \binits{R.}},
\bauthor{\bsnm{{Shuntov}}, \binits{M.}},
\bauthor{\bsnm{{Smit}}, \binits{R.}},
\bauthor{\bsnm{{Stefanon}}, \binits{M.}},
\bauthor{\bsnm{{Wyithe}}, \binits{J.S.B.}},
\bauthor{\bsnm{{Xiao}}, \binits{M.}}:
\batitle{{The JWST FRESCO survey: legacy NIRCam/grism spectroscopy and imaging in the two GOODS fields}}.
\bjtitle{\mnras}
\bvolume{525}(\bissue{2}),
\bfpage{2864}--\blpage{2874}
(\byear{2023})
\doiurl{10.1093/mnras/stad2411}
{\href{https://arxiv.org/abs/2304.02026}{{arXiv:2304.02026}}}
{[astro-ph.GA]}
\end{barticle}
\endbibitem

\bibitem[\protect\citeauthoryear{{Larson} et~al.}{2023}]{Larson23}
\begin{barticle}
\bauthor{\bsnm{{Larson}}, \binits{R.L.}},
\bauthor{\bsnm{{Hutchison}}, \binits{T.A.}},
\bauthor{\bsnm{{Bagley}}, \binits{M.}},
\bauthor{\bsnm{{Finkelstein}}, \binits{S.L.}},
\bauthor{\bsnm{{Yung}}, \binits{L.Y.A.}},
\bauthor{\bsnm{{Somerville}}, \binits{R.S.}},
\bauthor{\bsnm{{Hirschmann}}, \binits{M.}},
\bauthor{\bsnm{{Brammer}}, \binits{G.}},
\bauthor{\bsnm{{Holwerda}}, \binits{B.W.}},
\bauthor{\bsnm{{Papovich}}, \binits{C.}},
\bauthor{\bsnm{{Morales}}, \binits{A.M.}},
\bauthor{\bsnm{{Wilkins}}, \binits{S.M.}}:
\batitle{{Spectral Templates Optimal for Selecting Galaxies at z > 8 with the JWST}}.
\bjtitle{\apj}
\bvolume{958}(\bissue{2}),
\bfpage{141}
(\byear{2023})
\doiurl{10.3847/1538-4357/acfed4}
{\href{https://arxiv.org/abs/2211.10035}{{arXiv:2211.10035}}}
{[astro-ph.GA]}
\end{barticle}
\endbibitem

\bibitem[\protect\citeauthoryear{{Tepper-Garc{\'\i}a}}{2006}]{TepperGarcia06}
\begin{barticle}
\bauthor{\bsnm{{Tepper-Garc{\'\i}a}}, \binits{T.}}:
\batitle{{Voigt profile fitting to quasar absorption lines: an analytic approximation to the Voigt-Hjerting function}}.
\bjtitle{\mnras}
\bvolume{369}(\bissue{4}),
\bfpage{2025}--\blpage{2035}
(\byear{2006})
\doiurl{10.1111/j.1365-2966.2006.10450.x}
{\href{https://arxiv.org/abs/astro-ph/0602124}{{arXiv:astro-ph/0602124}}}
{[astro-ph]}
\end{barticle}
\endbibitem

\bibitem[\protect\citeauthoryear{{Bosman} et~al.}{2022}]{Bosman22}
\begin{barticle}
\bauthor{\bsnm{{Bosman}}, \binits{S.E.I.}},
\bauthor{\bsnm{{Davies}}, \binits{F.B.}},
\bauthor{\bsnm{{Becker}}, \binits{G.D.}},
\bauthor{\bsnm{{Keating}}, \binits{L.C.}},
\bauthor{\bsnm{{Davies}}, \binits{R.L.}},
\bauthor{\bsnm{{Zhu}}, \binits{Y.}},
\bauthor{\bsnm{{Eilers}}, \binits{A.-C.}},
\bauthor{\bsnm{{D'Odorico}}, \binits{V.}},
\bauthor{\bsnm{{Bian}}, \binits{F.}},
\bauthor{\bsnm{{Bischetti}}, \binits{M.}},
\bauthor{\bsnm{{Cristiani}}, \binits{S.V.}},
\bauthor{\bsnm{{Fan}}, \binits{X.}},
\bauthor{\bsnm{{Farina}}, \binits{E.P.}},
\bauthor{\bsnm{{Haehnelt}}, \binits{M.G.}},
\bauthor{\bsnm{{Hennawi}}, \binits{J.F.}},
\bauthor{\bsnm{{Kulkarni}}, \binits{G.}},
\bauthor{\bsnm{{Mesinger}}, \binits{A.}},
\bauthor{\bsnm{{Meyer}}, \binits{R.A.}},
\bauthor{\bsnm{{Onoue}}, \binits{M.}},
\bauthor{\bsnm{{Pallottini}}, \binits{A.}},
\bauthor{\bsnm{{Qin}}, \binits{Y.}},
\bauthor{\bsnm{{Ryan-Weber}}, \binits{E.}},
\bauthor{\bsnm{{Schindler}}, \binits{J.-T.}},
\bauthor{\bsnm{{Walter}}, \binits{F.}},
\bauthor{\bsnm{{Wang}}, \binits{F.}},
\bauthor{\bsnm{{Yang}}, \binits{J.}}:
\batitle{{Hydrogen reionization ends by z = 5.3: Lyman-{\ensuremath{\alpha}} optical depth measured by the XQR-30 sample}}.
\bjtitle{\mnras}
\bvolume{514}(\bissue{1}),
\bfpage{55}--\blpage{76}
(\byear{2022})
\doiurl{10.1093/mnras/stac1046}
{\href{https://arxiv.org/abs/2108.03699}{{arXiv:2108.03699}}}
{[astro-ph.CO]}
\end{barticle}
\endbibitem

\bibitem[\protect\citeauthoryear{{Walter} et~al.}{2020}]{Walter20}
\begin{barticle}
\bauthor{\bsnm{{Walter}}, \binits{F.}},
\bauthor{\bsnm{{Carilli}}, \binits{C.}},
\bauthor{\bsnm{{Neeleman}}, \binits{M.}},
\bauthor{\bsnm{{Decarli}}, \binits{R.}},
\bauthor{\bsnm{{Popping}}, \binits{G.}},
\bauthor{\bsnm{{Somerville}}, \binits{R.S.}},
\bauthor{\bsnm{{Aravena}}, \binits{M.}},
\bauthor{\bsnm{{Bertoldi}}, \binits{F.}},
\bauthor{\bsnm{{Boogaard}}, \binits{L.}},
\bauthor{\bsnm{{Cox}}, \binits{P.}},
\bauthor{\bsnm{{da Cunha}}, \binits{E.}},
\bauthor{\bsnm{{Magnelli}}, \binits{B.}},
\bauthor{\bsnm{{Obreschkow}}, \binits{D.}},
\bauthor{\bsnm{{Riechers}}, \binits{D.}},
\bauthor{\bsnm{{Rix}}, \binits{H.-W.}},
\bauthor{\bsnm{{Smail}}, \binits{I.}},
\bauthor{\bsnm{{Weiss}}, \binits{A.}},
\bauthor{\bsnm{{Assef}}, \binits{R.J.}},
\bauthor{\bsnm{{Bauer}}, \binits{F.}},
\bauthor{\bsnm{{Bouwens}}, \binits{R.}},
\bauthor{\bsnm{{Contini}}, \binits{T.}},
\bauthor{\bsnm{{Cortes}}, \binits{P.C.}},
\bauthor{\bsnm{{Daddi}}, \binits{E.}},
\bauthor{\bsnm{{Diaz-Santos}}, \binits{T.}},
\bauthor{\bsnm{{Gonz{\'a}lez-L{\'o}pez}}, \binits{J.}},
\bauthor{\bsnm{{Hennawi}}, \binits{J.}},
\bauthor{\bsnm{{Hodge}}, \binits{J.A.}},
\bauthor{\bsnm{{Inami}}, \binits{H.}},
\bauthor{\bsnm{{Ivison}}, \binits{R.}},
\bauthor{\bsnm{{Oesch}}, \binits{P.}},
\bauthor{\bsnm{{Sargent}}, \binits{M.}},
\bauthor{\bsnm{{van der Werf}}, \binits{P.}},
\bauthor{\bsnm{{Wagg}}, \binits{J.}},
\bauthor{\bsnm{{Yung}}, \binits{L.Y.A.}}:
\batitle{{The Evolution of the Baryons Associated with Galaxies Averaged over Cosmic Time and Space}}.
\bjtitle{\apj}
\bvolume{902}(\bissue{2}),
\bfpage{111}
(\byear{2020})
\doiurl{10.3847/1538-4357/abb82e}
{\href{https://arxiv.org/abs/2009.11126}{{arXiv:2009.11126}}}
{[astro-ph.GA]}
\end{barticle}
\endbibitem

\bibitem[\protect\citeauthoryear{{Bennett} and {Sijacki}}{2020}]{Bennett20}
\begin{barticle}
\bauthor{\bsnm{{Bennett}}, \binits{J.S.}},
\bauthor{\bsnm{{Sijacki}}, \binits{D.}}:
\batitle{{Resolving shocks and filaments in galaxy formation simulations: effects on gas properties and star formation in the circumgalactic medium}}.
\bjtitle{\mnras}
\bvolume{499}(\bissue{1}),
\bfpage{597}--\blpage{615}
(\byear{2020})
\doiurl{10.1093/mnras/staa2835}
{\href{https://arxiv.org/abs/2006.10058}{{arXiv:2006.10058}}}
{[astro-ph.GA]}
\end{barticle}
\endbibitem

\bibitem[\protect\citeauthoryear{{Rennehan}}{2024}]{Rennehan24}
\begin{botherref}
\oauthor{\bsnm{{Rennehan}}, \binits{D.}}:
{The Manhattan Suite: Accelerated galaxy evolution in the early Universe}.
arXiv e-prints,
2406--06672
(2024)
\doiurl{10.48550/arXiv.2406.06672}
{\href{https://arxiv.org/abs/2406.06672}{{arXiv:2406.06672}}}
{[astro-ph.GA]}
\end{botherref}
\endbibitem

\bibitem[\protect\citeauthoryear{{Keating} et~al.}{2024}]{Keating24}
\begin{barticle}
\bauthor{\bsnm{{Keating}}, \binits{L.C.}},
\bauthor{\bsnm{{Bolton}}, \binits{J.S.}},
\bauthor{\bsnm{{Cullen}}, \binits{F.}},
\bauthor{\bsnm{{Haehnelt}}, \binits{M.G.}},
\bauthor{\bsnm{{Puchwein}}, \binits{E.}},
\bauthor{\bsnm{{Kulkarni}}, \binits{G.}}:
\batitle{{JWST observations of galaxy damping wings during reionization interpreted with cosmological simulations}}.
\bjtitle{\mnras}
(\byear{2024})
\doiurl{10.1093/mnras/stae1530}
{\href{https://arxiv.org/abs/2308.05800}{{arXiv:2308.05800}}}
{[astro-ph.GA]}
\end{barticle}
\endbibitem

\bibitem[\protect\citeauthoryear{{Finlator} et~al.}{2018}]{Finlator18}
\begin{barticle}
\bauthor{\bsnm{{Finlator}}, \binits{K.}},
\bauthor{\bsnm{{Keating}}, \binits{L.}},
\bauthor{\bsnm{{Oppenheimer}}, \binits{B.D.}},
\bauthor{\bsnm{{Dav{\'e}}}, \binits{R.}},
\bauthor{\bsnm{{Zackrisson}}, \binits{E.}}:
\batitle{{Reionization in Technicolor}}.
\bjtitle{\mnras}
\bvolume{480}(\bissue{2}),
\bfpage{2628}--\blpage{2649}
(\byear{2018})
\doiurl{10.1093/mnras/sty1949}
{\href{https://arxiv.org/abs/1805.00099}{{arXiv:1805.00099}}}
{[astro-ph.CO]}
\end{barticle}
\endbibitem

\bibitem[\protect\citeauthoryear{{Witten} et~al.}{2024}]{Witten24}
\begin{barticle}
\bauthor{\bsnm{{Witten}}, \binits{C.}},
\bauthor{\bsnm{{Laporte}}, \binits{N.}},
\bauthor{\bsnm{{Martin-Alvarez}}, \binits{S.}},
\bauthor{\bsnm{{Sijacki}}, \binits{D.}},
\bauthor{\bsnm{{Yuan}}, \binits{Y.}},
\bauthor{\bsnm{{Haehnelt}}, \binits{M.G.}},
\bauthor{\bsnm{{Baker}}, \binits{W.M.}},
\bauthor{\bsnm{{Dunlop}}, \binits{J.S.}},
\bauthor{\bsnm{{Ellis}}, \binits{R.S.}},
\bauthor{\bsnm{{Grogin}}, \binits{N.A.}},
\bauthor{\bsnm{{Illingworth}}, \binits{G.}},
\bauthor{\bsnm{{Katz}}, \binits{H.}},
\bauthor{\bsnm{{Koekemoer}}, \binits{A.M.}},
\bauthor{\bsnm{{Magee}}, \binits{D.}},
\bauthor{\bsnm{{Maiolino}}, \binits{R.}},
\bauthor{\bsnm{{McClymont}}, \binits{W.}},
\bauthor{\bsnm{{P{\'e}rez-Gonz{\'a}lez}}, \binits{P.G.}},
\bauthor{\bsnm{{Pusk{\'a}s}}, \binits{D.}},
\bauthor{\bsnm{{Roberts-Borsani}}, \binits{G.}},
\bauthor{\bsnm{{Santini}}, \binits{P.}},
\bauthor{\bsnm{{Simmonds}}, \binits{C.}}:
\batitle{{Deciphering Lyman-{\ensuremath{\alpha}} emission deep into the epoch of reionization}}.
\bjtitle{Nature Astronomy}
\bvolume{8},
\bfpage{384}--\blpage{396}
(\byear{2024})
\doiurl{10.1038/s41550-023-02179-3}
{\href{https://arxiv.org/abs/2303.16225}{{arXiv:2303.16225}}}
{[astro-ph.GA]}
\end{barticle}
\endbibitem

\bibitem[\protect\citeauthoryear{{Angelinelli} et~al.}{2023}]{Angelinelli23}
\begin{barticle}
\bauthor{\bsnm{{Angelinelli}}, \binits{M.}},
\bauthor{\bsnm{{Ettori}}, \binits{S.}},
\bauthor{\bsnm{{Dolag}}, \binits{K.}},
\bauthor{\bsnm{{Vazza}}, \binits{F.}},
\bauthor{\bsnm{{Ragagnin}}, \binits{A.}}:
\batitle{{Redshift evolution of the baryon and gas fraction in simulated groups and clusters of galaxies}}.
\bjtitle{\aap}
\bvolume{675},
\bfpage{188}
(\byear{2023})
\doiurl{10.1051/0004-6361/202245782}
{\href{https://arxiv.org/abs/2305.09733}{{arXiv:2305.09733}}}
{[astro-ph.CO]}
\end{barticle}
\endbibitem

\bibitem[\protect\citeauthoryear{{Daddi} et~al.}{2022}]{Daddi22}
\begin{barticle}
\bauthor{\bsnm{{Daddi}}, \binits{E.}},
\bauthor{\bsnm{{Rich}}, \binits{R.M.}},
\bauthor{\bsnm{{Valentino}}, \binits{F.}},
\bauthor{\bsnm{{Jin}}, \binits{S.}},
\bauthor{\bsnm{{Delvecchio}}, \binits{I.}},
\bauthor{\bsnm{{Liu}}, \binits{D.}},
\bauthor{\bsnm{{Strazzullo}}, \binits{V.}},
\bauthor{\bsnm{{Neill}}, \binits{J.}},
\bauthor{\bsnm{{Gobat}}, \binits{R.}},
\bauthor{\bsnm{{Finoguenov}}, \binits{A.}},
\bauthor{\bsnm{{Bournaud}}, \binits{F.}},
\bauthor{\bsnm{{Elbaz}}, \binits{D.}},
\bauthor{\bsnm{{Kalita}}, \binits{B.S.}},
\bauthor{\bsnm{{O'Sullivan}}, \binits{D.}},
\bauthor{\bsnm{{Wang}}, \binits{T.}}:
\batitle{{Evidence for Cold-stream to Hot-accretion Transition as Traced by Ly{\ensuremath{\alpha}} Emission from Groups and Clusters at 2 < z < 3.3}}.
\bjtitle{\apjl}
\bvolume{926}(\bissue{2}),
\bfpage{21}
(\byear{2022})
\doiurl{10.3847/2041-8213/ac531f}
{\href{https://arxiv.org/abs/2202.03715}{{arXiv:2202.03715}}}
{[astro-ph.CO]}
\end{barticle}
\endbibitem

\bibitem[\protect\citeauthoryear{{Daddi} et~al.}{2021}]{Daddi21}
\begin{barticle}
\bauthor{\bsnm{{Daddi}}, \binits{E.}},
\bauthor{\bsnm{{Valentino}}, \binits{F.}},
\bauthor{\bsnm{{Rich}}, \binits{R.M.}},
\bauthor{\bsnm{{Neill}}, \binits{J.D.}},
\bauthor{\bsnm{{Gronke}}, \binits{M.}},
\bauthor{\bsnm{{O'Sullivan}}, \binits{D.}},
\bauthor{\bsnm{{Elbaz}}, \binits{D.}},
\bauthor{\bsnm{{Bournaud}}, \binits{F.}},
\bauthor{\bsnm{{Finoguenov}}, \binits{A.}},
\bauthor{\bsnm{{Marchal}}, \binits{A.}},
\bauthor{\bsnm{{Delvecchio}}, \binits{I.}},
\bauthor{\bsnm{{Jin}}, \binits{S.}},
\bauthor{\bsnm{{Liu}}, \binits{D.}},
\bauthor{\bsnm{{Strazzullo}}, \binits{V.}},
\bauthor{\bsnm{{Calabro}}, \binits{A.}},
\bauthor{\bsnm{{Coogan}}, \binits{R.}},
\bauthor{\bsnm{{D'Eugenio}}, \binits{C.}},
\bauthor{\bsnm{{Gobat}}, \binits{R.}},
\bauthor{\bsnm{{Kalita}}, \binits{B.S.}},
\bauthor{\bsnm{{Laursen}}, \binits{P.}},
\bauthor{\bsnm{{Martin}}, \binits{D.C.}},
\bauthor{\bsnm{{Puglisi}}, \binits{A.}},
\bauthor{\bsnm{{Schinnerer}}, \binits{E.}},
\bauthor{\bsnm{{Wang}}, \binits{T.}}:
\batitle{{Three Lyman-{\ensuremath{\alpha}}-emitting filaments converging to a massive galaxy group at z = 2.91: discussing the case for cold gas infall}}.
\bjtitle{\aap}
\bvolume{649},
\bfpage{78}
(\byear{2021})
\doiurl{10.1051/0004-6361/202038700}
{\href{https://arxiv.org/abs/2006.11089}{{arXiv:2006.11089}}}
{[astro-ph.GA]}
\end{barticle}
\endbibitem

\bibitem[\protect\citeauthoryear{{Leonova} et~al.}{2022}]{Leonova22}
\begin{barticle}
\bauthor{\bsnm{{Leonova}}, \binits{E.}},
\bauthor{\bsnm{{Oesch}}, \binits{P.A.}},
\bauthor{\bsnm{{Qin}}, \binits{Y.}},
\bauthor{\bsnm{{Naidu}}, \binits{R.P.}},
\bauthor{\bsnm{{Wyithe}}, \binits{J.S.B.}},
\bauthor{\bsnm{{de Barros}}, \binits{S.}},
\bauthor{\bsnm{{Bouwens}}, \binits{R.J.}},
\bauthor{\bsnm{{Ellis}}, \binits{R.S.}},
\bauthor{\bsnm{{Endsley}}, \binits{R.M.}},
\bauthor{\bsnm{{Hutter}}, \binits{A.}},
\bauthor{\bsnm{{Illingworth}}, \binits{G.D.}},
\bauthor{\bsnm{{Kerutt}}, \binits{J.}},
\bauthor{\bsnm{{Labb{\'e}}}, \binits{I.}},
\bauthor{\bsnm{{Laporte}}, \binits{N.}},
\bauthor{\bsnm{{Magee}}, \binits{D.}},
\bauthor{\bsnm{{Mutch}}, \binits{S.J.}},
\bauthor{\bsnm{{Roberts-Borsani}}, \binits{G.W.}},
\bauthor{\bsnm{{Smit}}, \binits{R.}},
\bauthor{\bsnm{{Stark}}, \binits{D.P.}},
\bauthor{\bsnm{{Stefanon}}, \binits{M.}},
\bauthor{\bsnm{{Tacchella}}, \binits{S.}},
\bauthor{\bsnm{{Zitrin}}, \binits{A.}}:
\batitle{{The prevalence of galaxy overdensities around UV-luminous Lyman-$\alpha$ emitters in the Epoch of Reionization}}.
\bjtitle{\mnras}
\bvolume{515}(\bissue{4}),
\bfpage{5790}--\blpage{5801}
(\byear{2022})
\doiurl{10.1093/mnras/stac1908}
{\href{https://arxiv.org/abs/2112.07675}{{arXiv:2112.07675}}}
{[astro-ph.GA]}
\end{barticle}
\endbibitem

\bibitem[\protect\citeauthoryear{{Brammer}}{2023a}]{Brammer23_grizli}
\begin{botherref}
\oauthor{\bsnm{{Brammer}}, \binits{G.}}:
{grizli}.
\doiurl{10.5281/zenodo.1146904}
\end{botherref}
\endbibitem

\bibitem[\protect\citeauthoryear{{Brammer}}{2023b}]{Brammer23_msaexp}
\begin{botherref}
\oauthor{\bsnm{{Brammer}}, \binits{G.}}:
{msaexp: NIRSpec Analyis Tools}.
\doiurl{10.5281/zenodo.7299500}
\end{botherref}
\endbibitem

\bibitem[\protect\citeauthoryear{{Planck Collaboration} et~al.}{2020}]{Planck18}
\begin{barticle}
\bauthor{\bsnm{{Planck Collaboration}}},
\bauthor{\bsnm{{Aghanim}}, \binits{N.}},
\bauthor{\bsnm{{Akrami}}, \binits{Y.}},
\bauthor{\bsnm{{Ashdown}}, \binits{M.}},
\bauthor{\bsnm{{Aumont}}, \binits{J.}},
\bauthor{\bsnm{{Baccigalupi}}, \binits{C.}},
\bauthor{\bsnm{{Ballardini}}, \binits{M.}},
\bauthor{\bsnm{{Banday}}, \binits{A.J.}},
\bauthor{\bsnm{{Barreiro}}, \binits{R.B.}},
\bauthor{\bsnm{{Bartolo}}, \binits{N.}},
\bauthor{\bsnm{{Basak}}, \binits{S.}},
\bauthor{\bsnm{{Battye}}, \binits{R.}},
\bauthor{\bsnm{{Benabed}}, \binits{K.}},
\bauthor{\bsnm{{Bernard}}, \binits{J.-P.}},
\bauthor{\bsnm{{Bersanelli}}, \binits{M.}},
\bauthor{\bsnm{{Bielewicz}}, \binits{P.}},
\bauthor{\bsnm{{Bock}}, \binits{J.J.}},
\bauthor{\bsnm{{Bond}}, \binits{J.R.}},
\bauthor{\bsnm{{Borrill}}, \binits{J.}},
\bauthor{\bsnm{{Bouchet}}, \binits{F.R.}},
\bauthor{\bsnm{{Boulanger}}, \binits{F.}},
\bauthor{\bsnm{{Bucher}}, \binits{M.}},
\bauthor{\bsnm{{Burigana}}, \binits{C.}},
\bauthor{\bsnm{{Butler}}, \binits{R.C.}},
\bauthor{\bsnm{{Calabrese}}, \binits{E.}},
\bauthor{\bsnm{{Cardoso}}, \binits{J.-F.}},
\bauthor{\bsnm{{Carron}}, \binits{J.}},
\bauthor{\bsnm{{Challinor}}, \binits{A.}},
\bauthor{\bsnm{{Chiang}}, \binits{H.C.}},
\bauthor{\bsnm{{Chluba}}, \binits{J.}},
\bauthor{\bsnm{{Colombo}}, \binits{L.P.L.}},
\bauthor{\bsnm{{Combet}}, \binits{C.}},
\bauthor{\bsnm{{Contreras}}, \binits{D.}},
\bauthor{\bsnm{{Crill}}, \binits{B.P.}},
\bauthor{\bsnm{{Cuttaia}}, \binits{F.}},
\bauthor{\bsnm{{de Bernardis}}, \binits{P.}},
\bauthor{\bsnm{{de Zotti}}, \binits{G.}},
\bauthor{\bsnm{{Delabrouille}}, \binits{J.}},
\bauthor{\bsnm{{Delouis}}, \binits{J.-M.}},
\bauthor{\bsnm{{Di Valentino}}, \binits{E.}},
\bauthor{\bsnm{{Diego}}, \binits{J.M.}},
\bauthor{\bsnm{{Dor{\'e}}}, \binits{O.}},
\bauthor{\bsnm{{Douspis}}, \binits{M.}},
\bauthor{\bsnm{{Ducout}}, \binits{A.}},
\bauthor{\bsnm{{Dupac}}, \binits{X.}},
\bauthor{\bsnm{{Dusini}}, \binits{S.}},
\bauthor{\bsnm{{Efstathiou}}, \binits{G.}},
\bauthor{\bsnm{{Elsner}}, \binits{F.}},
\bauthor{\bsnm{{En{\ss}lin}}, \binits{T.A.}},
\bauthor{\bsnm{{Eriksen}}, \binits{H.K.}},
\bauthor{\bsnm{{Fantaye}}, \binits{Y.}},
\bauthor{\bsnm{{Farhang}}, \binits{M.}},
\bauthor{\bsnm{{Fergusson}}, \binits{J.}},
\bauthor{\bsnm{{Fernandez-Cobos}}, \binits{R.}},
\bauthor{\bsnm{{Finelli}}, \binits{F.}},
\bauthor{\bsnm{{Forastieri}}, \binits{F.}},
\bauthor{\bsnm{{Frailis}}, \binits{M.}},
\bauthor{\bsnm{{Fraisse}}, \binits{A.A.}},
\bauthor{\bsnm{{Franceschi}}, \binits{E.}},
\bauthor{\bsnm{{Frolov}}, \binits{A.}},
\bauthor{\bsnm{{Galeotta}}, \binits{S.}},
\bauthor{\bsnm{{Galli}}, \binits{S.}},
\bauthor{\bsnm{{Ganga}}, \binits{K.}},
\bauthor{\bsnm{{G{\'e}nova-Santos}}, \binits{R.T.}},
\bauthor{\bsnm{{Gerbino}}, \binits{M.}},
\bauthor{\bsnm{{Ghosh}}, \binits{T.}},
\bauthor{\bsnm{{Gonz{\'a}lez-Nuevo}}, \binits{J.}},
\bauthor{\bsnm{{G{\'o}rski}}, \binits{K.M.}},
\bauthor{\bsnm{{Gratton}}, \binits{S.}},
\bauthor{\bsnm{{Gruppuso}}, \binits{A.}},
\bauthor{\bsnm{{Gudmundsson}}, \binits{J.E.}},
\bauthor{\bsnm{{Hamann}}, \binits{J.}},
\bauthor{\bsnm{{Handley}}, \binits{W.}},
\bauthor{\bsnm{{Hansen}}, \binits{F.K.}},
\bauthor{\bsnm{{Herranz}}, \binits{D.}},
\bauthor{\bsnm{{Hildebrandt}}, \binits{S.R.}},
\bauthor{\bsnm{{Hivon}}, \binits{E.}},
\bauthor{\bsnm{{Huang}}, \binits{Z.}},
\bauthor{\bsnm{{Jaffe}}, \binits{A.H.}},
\bauthor{\bsnm{{Jones}}, \binits{W.C.}},
\bauthor{\bsnm{{Karakci}}, \binits{A.}},
\bauthor{\bsnm{{Keih{\"a}nen}}, \binits{E.}},
\bauthor{\bsnm{{Keskitalo}}, \binits{R.}},
\bauthor{\bsnm{{Kiiveri}}, \binits{K.}},
\bauthor{\bsnm{{Kim}}, \binits{J.}},
\bauthor{\bsnm{{Kisner}}, \binits{T.S.}},
\bauthor{\bsnm{{Knox}}, \binits{L.}},
\bauthor{\bsnm{{Krachmalnicoff}}, \binits{N.}},
\bauthor{\bsnm{{Kunz}}, \binits{M.}},
\bauthor{\bsnm{{Kurki-Suonio}}, \binits{H.}},
\bauthor{\bsnm{{Lagache}}, \binits{G.}},
\bauthor{\bsnm{{Lamarre}}, \binits{J.-M.}},
\bauthor{\bsnm{{Lasenby}}, \binits{A.}},
\bauthor{\bsnm{{Lattanzi}}, \binits{M.}},
\bauthor{\bsnm{{Lawrence}}, \binits{C.R.}},
\bauthor{\bsnm{{Le Jeune}}, \binits{M.}},
\bauthor{\bsnm{{Lemos}}, \binits{P.}},
\bauthor{\bsnm{{Lesgourgues}}, \binits{J.}},
\bauthor{\bsnm{{Levrier}}, \binits{F.}},
\bauthor{\bsnm{{Lewis}}, \binits{A.}},
\bauthor{\bsnm{{Liguori}}, \binits{M.}},
\bauthor{\bsnm{{Lilje}}, \binits{P.B.}},
\bauthor{\bsnm{{Lilley}}, \binits{M.}},
\bauthor{\bsnm{{Lindholm}}, \binits{V.}},
\bauthor{\bsnm{{L{\'o}pez-Caniego}}, \binits{M.}},
\bauthor{\bsnm{{Lubin}}, \binits{P.M.}},
\bauthor{\bsnm{{Ma}}, \binits{Y.-Z.}},
\bauthor{\bsnm{{Mac{\'\i}as-P{\'e}rez}}, \binits{J.F.}},
\bauthor{\bsnm{{Maggio}}, \binits{G.}},
\bauthor{\bsnm{{Maino}}, \binits{D.}},
\bauthor{\bsnm{{Mandolesi}}, \binits{N.}},
\bauthor{\bsnm{{Mangilli}}, \binits{A.}},
\bauthor{\bsnm{{Marcos-Caballero}}, \binits{A.}},
\bauthor{\bsnm{{Maris}}, \binits{M.}},
\bauthor{\bsnm{{Martin}}, \binits{P.G.}},
\bauthor{\bsnm{{Martinelli}}, \binits{M.}},
\bauthor{\bsnm{{Mart{\'\i}nez-Gonz{\'a}lez}}, \binits{E.}},
\bauthor{\bsnm{{Matarrese}}, \binits{S.}},
\bauthor{\bsnm{{Mauri}}, \binits{N.}},
\bauthor{\bsnm{{McEwen}}, \binits{J.D.}},
\bauthor{\bsnm{{Meinhold}}, \binits{P.R.}},
\bauthor{\bsnm{{Melchiorri}}, \binits{A.}},
\bauthor{\bsnm{{Mennella}}, \binits{A.}},
\bauthor{\bsnm{{Migliaccio}}, \binits{M.}},
\bauthor{\bsnm{{Millea}}, \binits{M.}},
\bauthor{\bsnm{{Mitra}}, \binits{S.}},
\bauthor{\bsnm{{Miville-Desch{\^e}nes}}, \binits{M.-A.}},
\bauthor{\bsnm{{Molinari}}, \binits{D.}},
\bauthor{\bsnm{{Montier}}, \binits{L.}},
\bauthor{\bsnm{{Morgante}}, \binits{G.}},
\bauthor{\bsnm{{Moss}}, \binits{A.}},
\bauthor{\bsnm{{Natoli}}, \binits{P.}},
\bauthor{\bsnm{{N{\o}rgaard-Nielsen}}, \binits{H.U.}},
\bauthor{\bsnm{{Pagano}}, \binits{L.}},
\bauthor{\bsnm{{Paoletti}}, \binits{D.}},
\bauthor{\bsnm{{Partridge}}, \binits{B.}},
\bauthor{\bsnm{{Patanchon}}, \binits{G.}},
\bauthor{\bsnm{{Peiris}}, \binits{H.V.}},
\bauthor{\bsnm{{Perrotta}}, \binits{F.}},
\bauthor{\bsnm{{Pettorino}}, \binits{V.}},
\bauthor{\bsnm{{Piacentini}}, \binits{F.}},
\bauthor{\bsnm{{Polastri}}, \binits{L.}},
\bauthor{\bsnm{{Polenta}}, \binits{G.}},
\bauthor{\bsnm{{Puget}}, \binits{J.-L.}},
\bauthor{\bsnm{{Rachen}}, \binits{J.P.}},
\bauthor{\bsnm{{Reinecke}}, \binits{M.}},
\bauthor{\bsnm{{Remazeilles}}, \binits{M.}},
\bauthor{\bsnm{{Renzi}}, \binits{A.}},
\bauthor{\bsnm{{Rocha}}, \binits{G.}},
\bauthor{\bsnm{{Rosset}}, \binits{C.}},
\bauthor{\bsnm{{Roudier}}, \binits{G.}},
\bauthor{\bsnm{{Rubi{\~n}o-Mart{\'\i}n}}, \binits{J.A.}},
\bauthor{\bsnm{{Ruiz-Granados}}, \binits{B.}},
\bauthor{\bsnm{{Salvati}}, \binits{L.}},
\bauthor{\bsnm{{Sandri}}, \binits{M.}},
\bauthor{\bsnm{{Savelainen}}, \binits{M.}},
\bauthor{\bsnm{{Scott}}, \binits{D.}},
\bauthor{\bsnm{{Shellard}}, \binits{E.P.S.}},
\bauthor{\bsnm{{Sirignano}}, \binits{C.}},
\bauthor{\bsnm{{Sirri}}, \binits{G.}},
\bauthor{\bsnm{{Spencer}}, \binits{L.D.}},
\bauthor{\bsnm{{Sunyaev}}, \binits{R.}},
\bauthor{\bsnm{{Suur-Uski}}, \binits{A.-S.}},
\bauthor{\bsnm{{Tauber}}, \binits{J.A.}},
\bauthor{\bsnm{{Tavagnacco}}, \binits{D.}},
\bauthor{\bsnm{{Tenti}}, \binits{M.}},
\bauthor{\bsnm{{Toffolatti}}, \binits{L.}},
\bauthor{\bsnm{{Tomasi}}, \binits{M.}},
\bauthor{\bsnm{{Trombetti}}, \binits{T.}},
\bauthor{\bsnm{{Valenziano}}, \binits{L.}},
\bauthor{\bsnm{{Valiviita}}, \binits{J.}},
\bauthor{\bsnm{{Van Tent}}, \binits{B.}},
\bauthor{\bsnm{{Vibert}}, \binits{L.}},
\bauthor{\bsnm{{Vielva}}, \binits{P.}},
\bauthor{\bsnm{{Villa}}, \binits{F.}},
\bauthor{\bsnm{{Vittorio}}, \binits{N.}},
\bauthor{\bsnm{{Wandelt}}, \binits{B.D.}},
\bauthor{\bsnm{{Wehus}}, \binits{I.K.}},
\bauthor{\bsnm{{White}}, \binits{M.}},
\bauthor{\bsnm{{White}}, \binits{S.D.M.}},
\bauthor{\bsnm{{Zacchei}}, \binits{A.}},
\bauthor{\bsnm{{Zonca}}, \binits{A.}}:
\batitle{{Planck 2018 results. VI. Cosmological parameters}}.
\bjtitle{\aap}
\bvolume{641},
\bfpage{6}
(\byear{2020})
\doiurl{10.1051/0004-6361/201833910}
{\href{https://arxiv.org/abs/1807.06209}{{arXiv:1807.06209}}}
{[astro-ph.CO]}
\end{barticle}
\endbibitem

\bibitem[\protect\citeauthoryear{{Astropy Collaboration} et~al.}{2013}]{astropy}
\begin{barticle}
\bauthor{\bsnm{{Astropy Collaboration}}},
\bauthor{\bsnm{{Robitaille}}, \binits{T.P.}},
\bauthor{\bsnm{{Tollerud}}, \binits{E.J.}},
\bauthor{\bsnm{{Greenfield}}, \binits{P.}},
\bauthor{\bsnm{{Droettboom}}, \binits{M.}},
\bauthor{\bsnm{{Bray}}, \binits{E.}},
\bauthor{\bsnm{{Aldcroft}}, \binits{T.}},
\bauthor{\bsnm{{Davis}}, \binits{M.}},
\bauthor{\bsnm{{Ginsburg}}, \binits{A.}},
\bauthor{\bsnm{{Price-Whelan}}, \binits{A.M.}},
\bauthor{\bsnm{{Kerzendorf}}, \binits{W.E.}},
\bauthor{\bsnm{{Conley}}, \binits{A.}},
\bauthor{\bsnm{{Crighton}}, \binits{N.}},
\bauthor{\bsnm{{Barbary}}, \binits{K.}},
\bauthor{\bsnm{{Muna}}, \binits{D.}},
\bauthor{\bsnm{{Ferguson}}, \binits{H.}},
\bauthor{\bsnm{{Grollier}}, \binits{F.}},
\bauthor{\bsnm{{Parikh}}, \binits{M.M.}},
\bauthor{\bsnm{{Nair}}, \binits{P.H.}},
\bauthor{\bsnm{{Unther}}, \binits{H.M.}},
\bauthor{\bsnm{{Deil}}, \binits{C.}},
\bauthor{\bsnm{{Woillez}}, \binits{J.}},
\bauthor{\bsnm{{Conseil}}, \binits{S.}},
\bauthor{\bsnm{{Kramer}}, \binits{R.}},
\bauthor{\bsnm{{Turner}}, \binits{J.E.H.}},
\bauthor{\bsnm{{Singer}}, \binits{L.}},
\bauthor{\bsnm{{Fox}}, \binits{R.}},
\bauthor{\bsnm{{Weaver}}, \binits{B.A.}},
\bauthor{\bsnm{{Zabalza}}, \binits{V.}},
\bauthor{\bsnm{{Edwards}}, \binits{Z.I.}},
\bauthor{\bsnm{{Azalee Bostroem}}, \binits{K.}},
\bauthor{\bsnm{{Burke}}, \binits{D.J.}},
\bauthor{\bsnm{{Casey}}, \binits{A.R.}},
\bauthor{\bsnm{{Crawford}}, \binits{S.M.}},
\bauthor{\bsnm{{Dencheva}}, \binits{N.}},
\bauthor{\bsnm{{Ely}}, \binits{J.}},
\bauthor{\bsnm{{Jenness}}, \binits{T.}},
\bauthor{\bsnm{{Labrie}}, \binits{K.}},
\bauthor{\bsnm{{Lim}}, \binits{P.L.}},
\bauthor{\bsnm{{Pierfederici}}, \binits{F.}},
\bauthor{\bsnm{{Pontzen}}, \binits{A.}},
\bauthor{\bsnm{{Ptak}}, \binits{A.}},
\bauthor{\bsnm{{Refsdal}}, \binits{B.}},
\bauthor{\bsnm{{Servillat}}, \binits{M.}},
\bauthor{\bsnm{{Streicher}}, \binits{O.}}:
\batitle{{Astropy: A community Python package for astronomy}}.
\bjtitle{\aap}
\bvolume{558},
\bfpage{33}
(\byear{2013})
\doiurl{10.1051/0004-6361/201322068}
{\href{https://arxiv.org/abs/1307.6212}{{arXiv:1307.6212}}}
{[astro-ph.IM]}
\end{barticle}
\endbibitem

\bibitem[\protect\citeauthoryear{{Bunker} et~al.}{2023}]{Bunker23}
\begin{botherref}
\oauthor{\bsnm{{Bunker}}, \binits{A.J.}},
\oauthor{\bsnm{{Cameron}}, \binits{A.J.}},
\oauthor{\bsnm{{Curtis-Lake}}, \binits{E.}},
\oauthor{\bsnm{{Jakobsen}}, \binits{P.}},
\oauthor{\bsnm{{Carniani}}, \binits{S.}},
\oauthor{\bsnm{{Curti}}, \binits{M.}},
\oauthor{\bsnm{{Witstok}}, \binits{J.}},
\oauthor{\bsnm{{Maiolino}}, \binits{R.}},
\oauthor{\bsnm{{D'Eugenio}}, \binits{F.}},
\oauthor{\bsnm{{Looser}}, \binits{T.J.}},
\oauthor{\bsnm{{Willott}}, \binits{C.}},
\oauthor{\bsnm{{Bonaventura}}, \binits{N.}},
\oauthor{\bsnm{{Hainline}}, \binits{K.}},
\oauthor{\bsnm{{Uebler}}, \binits{H.}},
\oauthor{\bsnm{{Willmer}}, \binits{C.N.A.}},
\oauthor{\bsnm{{Saxena}}, \binits{A.}},
\oauthor{\bsnm{{Smit}}, \binits{R.}},
\oauthor{\bsnm{{Alberts}}, \binits{S.}},
\oauthor{\bsnm{{Arribas}}, \binits{S.}},
\oauthor{\bsnm{{Baker}}, \binits{W.M.}},
\oauthor{\bsnm{{Baum}}, \binits{S.}},
\oauthor{\bsnm{{Bhatawdekar}}, \binits{R.}},
\oauthor{\bsnm{{Bowler}}, \binits{R.A.A.}},
\oauthor{\bsnm{{Boyett}}, \binits{K.}},
\oauthor{\bsnm{{Charlot}}, \binits{S.}},
\oauthor{\bsnm{{Chen}}, \binits{Z.}},
\oauthor{\bsnm{{Chevallard}}, \binits{J.}},
\oauthor{\bsnm{{Circosta}}, \binits{C.}},
\oauthor{\bsnm{{DeCoursey}}, \binits{C.}},
\oauthor{\bsnm{{de Graaff}}, \binits{A.}},
\oauthor{\bsnm{{Egami}}, \binits{E.}},
\oauthor{\bsnm{{Eisenstein}}, \binits{D.J.}},
\oauthor{\bsnm{{Endsley}}, \binits{R.}},
\oauthor{\bsnm{{Ferruit}}, \binits{P.}},
\oauthor{\bsnm{{Giardino}}, \binits{G.}},
\oauthor{\bsnm{{Hausen}}, \binits{R.}},
\oauthor{\bsnm{{Helton}}, \binits{J.M.}},
\oauthor{\bsnm{{Hviding}}, \binits{R.E.}},
\oauthor{\bsnm{{Ji}}, \binits{Z.}},
\oauthor{\bsnm{{Johnson}}, \binits{B.D.}},
\oauthor{\bsnm{{Jones}}, \binits{G.C.}},
\oauthor{\bsnm{{Kumari}}, \binits{N.}},
\oauthor{\bsnm{{Laseter}}, \binits{I.}},
\oauthor{\bsnm{{Luetzgendorf}}, \binits{N.}},
\oauthor{\bsnm{{Maseda}}, \binits{M.V.}},
\oauthor{\bsnm{{Nelson}}, \binits{E.}},
\oauthor{\bsnm{{Parlanti}}, \binits{E.}},
\oauthor{\bsnm{{Perna}}, \binits{M.}},
\oauthor{\bsnm{{Rauscher}}, \binits{B.J.}},
\oauthor{\bsnm{{Rawle}}, \binits{T.}},
\oauthor{\bsnm{{Rix}}, \binits{H.-W.}},
\oauthor{\bsnm{{Rieke}}, \binits{M.}},
\oauthor{\bsnm{{Robertson}}, \binits{B.}},
\oauthor{\bsnm{{Rodriguez Del Pino}}, \binits{B.}},
\oauthor{\bsnm{{Sandles}}, \binits{L.}},
\oauthor{\bsnm{{Scholtz}}, \binits{J.}},
\oauthor{\bsnm{{Sharpe}}, \binits{K.}},
\oauthor{\bsnm{{Skarbinski}}, \binits{M.}},
\oauthor{\bsnm{{Stark}}, \binits{D.P.}},
\oauthor{\bsnm{{Sun}}, \binits{F.}},
\oauthor{\bsnm{{Tacchella}}, \binits{S.}},
\oauthor{\bsnm{{Topping}}, \binits{M.W.}},
\oauthor{\bsnm{{Villanueva}}, \binits{N.C.}},
\oauthor{\bsnm{{Wallace}}, \binits{I.E.B.}},
\oauthor{\bsnm{{Williams}}, \binits{C.C.}},
\oauthor{\bsnm{{Woodrum}}, \binits{C.}}:
{JADES NIRSpec Initial Data Release for the Hubble Ultra Deep Field: Redshifts and Line Fluxes of Distant Galaxies from the Deepest JWST Cycle 1 NIRSpec Multi-Object Spectroscopy}.
arXiv e-prints,
2306--02467
(2023)
\doiurl{10.48550/arXiv.2306.02467}
{\href{https://arxiv.org/abs/2306.02467}{{arXiv:2306.02467}}}
{[astro-ph.GA]}
\end{botherref}
\endbibitem

\bibitem[\protect\citeauthoryear{{Barrufet} et~al.}{2024}]{Barrufet24}
\begin{botherref}
\oauthor{\bsnm{{Barrufet}}, \binits{L.}},
\oauthor{\bsnm{{Oesch}}, \binits{P.}},
\oauthor{\bsnm{{Marques-Chaves}}, \binits{R.}},
\oauthor{\bsnm{{Arellano-Cordova}}, \binits{K.}},
\oauthor{\bsnm{{Baggen}}, \binits{J.F.W.}},
\oauthor{\bsnm{{Carnall}}, \binits{A.C.}},
\oauthor{\bsnm{{Cullen}}, \binits{F.}},
\oauthor{\bsnm{{Dunlop}}, \binits{J.S.}},
\oauthor{\bsnm{{Gottumukkala}}, \binits{R.}},
\oauthor{\bsnm{{Fudamoto}}, \binits{Y.}},
\oauthor{\bsnm{{Illingworth}}, \binits{G.D.}},
\oauthor{\bsnm{{Magee}}, \binits{D.}},
\oauthor{\bsnm{{McLure}}, \binits{R.J.}},
\oauthor{\bsnm{{McLeod}}, \binits{D.J.}},
\oauthor{\bsnm{{Micha{\l}owski}}, \binits{M.J.}},
\oauthor{\bsnm{{Stefanon}}, \binits{M.}},
\oauthor{\bsnm{{van Dokkum}}, \binits{P.G.}},
\oauthor{\bsnm{{Weibel}}, \binits{A.}}:
{Quiescent or dusty? Unveiling the nature of extremely red galaxies at $z>3$}.
arXiv e-prints,
2404--08052
(2024)
\doiurl{10.48550/arXiv.2404.08052}
{\href{https://arxiv.org/abs/2404.08052}{{arXiv:2404.08052}}}
{[astro-ph.GA]}
\end{botherref}
\endbibitem

\bibitem[\protect\citeauthoryear{{Jakobsen} et~al.}{2022}]{Jakobsen22}
\begin{barticle}
\bauthor{\bsnm{{Jakobsen}}, \binits{P.}},
\bauthor{\bsnm{{Ferruit}}, \binits{P.}},
\bauthor{\bsnm{{Alves de Oliveira}}, \binits{C.}},
\bauthor{\bsnm{{Arribas}}, \binits{S.}},
\bauthor{\bsnm{{Bagnasco}}, \binits{G.}},
\bauthor{\bsnm{{Barho}}, \binits{R.}},
\bauthor{\bsnm{{Beck}}, \binits{T.L.}},
\bauthor{\bsnm{{Birkmann}}, \binits{S.}},
\bauthor{\bsnm{{B{\"o}ker}}, \binits{T.}},
\bauthor{\bsnm{{Bunker}}, \binits{A.J.}},
\bauthor{\bsnm{{Charlot}}, \binits{S.}},
\bauthor{\bsnm{{de Jong}}, \binits{P.}},
\bauthor{\bsnm{{de Marchi}}, \binits{G.}},
\bauthor{\bsnm{{Ehrenwinkler}}, \binits{R.}},
\bauthor{\bsnm{{Falcolini}}, \binits{M.}},
\bauthor{\bsnm{{Fels}}, \binits{R.}},
\bauthor{\bsnm{{Franx}}, \binits{M.}},
\bauthor{\bsnm{{Franz}}, \binits{D.}},
\bauthor{\bsnm{{Funke}}, \binits{M.}},
\bauthor{\bsnm{{Giardino}}, \binits{G.}},
\bauthor{\bsnm{{Gnata}}, \binits{X.}},
\bauthor{\bsnm{{Holota}}, \binits{W.}},
\bauthor{\bsnm{{Honnen}}, \binits{K.}},
\bauthor{\bsnm{{Jensen}}, \binits{P.L.}},
\bauthor{\bsnm{{Jentsch}}, \binits{M.}},
\bauthor{\bsnm{{Johnson}}, \binits{T.}},
\bauthor{\bsnm{{Jollet}}, \binits{D.}},
\bauthor{\bsnm{{Karl}}, \binits{H.}},
\bauthor{\bsnm{{Kling}}, \binits{G.}},
\bauthor{\bsnm{{K{\"o}hler}}, \binits{J.}},
\bauthor{\bsnm{{Kolm}}, \binits{M.-G.}},
\bauthor{\bsnm{{Kumari}}, \binits{N.}},
\bauthor{\bsnm{{Lander}}, \binits{M.E.}},
\bauthor{\bsnm{{Lemke}}, \binits{R.}},
\bauthor{\bsnm{{L{\'o}pez-Caniego}}, \binits{M.}},
\bauthor{\bsnm{{L{\"u}tzgendorf}}, \binits{N.}},
\bauthor{\bsnm{{Maiolino}}, \binits{R.}},
\bauthor{\bsnm{{Manjavacas}}, \binits{E.}},
\bauthor{\bsnm{{Marston}}, \binits{A.}},
\bauthor{\bsnm{{Maschmann}}, \binits{M.}},
\bauthor{\bsnm{{Maurer}}, \binits{R.}},
\bauthor{\bsnm{{Messerschmidt}}, \binits{B.}},
\bauthor{\bsnm{{Moseley}}, \binits{S.H.}},
\bauthor{\bsnm{{Mosner}}, \binits{P.}},
\bauthor{\bsnm{{Mott}}, \binits{D.B.}},
\bauthor{\bsnm{{Muzerolle}}, \binits{J.}},
\bauthor{\bsnm{{Pirzkal}}, \binits{N.}},
\bauthor{\bsnm{{Pittet}}, \binits{J.-F.}},
\bauthor{\bsnm{{Plitzke}}, \binits{A.}},
\bauthor{\bsnm{{Posselt}}, \binits{W.}},
\bauthor{\bsnm{{Rapp}}, \binits{B.}},
\bauthor{\bsnm{{Rauscher}}, \binits{B.J.}},
\bauthor{\bsnm{{Rawle}}, \binits{T.}},
\bauthor{\bsnm{{Rix}}, \binits{H.-W.}},
\bauthor{\bsnm{{R{\"o}del}}, \binits{A.}},
\bauthor{\bsnm{{Rumler}}, \binits{P.}},
\bauthor{\bsnm{{Sabbi}}, \binits{E.}},
\bauthor{\bsnm{{Salvignol}}, \binits{J.-C.}},
\bauthor{\bsnm{{Schmid}}, \binits{T.}},
\bauthor{\bsnm{{Sirianni}}, \binits{M.}},
\bauthor{\bsnm{{Smith}}, \binits{C.}},
\bauthor{\bsnm{{Strada}}, \binits{P.}},
\bauthor{\bsnm{{te Plate}}, \binits{M.}},
\bauthor{\bsnm{{Valenti}}, \binits{J.}},
\bauthor{\bsnm{{Wettemann}}, \binits{T.}},
\bauthor{\bsnm{{Wiehe}}, \binits{T.}},
\bauthor{\bsnm{{Wiesmayer}}, \binits{M.}},
\bauthor{\bsnm{{Willott}}, \binits{C.J.}},
\bauthor{\bsnm{{Wright}}, \binits{R.}},
\bauthor{\bsnm{{Zeidler}}, \binits{P.}},
\bauthor{\bsnm{{Zincke}}, \binits{C.}}:
\batitle{{The Near-Infrared Spectrograph (NIRSpec) on the James Webb Space Telescope. I. Overview of the instrument and its capabilities}}.
\bjtitle{\aap}
\bvolume{661},
\bfpage{80}
(\byear{2022})
\doiurl{10.1051/0004-6361/202142663}
{\href{https://arxiv.org/abs/2202.03305}{{arXiv:2202.03305}}}
{[astro-ph.IM]}
\end{barticle}
\endbibitem

\bibitem[\protect\citeauthoryear{{Schaerer} et~al.}{2024}]{Schaerer24}
\begin{botherref}
\oauthor{\bsnm{{Schaerer}}, \binits{D.}},
\oauthor{\bsnm{{Marques-Chaves}}, \binits{R.}},
\oauthor{\bsnm{{Xiao}}, \binits{M.}},
\oauthor{\bsnm{{Korber}}, \binits{D.}}:
{Discovery of a new N-emitter in the epoch of reionization}.
arXiv e-prints,
2406--08408
(2024)
\doiurl{10.48550/arXiv.2406.08408}
{\href{https://arxiv.org/abs/2406.08408}{{arXiv:2406.08408}}}
{[astro-ph.GA]}
\end{botherref}
\endbibitem

\bibitem[\protect\citeauthoryear{{Horne}}{1986}]{Horne86}
\begin{barticle}
\bauthor{\bsnm{{Horne}}, \binits{K.}}:
\batitle{{An optimal extraction algorithm for CCD spectroscopy.}}
\bjtitle{\pasp}
\bvolume{98},
\bfpage{609}--\blpage{617}
(\byear{1986})
\doiurl{10.1086/131801}
\end{barticle}
\endbibitem

\bibitem[\protect\citeauthoryear{{Miralda-Escud{\'e}}}{1998}]{MiraldaEscude98}
\begin{barticle}
\bauthor{\bsnm{{Miralda-Escud{\'e}}}, \binits{J.}}:
\batitle{{Reionization of the Intergalactic Medium and the Damping Wing of the Gunn-Peterson Trough}}.
\bjtitle{\apj}
\bvolume{501}(\bissue{1}),
\bfpage{15}--\blpage{22}
(\byear{1998})
\doiurl{10.1086/305799}
{\href{https://arxiv.org/abs/astro-ph/9708253}{{arXiv:astro-ph/9708253}}}
{[astro-ph]}
\end{barticle}
\endbibitem

\bibitem[\protect\citeauthoryear{{McQuinn} et~al.}{2008}]{McQuinn08}
\begin{barticle}
\bauthor{\bsnm{{McQuinn}}, \binits{M.}},
\bauthor{\bsnm{{Lidz}}, \binits{A.}},
\bauthor{\bsnm{{Zaldarriaga}}, \binits{M.}},
\bauthor{\bsnm{{Hernquist}}, \binits{L.}},
\bauthor{\bsnm{{Dutta}}, \binits{S.}}:
\batitle{{Probing the neutral fraction of the IGM with GRBs during the epoch of reionization}}.
\bjtitle{\mnras}
\bvolume{388}(\bissue{3}),
\bfpage{1101}--\blpage{1110}
(\byear{2008})
\doiurl{10.1111/j.1365-2966.2008.13271.x}
{\href{https://arxiv.org/abs/0710.1018}{{arXiv:0710.1018}}}
{[astro-ph]}
\end{barticle}
\endbibitem

\bibitem[\protect\citeauthoryear{{McGreer} et~al.}{2015}]{McGreer15}
\begin{barticle}
\bauthor{\bsnm{{McGreer}}, \binits{I.D.}},
\bauthor{\bsnm{{Mesinger}}, \binits{A.}},
\bauthor{\bsnm{{D'Odorico}}, \binits{V.}}:
\batitle{{Model-independent evidence in favour of an end to reionization by z {\ensuremath{\approx}} 6}}.
\bjtitle{\mnras}
\bvolume{447}(\bissue{1}),
\bfpage{499}--\blpage{505}
(\byear{2015})
\doiurl{10.1093/mnras/stu2449}
{\href{https://arxiv.org/abs/1411.5375}{{arXiv:1411.5375}}}
{[astro-ph.CO]}
\end{barticle}
\endbibitem

\bibitem[\protect\citeauthoryear{{Fan} et~al.}{2023}]{Fan23}
\begin{barticle}
\bauthor{\bsnm{{Fan}}, \binits{X.}},
\bauthor{\bsnm{{Ba{\~n}ados}}, \binits{E.}},
\bauthor{\bsnm{{Simcoe}}, \binits{R.A.}}:
\batitle{{Quasars and the Intergalactic Medium at Cosmic Dawn}}.
\bjtitle{\araa}
\bvolume{61},
\bfpage{373}--\blpage{426}
(\byear{2023})
\doiurl{10.1146/annurev-astro-052920-102455}
{\href{https://arxiv.org/abs/2212.06907}{{arXiv:2212.06907}}}
{[astro-ph.GA]}
\end{barticle}
\endbibitem

\bibitem[\protect\citeauthoryear{{Totani} et~al.}{2006}]{Totani06}
\begin{barticle}
\bauthor{\bsnm{{Totani}}, \binits{T.}},
\bauthor{\bsnm{{Kawai}}, \binits{N.}},
\bauthor{\bsnm{{Kosugi}}, \binits{G.}},
\bauthor{\bsnm{{Aoki}}, \binits{K.}},
\bauthor{\bsnm{{Yamada}}, \binits{T.}},
\bauthor{\bsnm{{Iye}}, \binits{M.}},
\bauthor{\bsnm{{Ohta}}, \binits{K.}},
\bauthor{\bsnm{{Hattori}}, \binits{T.}}:
\batitle{{Implications for Cosmic Reionization from the Optical Afterglow Spectrum of the Gamma-Ray Burst 050904 at z = 6.3$^{*}$}}.
\bjtitle{\pasj}
\bvolume{58}(\bissue{3}),
\bfpage{485}--\blpage{498}
(\byear{2006})
\doiurl{10.1093/pasj/58.3.485}
{\href{https://arxiv.org/abs/astro-ph/0512154}{{arXiv:astro-ph/0512154}}}
{[astro-ph]}
\end{barticle}
\endbibitem

\bibitem[\protect\citeauthoryear{{Umeda} et~al.}{2023}]{Umeda23}
\begin{botherref}
\oauthor{\bsnm{{Umeda}}, \binits{H.}},
\oauthor{\bsnm{{Ouchi}}, \binits{M.}},
\oauthor{\bsnm{{Nakajima}}, \binits{K.}},
\oauthor{\bsnm{{Harikane}}, \binits{Y.}},
\oauthor{\bsnm{{Ono}}, \binits{Y.}},
\oauthor{\bsnm{{Xu}}, \binits{Y.}},
\oauthor{\bsnm{{Isobe}}, \binits{Y.}},
\oauthor{\bsnm{{Zhang}}, \binits{Y.}}:
{JWST Measurements of Neutral Hydrogen Fractions and Ionized Bubble Sizes at $z=7-12$ Obtained with Ly$\alpha$ Damping Wing Absorptions in 27 Bright Continuum Galaxies}.
arXiv e-prints,
2306--00487
(2023)
\doiurl{10.48550/arXiv.2306.00487}
{\href{https://arxiv.org/abs/2306.00487}{{arXiv:2306.00487}}}
{[astro-ph.GA]}
\end{botherref}
\endbibitem

\bibitem[\protect\citeauthoryear{{D'Eugenio} et~al.}{2023}]{DEugenio23}
\begin{botherref}
\oauthor{\bsnm{{D'Eugenio}}, \binits{F.}},
\oauthor{\bsnm{{Maiolino}}, \binits{R.}},
\oauthor{\bsnm{{Carniani}}, \binits{S.}},
\oauthor{\bsnm{{Curtis-Lake}}, \binits{E.}},
\oauthor{\bsnm{{Witstok}}, \binits{J.}},
\oauthor{\bsnm{{Chevallard}}, \binits{J.}},
\oauthor{\bsnm{{Charlot}}, \binits{S.}},
\oauthor{\bsnm{{Baker}}, \binits{W.M.}},
\oauthor{\bsnm{{Arribas}}, \binits{S.}},
\oauthor{\bsnm{{Boyett}}, \binits{K.}},
\oauthor{\bsnm{{Bunker}}, \binits{A.J.}},
\oauthor{\bsnm{{Curti}}, \binits{M.}},
\oauthor{\bsnm{{Eisenstein}}, \binits{D.J.}},
\oauthor{\bsnm{{Hainline}}, \binits{K.}},
\oauthor{\bsnm{{Ji}}, \binits{Z.}},
\oauthor{\bsnm{{Johnson}}, \binits{B.D.}},
\oauthor{\bsnm{{Looser}}, \binits{T.J.}},
\oauthor{\bsnm{{Nakajima}}, \binits{K.}},
\oauthor{\bsnm{{Nelson}}, \binits{E.}},
\oauthor{\bsnm{{Rieke}}, \binits{M.}},
\oauthor{\bsnm{{Robertson}}, \binits{B.}},
\oauthor{\bsnm{{Scholtz}}, \binits{J.}},
\oauthor{\bsnm{{Smit}}, \binits{R.}},
\oauthor{\bsnm{{Venturi}}, \binits{G.}},
\oauthor{\bsnm{{Tacchella}}, \binits{S.}},
\oauthor{\bsnm{{Uebler}}, \binits{H.}},
\oauthor{\bsnm{{Willmer}}, \binits{C.N.A.}},
\oauthor{\bsnm{{Willott}}, \binits{C.}}:
{JADES: Carbon enrichment 350 Myr after the Big Bang in a gas-rich galaxy}.
arXiv e-prints,
2311--09908
(2023)
\doiurl{10.48550/arXiv.2311.09908}
{\href{https://arxiv.org/abs/2311.09908}{{arXiv:2311.09908}}}
{[astro-ph.GA]}
\end{botherref}
\endbibitem

\bibitem[\protect\citeauthoryear{{Hainline} et~al.}{2024}]{Hainline24}
\begin{botherref}
\oauthor{\bsnm{{Hainline}}, \binits{K.N.}},
\oauthor{\bsnm{{D'Eugenio}}, \binits{F.}},
\oauthor{\bsnm{{Jakobsen}}, \binits{P.}},
\oauthor{\bsnm{{Chevallard}}, \binits{J.}},
\oauthor{\bsnm{{Carniani}}, \binits{S.}},
\oauthor{\bsnm{{Witstok}}, \binits{J.}},
\oauthor{\bsnm{{Ji}}, \binits{Z.}},
\oauthor{\bsnm{{Curtis-Lake}}, \binits{E.}},
\oauthor{\bsnm{{Johnson}}, \binits{B.D.}},
\oauthor{\bsnm{{Robertson}}, \binits{B.}},
\oauthor{\bsnm{{Tacchella}}, \binits{S.}},
\oauthor{\bsnm{{Curti}}, \binits{M.}},
\oauthor{\bsnm{{Charlot}}, \binits{S.}},
\oauthor{\bsnm{{Helton}}, \binits{J.M.}},
\oauthor{\bsnm{{Arribas}}, \binits{S.}},
\oauthor{\bsnm{{Bhatawdekar}}, \binits{R.}},
\oauthor{\bsnm{{Bunker}}, \binits{A.J.}},
\oauthor{\bsnm{{Cameron}}, \binits{A.J.}},
\oauthor{\bsnm{{Egami}}, \binits{E.}},
\oauthor{\bsnm{{Eisenstein}}, \binits{D.J.}},
\oauthor{\bsnm{{Hausen}}, \binits{R.}},
\oauthor{\bsnm{{Kumari}}, \binits{N.}},
\oauthor{\bsnm{{Maiolino}}, \binits{R.}},
\oauthor{\bsnm{{Perez-Gonzalez}}, \binits{P.G.}},
\oauthor{\bsnm{{Rieke}}, \binits{M.}},
\oauthor{\bsnm{{Saxena}}, \binits{A.}},
\oauthor{\bsnm{{Scholtz}}, \binits{J.}},
\oauthor{\bsnm{{Smit}}, \binits{R.}},
\oauthor{\bsnm{{Sun}}, \binits{F.}},
\oauthor{\bsnm{{Williams}}, \binits{C.C.}},
\oauthor{\bsnm{{Willmer}}, \binits{C.N.A.}},
\oauthor{\bsnm{{Willott}}, \binits{C.}}:
{Searching for Emission Lines at $z>11$: The Role of Damped Lyman-$\alpha$ and Hints About the Escape of Ionizing Photons}.
arXiv e-prints,
2404--04325
(2024)
\doiurl{10.48550/arXiv.2404.04325}
{\href{https://arxiv.org/abs/2404.04325}{{arXiv:2404.04325}}}
{[astro-ph.GA]}
\end{botherref}
\endbibitem

\bibitem[\protect\citeauthoryear{{Carniani} et~al.}{2024}]{Carniani24}
\begin{botherref}
\oauthor{\bsnm{{Carniani}}, \binits{S.}},
\oauthor{\bsnm{{Hainline}}, \binits{K.}},
\oauthor{\bsnm{{D'Eugenio}}, \binits{F.}},
\oauthor{\bsnm{{Eisenstein}}, \binits{D.J.}},
\oauthor{\bsnm{{Jakobsen}}, \binits{P.}},
\oauthor{\bsnm{{Witstok}}, \binits{J.}},
\oauthor{\bsnm{{Johnson}}, \binits{B.D.}},
\oauthor{\bsnm{{Chevallard}}, \binits{J.}},
\oauthor{\bsnm{{Maiolino}}, \binits{R.}},
\oauthor{\bsnm{{Helton}}, \binits{J.M.}},
\oauthor{\bsnm{{Willott}}, \binits{C.}},
\oauthor{\bsnm{{Robertson}}, \binits{B.}},
\oauthor{\bsnm{{Alberts}}, \binits{S.}},
\oauthor{\bsnm{{Arribas}}, \binits{S.}},
\oauthor{\bsnm{{Baker}}, \binits{W.M.}},
\oauthor{\bsnm{{Bhatawdekar}}, \binits{R.}},
\oauthor{\bsnm{{Boyett}}, \binits{K.}},
\oauthor{\bsnm{{Bunker}}, \binits{A.J.}},
\oauthor{\bsnm{{Cameron}}, \binits{A.J.}},
\oauthor{\bsnm{{Cargile}}, \binits{P.A.}},
\oauthor{\bsnm{{Charlot}}, \binits{S.}},
\oauthor{\bsnm{{Curti}}, \binits{M.}},
\oauthor{\bsnm{{Curtis-Lake}}, \binits{E.}},
\oauthor{\bsnm{{Egami}}, \binits{E.}},
\oauthor{\bsnm{{Giardino}}, \binits{G.}},
\oauthor{\bsnm{{Isaak}}, \binits{K.}},
\oauthor{\bsnm{{Ji}}, \binits{Z.}},
\oauthor{\bsnm{{Jones}}, \binits{G.C.}},
\oauthor{\bsnm{{Maseda}}, \binits{M.V.}},
\oauthor{\bsnm{{Parlanti}}, \binits{E.}},
\oauthor{\bsnm{{Rawle}}, \binits{T.}},
\oauthor{\bsnm{{Rieke}}, \binits{G.}},
\oauthor{\bsnm{{Rieke}}, \binits{M.}},
\oauthor{\bsnm{{Rodr{\'\i}guez Del Pino}}, \binits{B.}},
\oauthor{\bsnm{{Saxena}}, \binits{A.}},
\oauthor{\bsnm{{Scholtz}}, \binits{J.}},
\oauthor{\bsnm{{Smit}}, \binits{R.}},
\oauthor{\bsnm{{Sun}}, \binits{F.}},
\oauthor{\bsnm{{Tacchella}}, \binits{S.}},
\oauthor{\bsnm{{{\"U}bler}}, \binits{H.}},
\oauthor{\bsnm{{Venturi}}, \binits{G.}},
\oauthor{\bsnm{{Williams}}, \binits{C.C.}},
\oauthor{\bsnm{{Willmer}}, \binits{C.N.A.}}:
{A shining cosmic dawn: spectroscopic confirmation of two luminous galaxies at $z\sim14$}.
arXiv e-prints,
2405--18485
(2024)
\doiurl{10.48550/arXiv.2405.18485}
{\href{https://arxiv.org/abs/2405.18485}{{arXiv:2405.18485}}}
{[astro-ph.GA]}
\end{botherref}
\endbibitem

\bibitem[\protect\citeauthoryear{{Brammer} et~al.}{2008}]{Brammer08_Eazy}
\begin{barticle}
\bauthor{\bsnm{{Brammer}}, \binits{G.B.}},
\bauthor{\bsnm{{van Dokkum}}, \binits{P.G.}},
\bauthor{\bsnm{{Coppi}}, \binits{P.}}:
\batitle{{EAZY: A Fast, Public Photometric Redshift Code}}.
\bjtitle{\apj}
\bvolume{686}(\bissue{2}),
\bfpage{1503}--\blpage{1513}
(\byear{2008})
\doiurl{10.1086/591786}
{\href{https://arxiv.org/abs/0807.1533}{{arXiv:0807.1533}}}
{[astro-ph]}
\end{barticle}
\endbibitem

\bibitem[\protect\citeauthoryear{{Feroz} and {Hobson}}{2008}]{Feroz08}
\begin{barticle}
\bauthor{\bsnm{{Feroz}}, \binits{F.}},
\bauthor{\bsnm{{Hobson}}, \binits{M.P.}}:
\batitle{{Multimodal nested sampling: an efficient and robust alternative to Markov Chain Monte Carlo methods for astronomical data analyses}}.
\bjtitle{\mnras}
\bvolume{384}(\bissue{2}),
\bfpage{449}--\blpage{463}
(\byear{2008})
\doiurl{10.1111/j.1365-2966.2007.12353.x}
{\href{https://arxiv.org/abs/0704.3704}{{arXiv:0704.3704}}}
{[astro-ph]}
\end{barticle}
\endbibitem

\bibitem[\protect\citeauthoryear{{Buchner} et~al.}{2014}]{Buchner14}
\begin{barticle}
\bauthor{\bsnm{{Buchner}}, \binits{J.}},
\bauthor{\bsnm{{Georgakakis}}, \binits{A.}},
\bauthor{\bsnm{{Nandra}}, \binits{K.}},
\bauthor{\bsnm{{Hsu}}, \binits{L.}},
\bauthor{\bsnm{{Rangel}}, \binits{C.}},
\bauthor{\bsnm{{Brightman}}, \binits{M.}},
\bauthor{\bsnm{{Merloni}}, \binits{A.}},
\bauthor{\bsnm{{Salvato}}, \binits{M.}},
\bauthor{\bsnm{{Donley}}, \binits{J.}},
\bauthor{\bsnm{{Kocevski}}, \binits{D.}}:
\batitle{{X-ray spectral modelling of the AGN obscuring region in the CDFS: Bayesian model selection and catalogue}}.
\bjtitle{\aap}
\bvolume{564},
\bfpage{125}
(\byear{2014})
\doiurl{10.1051/0004-6361/201322971}
{\href{https://arxiv.org/abs/1402.0004}{{arXiv:1402.0004}}}
{[astro-ph.HE]}
\end{barticle}
\endbibitem

\bibitem[\protect\citeauthoryear{{Frye} et~al.}{2008}]{Frye08}
\begin{barticle}
\bauthor{\bsnm{{Frye}}, \binits{B.L.}},
\bauthor{\bsnm{{Bowen}}, \binits{D.V.}},
\bauthor{\bsnm{{Hurley}}, \binits{M.}},
\bauthor{\bsnm{{Tripp}}, \binits{T.M.}},
\bauthor{\bsnm{{Fan}}, \binits{X.}},
\bauthor{\bsnm{{Holden}}, \binits{B.}},
\bauthor{\bsnm{{Guhathakurta}}, \binits{P.}},
\bauthor{\bsnm{{Coe}}, \binits{D.}},
\bauthor{\bsnm{{Broadhurst}}, \binits{T.}},
\bauthor{\bsnm{{Egami}}, \binits{E.}},
\bauthor{\bsnm{{Meylan}}, \binits{G.}}:
\batitle{{Observations of the Gas Reservoir around a Star-Forming Galaxy in the Early Universe}}.
\bjtitle{\apjl}
\bvolume{685}(\bissue{1}),
\bfpage{5}
(\byear{2008})
\doiurl{10.1086/592273}
{\href{https://arxiv.org/abs/0808.0921}{{arXiv:0808.0921}}}
{[astro-ph]}
\end{barticle}
\endbibitem

\bibitem[\protect\citeauthoryear{{Cai} et~al.}{2017}]{Cai17}
\begin{barticle}
\bauthor{\bsnm{{Cai}}, \binits{Z.}},
\bauthor{\bsnm{{Fan}}, \binits{X.}},
\bauthor{\bsnm{{Bian}}, \binits{F.}},
\bauthor{\bsnm{{Zabludoff}}, \binits{A.}},
\bauthor{\bsnm{{Yang}}, \binits{Y.}},
\bauthor{\bsnm{{Prochaska}}, \binits{J.X.}},
\bauthor{\bsnm{{McGreer}}, \binits{I.}},
\bauthor{\bsnm{{Zheng}}, \binits{Z.-Y.}},
\bauthor{\bsnm{{Kashikawa}}, \binits{N.}},
\bauthor{\bsnm{{Wang}}, \binits{R.}},
\bauthor{\bsnm{{Frye}}, \binits{B.}},
\bauthor{\bsnm{{Green}}, \binits{R.}},
\bauthor{\bsnm{{Jiang}}, \binits{L.}}:
\batitle{{Mapping the Most Massive Overdensities through Hydrogen (MAMMOTH). II. Discovery of the Extremely Massive Overdensity BOSS1441 at z = 2.32}}.
\bjtitle{\apj}
\bvolume{839}(\bissue{2}),
\bfpage{131}
(\byear{2017})
\doiurl{10.3847/1538-4357/aa6a1a}
{\href{https://arxiv.org/abs/1609.02913}{{arXiv:1609.02913}}}
{[astro-ph.GA]}
\end{barticle}
\endbibitem

\bibitem[\protect\citeauthoryear{{Lee} et~al.}{2018}]{Lee18}
\begin{barticle}
\bauthor{\bsnm{{Lee}}, \binits{K.-G.}},
\bauthor{\bsnm{{Krolewski}}, \binits{A.}},
\bauthor{\bsnm{{White}}, \binits{M.}},
\bauthor{\bsnm{{Schlegel}}, \binits{D.}},
\bauthor{\bsnm{{Nugent}}, \binits{P.E.}},
\bauthor{\bsnm{{Hennawi}}, \binits{J.F.}},
\bauthor{\bsnm{{M{\"u}ller}}, \binits{T.}},
\bauthor{\bsnm{{Pan}}, \binits{R.}},
\bauthor{\bsnm{{Prochaska}}, \binits{J.X.}},
\bauthor{\bsnm{{Font-Ribera}}, \binits{A.}},
\bauthor{\bsnm{{Suzuki}}, \binits{N.}},
\bauthor{\bsnm{{Glazebrook}}, \binits{K.}},
\bauthor{\bsnm{{Kacprzak}}, \binits{G.G.}},
\bauthor{\bsnm{{Kartaltepe}}, \binits{J.S.}},
\bauthor{\bsnm{{Koekemoer}}, \binits{A.M.}},
\bauthor{\bsnm{{Le F{\`e}vre}}, \binits{O.}},
\bauthor{\bsnm{{Lemaux}}, \binits{B.C.}},
\bauthor{\bsnm{{Maier}}, \binits{C.}},
\bauthor{\bsnm{{Nanayakkara}}, \binits{T.}},
\bauthor{\bsnm{{Rich}}, \binits{R.M.}},
\bauthor{\bsnm{{Sanders}}, \binits{D.B.}},
\bauthor{\bsnm{{Salvato}}, \binits{M.}},
\bauthor{\bsnm{{Tasca}}, \binits{L.}},
\bauthor{\bsnm{{Tran}}, \binits{K.-V.H.}}:
\batitle{{First Data Release of the COSMOS Ly{\ensuremath{\alpha}} Mapping and Tomography Observations: 3D Ly{\ensuremath{\alpha}} Forest Tomography at 2.05 < z < 2.55}}.
\bjtitle{\apjs}
\bvolume{237}(\bissue{2}),
\bfpage{31}
(\byear{2018})
\doiurl{10.3847/1538-4365/aace58}
{\href{https://arxiv.org/abs/1710.02894}{{arXiv:1710.02894}}}
{[astro-ph.CO]}
\end{barticle}
\endbibitem

\bibitem[\protect\citeauthoryear{{Hayashino} et~al.}{2019}]{Hayashino19}
\begin{barticle}
\bauthor{\bsnm{{Hayashino}}, \binits{T.}},
\bauthor{\bsnm{{Inoue}}, \binits{A.K.}},
\bauthor{\bsnm{{Kousai}}, \binits{K.}},
\bauthor{\bsnm{{Kashikawa}}, \binits{N.}},
\bauthor{\bsnm{{Mawatari}}, \binits{K.}},
\bauthor{\bsnm{{Matsuda}}, \binits{Y.}},
\bauthor{\bsnm{{Tejos}}, \binits{N.}},
\bauthor{\bsnm{{Prochaska}}, \binits{J.X.}},
\bauthor{\bsnm{{Iwata}}, \binits{I.}},
\bauthor{\bsnm{{Noll}}, \binits{S.}},
\bauthor{\bsnm{{Burgarella}}, \binits{D.}},
\bauthor{\bsnm{{Yamada}}, \binits{T.}},
\bauthor{\bsnm{{Akiyama}}, \binits{M.}}:
\batitle{{Enhancement of H I absorption associated with the z = 3.1 large-scale proto-cluster and characteristic structures with AGNs sculptured over Gpc scale in the SSA22 field}}.
\bjtitle{\mnras}
\bvolume{484}(\bissue{4}),
\bfpage{5868}--\blpage{5887}
(\byear{2019})
\doiurl{10.1093/mnras/stz388}
{\href{https://arxiv.org/abs/1901.11242}{{arXiv:1901.11242}}}
{[astro-ph.GA]}
\end{barticle}
\endbibitem

\bibitem[\protect\citeauthoryear{{Newman} et~al.}{2022}]{Newman22}
\begin{barticle}
\bauthor{\bsnm{{Newman}}, \binits{A.B.}},
\bauthor{\bsnm{{Rudie}}, \binits{G.C.}},
\bauthor{\bsnm{{Blanc}}, \binits{G.A.}},
\bauthor{\bsnm{{Qezlou}}, \binits{M.}},
\bauthor{\bsnm{{Bird}}, \binits{S.}},
\bauthor{\bsnm{{Kelson}}, \binits{D.D.}},
\bauthor{\bsnm{{P{\'e}rez}}, \binits{V.}},
\bauthor{\bsnm{{Congiu}}, \binits{E.}},
\bauthor{\bsnm{{Lemaux}}, \binits{B.C.}},
\bauthor{\bsnm{{Dressler}}, \binits{A.}},
\bauthor{\bsnm{{Mulchaey}}, \binits{J.S.}}:
\batitle{{A population of ultraviolet-dim protoclusters detected in absorption}}.
\bjtitle{\nat}
\bvolume{606}(\bissue{7914}),
\bfpage{475}--\blpage{478}
(\byear{2022})
\doiurl{10.1038/s41586-022-04681-6}
{\href{https://arxiv.org/abs/2206.07661}{{arXiv:2206.07661}}}
{[astro-ph.GA]}
\end{barticle}
\endbibitem

\bibitem[\protect\citeauthoryear{{Cameron} et~al.}{2023}]{Cameron23}
\begin{botherref}
\oauthor{\bsnm{{Cameron}}, \binits{A.J.}},
\oauthor{\bsnm{{Katz}}, \binits{H.}},
\oauthor{\bsnm{{Witten}}, \binits{C.}},
\oauthor{\bsnm{{Saxena}}, \binits{A.}},
\oauthor{\bsnm{{Laporte}}, \binits{N.}},
\oauthor{\bsnm{{Bunker}}, \binits{A.J.}}:
{Nebular dominated galaxies: insights into the stellar initial mass function at high redshift}.
arXiv e-prints,
2311--02051
(2023)
\doiurl{10.48550/arXiv.2311.02051}
{\href{https://arxiv.org/abs/2311.02051}{{arXiv:2311.02051}}}
{[astro-ph.GA]}
\end{botherref}
\endbibitem

\bibitem[\protect\citeauthoryear{{Chen} et~al.}{2024}]{Chen24}
\begin{barticle}
\bauthor{\bsnm{{Chen}}, \binits{Z.}},
\bauthor{\bsnm{{Stark}}, \binits{D.P.}},
\bauthor{\bsnm{{Mason}}, \binits{C.}},
\bauthor{\bsnm{{Topping}}, \binits{M.W.}},
\bauthor{\bsnm{{Whitler}}, \binits{L.}},
\bauthor{\bsnm{{Tang}}, \binits{M.}},
\bauthor{\bsnm{{Endsley}}, \binits{R.}},
\bauthor{\bsnm{{Charlot}}, \binits{S.}}:
\batitle{{JWST spectroscopy of z 5-8 UV-selected galaxies: new constraints on the evolution of the Ly {\ensuremath{\alpha}} escape fraction in the reionization era}}.
\bjtitle{\mnras}
\bvolume{528}(\bissue{4}),
\bfpage{7052}--\blpage{7075}
(\byear{2024})
\doiurl{10.1093/mnras/stae455}
{\href{https://arxiv.org/abs/2311.13683}{{arXiv:2311.13683}}}
{[astro-ph.GA]}
\end{barticle}
\endbibitem

\bibitem[\protect\citeauthoryear{{Li} et~al.}{2024}]{Li24}
\begin{botherref}
\oauthor{\bsnm{{Li}}, \binits{Q.}},
\oauthor{\bsnm{{Conselice}}, \binits{C.J.}},
\oauthor{\bsnm{{Sarron}}, \binits{F.}},
\oauthor{\bsnm{{Harvey}}, \binits{T.}},
\oauthor{\bsnm{{Austin}}, \binits{D.}},
\oauthor{\bsnm{{Adams}}, \binits{N.}},
\oauthor{\bsnm{{Trussler}}, \binits{J.A.A.}},
\oauthor{\bsnm{{Duan}}, \binits{Q.}},
\oauthor{\bsnm{{Ferreira}}, \binits{L.}},
\oauthor{\bsnm{{Westcott}}, \binits{L.}},
\oauthor{\bsnm{{Harris}}, \binits{H.}},
\oauthor{\bsnm{{Dole}}, \binits{H.}},
\oauthor{\bsnm{{Grogin}}, \binits{N.A.}},
\oauthor{\bsnm{{Frye}}, \binits{B.}},
\oauthor{\bsnm{{Koekemoer}}, \binits{A.M.}},
\oauthor{\bsnm{{Robertson}}, \binits{C.}},
\oauthor{\bsnm{{Windhorst}}, \binits{R.A.}},
\oauthor{\bsnm{{Polletta}}, \binits{M.d.C.}},
\oauthor{\bsnm{{Hathi}}, \binits{N.P.}}:
{EPOCHS Paper X: Environmental effects on Galaxy Formation and Protocluster Galaxy candidates at $4.5<z<10$ from JWST observations}.
arXiv e-prints,
2405--17359
(2024)
\doiurl{10.48550/arXiv.2405.17359}
{\href{https://arxiv.org/abs/2405.17359}{{arXiv:2405.17359}}}
{[astro-ph.GA]}
\end{botherref}
\endbibitem

\bibitem[\protect\citeauthoryear{{Henden} et~al.}{2018}]{Henden18}
\begin{barticle}
\bauthor{\bsnm{{Henden}}, \binits{N.A.}},
\bauthor{\bsnm{{Puchwein}}, \binits{E.}},
\bauthor{\bsnm{{Shen}}, \binits{S.}},
\bauthor{\bsnm{{Sijacki}}, \binits{D.}}:
\batitle{{The FABLE simulations: a feedback model for galaxies, groups, and clusters}}.
\bjtitle{\mnras}
\bvolume{479}(\bissue{4}),
\bfpage{5385}--\blpage{5412}
(\byear{2018})
\doiurl{10.1093/mnras/sty1780}
{\href{https://arxiv.org/abs/1804.05064}{{arXiv:1804.05064}}}
{[astro-ph.GA]}
\end{barticle}
\endbibitem

\bibitem[\protect\citeauthoryear{{Angulo} et~al.}{2012}]{Angulo12}
\begin{barticle}
\bauthor{\bsnm{{Angulo}}, \binits{R.E.}},
\bauthor{\bsnm{{Springel}}, \binits{V.}},
\bauthor{\bsnm{{White}}, \binits{S.D.M.}},
\bauthor{\bsnm{{Jenkins}}, \binits{A.}},
\bauthor{\bsnm{{Baugh}}, \binits{C.M.}},
\bauthor{\bsnm{{Frenk}}, \binits{C.S.}}:
\batitle{{Scaling relations for galaxy clusters in the Millennium-XXL simulation}}.
\bjtitle{\mnras}
\bvolume{426}(\bissue{3}),
\bfpage{2046}--\blpage{2062}
(\byear{2012})
\doiurl{10.1111/j.1365-2966.2012.21830.x}
{\href{https://arxiv.org/abs/1203.3216}{{arXiv:1203.3216}}}
{[astro-ph.CO]}
\end{barticle}
\endbibitem

\bibitem[\protect\citeauthoryear{{Bird} et~al.}{2013}]{Bird2013}
\begin{barticle}
\bauthor{\bsnm{{Bird}}, \binits{S.}},
\bauthor{\bsnm{{Vogelsberger}}, \binits{M.}},
\bauthor{\bsnm{{Sijacki}}, \binits{D.}},
\bauthor{\bsnm{{Zaldarriaga}}, \binits{M.}},
\bauthor{\bsnm{{Springel}}, \binits{V.}},
\bauthor{\bsnm{{Hernquist}}, \binits{L.}}:
\batitle{{Moving-mesh cosmology: properties of neutral hydrogen in absorption}}.
\bjtitle{\mnras}
\bvolume{429}(\bissue{4}),
\bfpage{3341}--\blpage{3352}
(\byear{2013})
\doiurl{10.1093/mnras/sts590}
{\href{https://arxiv.org/abs/1209.2118}{{arXiv:1209.2118}}}
{[astro-ph.CO]}
\end{barticle}
\endbibitem

\bibitem[\protect\citeauthoryear{{Madau} and {Dickinson}}{2014}]{Madau14}
\begin{barticle}
\bauthor{\bsnm{{Madau}}, \binits{P.}},
\bauthor{\bsnm{{Dickinson}}, \binits{M.}}:
\batitle{{Cosmic Star-Formation History}}.
\bjtitle{\araa}
\bvolume{52},
\bfpage{415}--\blpage{486}
(\byear{2014})
\doiurl{10.1146/annurev-astro-081811-125615}
{\href{https://arxiv.org/abs/1403.0007}{{arXiv:1403.0007}}}
{[astro-ph.CO]}
\end{barticle}
\endbibitem

\end{thebibliography}

\end{document}